\documentclass[twoside,twocolumn,9pt]{extarticle}
\usepackage[super,sort&compress,comma]{natbib} 
\usepackage[left=1.5cm, right=1.5cm, top=1.785cm, bottom=2.0cm]{geometry}
\usepackage{balance}
\usepackage{mathptmx}
\usepackage{graphicx}
\usepackage{lastpage}
\usepackage[format=plain,justification=justified,singlelinecheck=false,font={stretch=1.125,small,sf},labelfont=bf,labelsep=space]{caption}
\usepackage{float}
\usepackage{fancyhdr}
\usepackage{fnpos}
\usepackage{silence}
\usepackage{microtype}
\usepackage[english]{babel}
\usepackage{array}
\usepackage{droidsans}
\usepackage{charter}
\usepackage[T1]{fontenc}

\usepackage[dvipsnames]{xcolor}
\usepackage{setspace}
\usepackage[compact]{titlesec}
\usepackage[hyperfootnotes=false]{hyperref}
\usepackage{amssymb}
\input{xref-esi}

\definecolor{cream}{RGB}{222,217,201}

\usepackage{tikz}
\usepackage{pgfplots}
\pgfplotsset{compat=1.18}
\usetikzlibrary{positioning, arrows.meta, calc, shapes.geometric, plotmarks}

\usepackage{subcaption}

\usepackage{amsmath}
\usepackage{cleveref}

\usepackage[version=4]{mhchem}
\usepackage{siunitx}
\usepackage{xspace}

\DeclareSIUnit\molatom{mol\,atom}
\DeclareSIUnit{\atom}{atom}
\DeclareSIUnit{\bohrmagneton}{\ensuremath{\mu_\mathrm{B}}}
\DeclareSIUnit{\states}{states}
\DeclareSIUnit{\ppm}{ppm}
\DeclareSIUnit{\bar}{bar}
\DeclareSIUnit{\angstrom}{\text{\AA}}

\newcommand{\Ehull}{\ensuremath{E_{\mathrm{hull}}}\xspace}
\newcommand{\XCe}{\ensuremath{X_{\ce{Ce}}}\xspace}
\newcommand{\XMn}{\ensuremath{X_{\ce{Mn}}}\xspace}
\newcommand{\Ev}{\ensuremath{E_{\mathrm{v}}}\xspace}
\newcommand{\xCe}{\ensuremath{x_{\ce{Ce}}}\xspace}
\newcommand{\xMn}{\ensuremath{x_{\ce{Mn}}}\xspace}
\newcommand{\xCa}{\ensuremath{x_{\ce{Ca}}}\xspace}
\newcommand{\xTi}{\ensuremath{x_{\ce{Ti}}}\xspace}

\newcommand{\kB}{\ensuremath{k_{\mathrm{B}}}\xspace}
\newcommand{\bb}{\ensuremath{\beta_{\mathrm{b}}}\xspace}
\newcommand{\br}{\ensuremath{\beta_{\mathrm{r}}}\xspace}

\newcommand{\Eb}{\ensuremath{\langle E_{\mathrm{b}} \rangle}\xspace}
\newcommand{\Vr}{\ensuremath{\langle V_{\mathrm{r}} \rangle}\xspace}

\newcommand{\CeIV}{\ce{Ce^4+}\xspace}
\newcommand{\CeIII}{\ce{Ce^3+}\xspace}
\newcommand{\MnIV}{\ce{Mn^4+}\xspace}
\newcommand{\MnIII}{\ce{Mn^3+}\xspace}
\newcommand{\MnII}{\ce{Mn^2+}\xspace}
\newcommand{\TiIV}{\ce{Ti^4+}\xspace}
\newcommand{\TiIII}{\ce{Ti^3+}\xspace}
\newcommand{\Tred}{\ensuremath{T_{\mathrm{red}}}}
\newcommand{\POtwo}{\ensuremath{P_{\ce{O2}}}}
\newcommand{\Tox}{\ensuremath{T_{\mathrm{ox}}}}
\newcommand{\PHtwo}{\ensuremath{P_{\ce{H2}}}}
\newcommand{\PHtwoO}{\ensuremath{P_{\ce{H2O}}}}
\newcommand{\muO}{\ensuremath{\mu_{\ce{O}}}}
\newcommand{\Dd}{\ensuremath{\Delta\delta}\xspace}

\crefname{figure}{Fig.}{Fig.}
\Crefname{figure}{Fig.}{Fig.}
\crefformat{figure}{Fig.~#2#1#3}
\Crefformat{figure}{Fig.~#2#1#3}
\crefrangeformat{figure}{Fig.~#3#1#4--#5#2#6}
\Crefrangeformat{figure}{Fig.~#3#1#4--#5#2#6}
\crefmultiformat{figure}%
  {Fig.~#2#1#3}{ and~#2#1#3}{, #2#1#3}{, and~#2#1#3}
\Crefmultiformat{figure}%
  {Fig.~#2#1#3}{ and~#2#1#3}{, #2#1#3}{, and~#2#1#3}

\crefname{equation}{eqn}{eqn}
\Crefname{equation}{Eqn}{Eqn}
\crefformat{equation}{eqn~#2(#1)#3}
\Crefformat{equation}{Eqn~#2(#1)#3}
\crefrangeformat{equation}{eqn~#3(#1)#4--#5(#2)#6}
\Crefrangeformat{equation}{Eqn~#3(#1)#4--#5(#2)#6}
\crefmultiformat{equation}%
  {eqn~#2(#1)#3}{ and~#2(#1)#3}{, #2(#1)#3}{, and~#2(#1)#3}
\Crefmultiformat{equation}%
  {Eqn~#2(#1)#3}{ and~#2(#1)#3}{, #2(#1)#3}{, and~#2(#1)#3}
\crefrangemultiformat{equation}%
  {eqn~#3(#1)#4--#5(#2)#6}{ and~#3(#1)#4--#5(#2)#6}%
  {, #3(#1)#4--#5(#2)#6}{, and~#3(#1)#4--#5(#2)#6}
\Crefrangemultiformat{equation}%
  {Eqn~#3(#1)#4--#5(#2)#6}{ and~#3(#1)#4--#5(#2)#6}%
  {, #3(#1)#4--#5(#2)#6}{, and~#3(#1)#4--#5(#2)#6}

\crefname{table}{Table}{Tables}
\Crefname{table}{Table}{Tables}
\crefname{scheme}{Scheme}{Schemes}
\Crefname{scheme}{Scheme}{Schemes}
\crefname{section}{Section}{Sections}
\Crefname{section}{Section}{Sections}
\crefname{subsection}{Section}{Sections}
\Crefname{subsection}{Section}{Sections}
\crefname{subsubsection}{Section}{Sections}
\Crefname{subsubsection}{Section}{Sections}
\crefname{appendix}{Appendix}{Appendices}
\Crefname{appendix}{Appendix}{Appendices}

\begin{document}

\pagestyle{fancy}
\thispagestyle{plain}
\fancypagestyle{plain}{
\renewcommand{\headrulewidth}{0pt}
}

\makeFNbottom
\makeatletter
\renewcommand\LARGE{\@setfontsize\LARGE{15pt}{17}}
\renewcommand\Large{\@setfontsize\Large{12pt}{14}}
\renewcommand\large{\@setfontsize\large{10pt}{12}}
\renewcommand\footnotesize{\@setfontsize\footnotesize{7pt}{10}}
\makeatother

\renewcommand{\thefootnote}{\fnsymbol{footnote}}
\renewcommand\footnoterule{\vspace*{1pt}%
\color{cream}\hrule width 3.5in height 0.4pt \color{black}\vspace*{5pt}} 
\setcounter{secnumdepth}{5}

\makeatletter 
\renewcommand\@biblabel[1]{#1}            
\renewcommand\@makefntext[1]%
{\noindent\makebox[0pt][r]{\@thefnmark\,}#1}
\makeatother 
\renewcommand{\figurename}{\small{Fig.}~}
\titleformat{\section}{\sffamily\Large\bfseries}{\thesection}{1em}{}
\titleformat{\subsection}{\normalsize\bfseries}{\thesubsection}{1em}{}
\titleformat{\subsubsection}{\normalsize\bfseries}{\thesubsubsection}{1em}{}
\setstretch{1.125} 
\setlength{\skip\footins}{0.8cm}
\setlength{\footnotesep}{0.25cm}
\setlength{\jot}{10pt}
\titlespacing*{\section}{0pt}{4pt}{4pt}
\titlespacing*{\subsection}{0pt}{15pt}{1pt}

\fancyfoot{}
\fancyfoot[RO]{\footnotesize{\sffamily{1--\pageref{LastPage} ~\textbar  \hspace{2pt}\thepage}}}
\fancyfoot[LE]{\footnotesize{\sffamily{\thepage~\textbar\hspace{3.45cm} 1--\pageref{LastPage}}}}
\fancyhead{}
\renewcommand{\headrulewidth}{0pt} 
\renewcommand{\footrulewidth}{0pt}
\setlength{\arrayrulewidth}{1pt}
\setlength{\columnsep}{6.5mm}
\setlength\bibsep{1pt plus 0pt minus 0.5pt}
\Urlmuskip=0mu plus 1mu

\makeatletter 
\newlength{\figrulesep} 
\setlength{\figrulesep}{0.5\textfloatsep} 

\newcommand{\topfigrule}{\vspace*{-1pt}%
\noindent{\color{cream}\rule[-\figrulesep]{\columnwidth}{1.5pt}} }

\newcommand{\botfigrule}{\vspace*{-2pt}%
\noindent{\color{cream}\rule[\figrulesep]{\columnwidth}{1.5pt}} }

\newcommand{\dblfigrule}{\vspace*{-1pt}%
\noindent{\color{cream}\rule[-\figrulesep]{\textwidth}{1.5pt}} }

\makeatother

\twocolumn[
  \begin{@twocolumnfalse}
\vspace{1em}
\sffamily
\begin{tabular}{m{4.5cm} p{13.5cm} }

 & \noindent\LARGE{\textbf{Local B-site chemistry controls oxygen-vacancy energetics in Ca--Ce--Ti--Mn perovskites for thermochemical hydrogen production$^\dag$}} \\
\vspace{0.3cm} & \vspace{0.3cm} \\

& \noindent\large{Manish Kumar,$^{\ddag}$\textit{$^{a}$} Natalia Ali,$^{\ddag}$\textit{$^{b}$} Matthew D.\ Witman,\textit{$^{c}$} Shang Zhai,\textit{$^{d,e}$} James E.\ Miller,\textit{$^{f}$} Ivan Ermanoski,\textit{$^{f}$} Ellen B.\ Stechel,\textit{$^{g}$} and Robert B.\ Wexler$^{\ast}$\textit{$^{a}$}} \\

 & \noindent\normalsize{
Two-step thermochemical water splitting driven by concentrated solar heat offers a scalable route to renewable hydrogen, but practical deployment requires oxide materials whose oxygen-vacancy formation energies balance facile reduction with favorable reoxidation.
Perovskite solid solutions provide compositional flexibility to tune this balance, yet the relationship between bulk stoichiometry and local defect energetics remains poorly understood.
Here we map oxygen-vacancy formation energetics across Ca--Ce--Ti--Mn (CCTM) perovskites using first-principles calculations combined with two complementary fitted models.
A coverage-constrained special quasirandom structure approach is designed to sample all fifteen symmetry-distinct oxygen nearest-neighbor environments.
We develop a crystal-feature model whose fitted coefficients directly encode the underlying Born--Haber thermochemistry, and we fine-tune a defect graph neural network to model potential nonlinear structure--composition--defect coupling beyond the linear form of the crystal-feature model.
Local B-site chemistry dominates the oxygen-vacancy formation energy (\Ev): varying the nearest-neighbor Mn fraction (B site) shifts \Ev by \qtyrange[range-phrase = --, range-units = single]{1.0}{1.5}{\eV} depending on local Ce content, whereas A-site Ce variation contributes a smaller, Mn-dependent shift of \qtyrange[range-phrase = --, range-units = single]{0.2}{0.6}{\eV}.
This hierarchy indicates that short-range B-site cation order, if it can be established and kinetically retained through targeted processing, is a candidate means of tuning redox performance without changing bulk composition.
Composition-space maps show a Ce/Mn-balanced region (approximately \qtyrange[range-phrase = --, range-units = single]{29}{33}{\percent} Ce on the A site and \qtyrange[range-phrase = --, range-units = single]{58}{67}{\percent} Mn on the B site) that combines a high fraction of vacancy sites within the targeted \Ev window with phase stability and solubility, where the predicted redox cycle capacity \Dd is comparable to or exceeds the ceria benchmark at substantially lower reduction temperatures ($\Tred = \qty{1350}{\degreeCelsius}$~vs.~\qty{\sim 1600}{\degreeCelsius} for ceria).
Experimental measurements on three CCTM compositions show \Dd increasing monotonically with Ce content under cycling protocols close to the model conditions, consistent with the qualitative trend from the fitted models.
The interpretable framework provides both compositional screening capability and mechanistic design rules expected to transfer to related perovskite families for solar thermochemical hydrogen production.
} \\

\end{tabular}

 \end{@twocolumnfalse} \vspace{0.6cm}

  ]

\renewcommand*\rmdefault{bch}\normalfont\upshape
\rmfamily
\section*{}
\vspace{-1cm}


\footnotetext{\textit{$^{a}$~Department of Chemistry and Institute of Materials Science and Engineering, Washington University in St.\ Louis, St.\ Louis, MO 63130, USA. E-mail: wexler@wustl.edu}}
\footnotetext{\textit{$^{b}$~ASU LightWorks\textregistered{} and the School for Engineering of Matter, Transport and Energy, Arizona State University, Tempe, AZ 85287-8204, USA.}}
\footnotetext{\textit{$^{c}$~Sandia National Laboratories, Livermore, CA 94550, USA.}}
\footnotetext{\textit{$^{d}$~Department of Mechanical and Aerospace Engineering, The Ohio State University, Columbus, OH 43210, USA.}}
\footnotetext{\textit{$^{e}$~School of Earth Sciences, The Ohio State University, Columbus, OH 43210, USA.}}
\footnotetext{\textit{$^{f}$~ASU LightWorks\textregistered{} and the School of Sustainability, Arizona State University, Tempe, AZ 85287-8204, USA.}}
\footnotetext{\textit{$^{g}$~ASU LightWorks\textregistered{} and the School of Molecular Sciences, Arizona State University, Tempe, AZ 85287-8204, USA.}}

\footnotetext{\ddag~These authors contributed equally to this work.}
\footnotetext{\dag~Electronic Supplementary Information (ESI) available: magnetic moments and oxidation-state assignments; density of states and metallic character; crystal-feature-model parity; the vacancy--vacancy electrostatic upper bound; powder X-ray diffraction; thermogravimetric analysis; the finite-temperature thermodynamic formalism with a worked example; the defect graph neural network model; configurational sensitivity of bulk energetics under SQS sampling; Brillouin-zone sampling convergence; and the atomic-property values used in the crystal-feature-model descriptors.}


\section*{Introduction}

Thermochemical water splitting driven by concentrated solar heat offers a renewable route to hydrogen production that can help decarbonize transportation, manufacturing, and chemical synthesis.\cite{tran_updated_2024, wexler_materials_2023}
In the canonical two-step redox cycle, a metal oxide is thermally reduced at high temperature, then reoxidized with steam at a lower temperature to produce hydrogen.\cite{wexler_materials_2023, chueh_highflux_2010} For the oxides considered here, which accommodate oxygen off-stoichiometry, thermal reduction proceeds by oxygen-vacancy formation.
Ceria remains the benchmark material: it exhibits fast kinetics, stable oxygen-exchange capacity over repeated cycling, and a favorable reduction entropy that lowers the temperature at which thermal reduction becomes thermodynamically favorable.\cite{wexler_materials_2023, chueh_highflux_2010, lany_chemical_2024, panlener_thermodynamic_1975} This favorable reduction entropy is largely electronic in origin: the onsite electronic configurational entropy of reduction is largest for the \CeIV/\CeIII\ couple among the $f$ elements.\cite{naghavi_giant_2017} In addition, the charged (ionized) nature of ceria's oxygen-vacancy defects underlies its ability to sustain hydrogen production even at high \ce{H2}/\ce{H2O} ratios, the most demanding reoxidation regime for practical operation, whereas many candidate oxides are limited to far more dilute ratios.\cite{lany_chemical_2024}
However, ceria's high reduction enthalpy requires operating temperatures above \qty{1500}{\degreeCelsius}, imposing stringent constraints on reactor design and limiting cost competitiveness.\cite{wexler_materials_2023, lany_chemical_2024, ma_analysis_2019, budama_potential_2023, cheng_hydrogen_2021, ma_system_2022}
Alternative materials that reduce at lower temperatures while maintaining comparable hydrogen yields are therefore essential for practical deployment.\cite{millican_redox_2022}

Perovskite oxides (\ce{ABO_{3-\delta}}) have emerged as promising alternatives because their compositional flexibility enables systematic tuning of vacancy formation energetics.\cite{tran_updated_2024, wexler_materials_2023, liu_perovskite_2024}
McDaniel et al.\ introduced the \ce{LaAlO3}-derived series \ce{(Sr_xLa_{1-x})(Mn_yAl_{1-y})O_{3-\delta}} (SLMA), in which \ce{Sr^2+} substitution on the A site introduces charge imbalance that promotes vacancy formation, while Mn on the B site provides multivalent redox centers (\MnIV/\MnIII/\MnII) for reversible oxygen exchange.\cite{mcdanielSrMndopedLaAlO3d2013}
Subsequent work expanded this design space through A-site entropy engineering,\cite{liu_manganesebased_2023} size-mismatch and disorder effects,\cite{arzumanyan_structural_2025} and B-site substitution with Cr, Co, Ni, Fe, or Ga.\cite{zhang_unusual_2022, fernandescauduro_stabilization_2024, xu_local_2024, morelock_computationally_2024, mccord_solar_2024, mccord_thermodynamic_2025, tran_new_2025, wang_lamno3_2023, zhang_compositionally_2023}
Complementary processing innovations, including foamed morphologies, have improved heat and mass transport.\cite{mccord_assessment_2023}

A complementary design strategy uses mixed Ce--Mn B-site occupancy in \ce{BaCe_xMn_{1-x}O_{3-\delta}} (BCM).\cite{barcellos_bace_2018}
The large \ce{Ba^2+} cation stabilizes a lattice that tolerates significant oxygen off-stoichiometry, while the vacancy formation energy, set jointly by the \CeIV/\CeIII and \MnIV/\MnIII redox couples, falls within the range required for efficient cycling.
The \ce{Ba(Ce_{0.25}Mn_{0.75})O_{3-\delta}} composition reduces below \qty{1400}{\degreeCelsius} with fast kinetics, and Barcellos et al.\ report approximately three times the hydrogen yield of ceria when both are reduced at \qty{1350}{\degreeCelsius}, a temperature below ceria's favorable reduction window.\cite{barcellos_bace_2018, roychoudhury_investigating_2023}
More recently, Perry et al.\ combined machine learning with density functional theory (DFT) to design \ce{Ba_{0.875}Ca_{0.125}Zr_{0.875}Mn_{0.125}O_{3-\delta}}, predicted by thermodynamic system-model simulations to reduce at a maximum temperature of \qty{1523}{\degreeCelsius} versus \qty{1734}{\degreeCelsius} for ceria.\cite{perry_discovery_2025}

Among these materials, Ca--Ce--Ti--Mn (CCTM) perovskites have been identified as promising candidates.
Initial DFT calculations predicted favorable vacancy energetics with redox activity on both Ce and Mn,\cite{saigautam_exploring_2020} and subsequent work demonstrated that Ti substitution enhances phase stability while placing vacancy formation energies near the target \qtyrange[range-phrase = --, range-units = single]{3.4}{3.9}{\eV} range established in our earlier thermodynamic analysis.\cite{wexler_multiple_2023, wexler_materials_2023}
Experimental studies on the reference composition \ce{Ca2/3Ce1/3Ti1/3Mn2/3O3} (CCTM2112) confirmed hydrogen productivities of \qty[per-mode = symbol, qualifier-mode = phrase, qualifier-phrase = \,]{\sim 10}{\mmol\of{\ce{H2}}\per\molatom\of{CCTM2112}} per cycle at \qty{1350}{\degreeCelsius}, with rapid and reversible kinetics.\cite{wexler_multiple_2023}
Ce reduction on the A site, in addition to conventional B-site Mn reduction, was identified as a contributor to the redox activity, providing a design variable absent in B-site-only Ce perovskites such as BCM.\cite{wexler_multiple_2023}

Despite this progress, the relationship between bulk stoichiometry and local defect energetics in CCTM perovskites remains poorly understood, limiting rational optimization.
Prior computational studies examined single compositions using representative supercells that do not systematically sample the diversity of local oxygen environments present in disordered solid solutions.\cite{saigautam_exploring_2020, wexler_multiple_2023}
Complementary high-throughput first-principles screening has mapped oxygen-vacancy formation energies and phase stability across large perovskite and oxide sets, generally without resolving the distribution of distinct local oxygen environments within a disordered composition.\cite{emery_highthroughput_2016, kumagai_insights_2021a, baldassarri_oxygen_2023}
Furthermore, although machine-learning models have been applied to screen perovskite compositions,\cite{wexler_factors_2021, witman_defect_2023, witman_metal_2026} many of these models lack physical transparency, obscuring the factors that control vacancy formation.
A predictive framework connecting local cation chemistry to macroscopic redox performance through interpretable descriptors would enable both rational composition optimization and transferable design rules for related perovskite families.

The principal advance of this work is the systematic resolution of all fifteen symmetry-distinct oxygen nearest-neighbor environments in the CCTM solid solution and the demonstration that local B-site chemistry, rather than bulk stoichiometry, dominates oxygen-vacancy energetics; the composition-space screening that follows is an application of this framework rather than its central result. To establish this, we combine first-principles defect thermochemistry with two complementary fitted models to map oxygen-vacancy energetics across the CCTM composition space.
To ensure complete sampling of local chemical diversity, we introduce a coverage-constrained special quasirandom structure (SQS) methodology in which each supercell realizes all fifteen symmetry-distinct oxygen nearest-neighbor environments.
We develop an interpretable crystal-feature model (CFM) whose fitted coefficients directly encode the underlying Born--Haber thermochemistry, and we benchmark these predictions against a retrained defect graph neural network (dGNN).

This analysis shows that local B-site chemistry dominates vacancy energetics: varying the Mn fraction among nearest-neighbor B sites shifts \Ev by \qtyrange[range-phrase = --, range-units = single]{1.0}{1.5}{\eV} depending on local Ce content, whereas A-site Ce variation contributes a smaller, Mn-dependent shift of \qtyrange[range-phrase = --, range-units = single]{0.2}{0.6}{\eV}; this hierarchy indicates that short-range B-site cation order, if it can be kinetically retained, is a candidate means of tuning redox performance without changing bulk composition.
Combining the in-window \Ev fraction with phase stability, we identify \ce{Ca_{0.67}Ce_{0.33}Ti_{0.33}Mn_{0.67}O3} (CCTM2112) and \ce{Ca_{0.71}Ce_{0.29}Ti_{0.42}Mn_{0.58}O3} as the most promising candidates among the six compositions examined (\cref{tbl:cctm_compositions}), with predicted cycle capacities comparable to or exceeding ceria at substantially lower reduction temperatures; experimental measurements on three compositions are consistent with these predictions.
The interpretable framework provides both compositional screening capability and mechanistic design rules expected to transfer to related perovskite families.

\section*{Results\label{sec:results}}

\subsection*{Compositional design space}

\begin{figure*}
    \centering
    \includegraphics{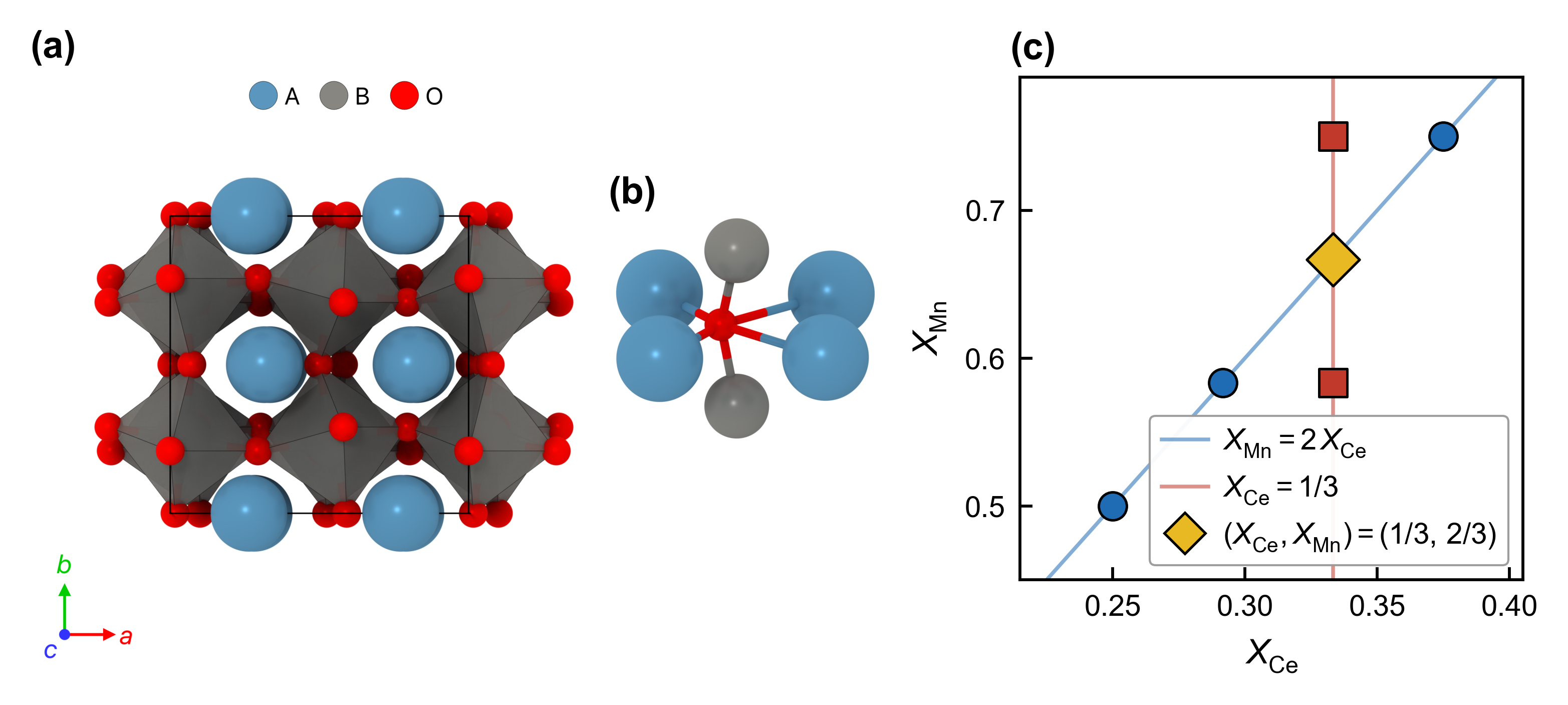}
    \caption{
(a) The \ce{ABO3} perovskite host structure common to the CCTM compositions (\ce{CaTiO3} cell, [001] projection), with A-site cations (blue) between corner-sharing \ce{BO6} octahedra (B-site cations gray, O red).
(b) Zoom on the local nearest-neighbor environment of a representative oxygen site (four A-site and two B-site coordinating cations).
(c) Composition space explored, plotted as \XMn~vs.~\XCe.
The blue line marks the charge-compensation condition $\XMn = 2\XCe$; the red line fixes $\XCe = 1/3$.
Blue circles and red squares denote the two compositional series, and the gold diamond marks their intersection, the reference composition CCTM2112.
Compositions with $\XCe > 1/3$ lie beyond the empirical Ce solubility limit, where \ce{CeO2} exsolution is expected.\cite{dukic_crystal_2009, naik_cationdeficient_2023}
    }
    \label{fgr:1}
\end{figure*}

Maximizing hydrogen yield in thermochemical water splitting requires oxygen-vacancy formation energies, \Ev, that are low enough to permit thermal reduction yet high enough to drive steam reoxidation, together with a phase-stable host that withstands repeated cycling.
Guided by the favorable hydrogen productivity of CCTM2112 (\qty[per-mode = symbol, qualifier-mode = phrase, qualifier-phrase = \ ]{\sim 10}{\mmol\of{\ce{H2}}\per\molatom\of{CCTM2112}} per cycle at \qty{1350}{\degreeCelsius}; reduction and reoxidation conditions comparable to those in \cref{tbl:cctm_delta_delta}),\cite{wexler_multiple_2023} we examined six Ca--Ce--Ti--Mn perovskites (host structure in \cref{fgr:1}a; local oxygen coordination in \cref{fgr:1}b) in which only the A-site Ce fraction and B-site Mn fraction vary, with Ca retained as the majority A-site cation (\cref{tbl:cctm_compositions}).
The Ce mole fraction \XCe spans \numrange[range-phrase = --]{0.25}{0.37}, bracketing the empirically observed solubility limit ($\sim 1/3$) above which \ce{CeO2} segregation is expected.\cite{dukic_crystal_2009, naik_cationdeficient_2023, wexler_multiple_2023}
Four compositions lie along the line $\XMn = 2\XCe$ (\cref{fgr:1}c); this relation enforces nominal charge compensation between Ce and Mn (each \CeIV\ substituting for \ce{Ca^2+} on the A site is balanced by two \MnIII\ substituting for \TiIV\ on the B site) and enables a test of whether maintaining this balance promotes reversible oxygen exchange.
Two additional compositions hold $\XCe = 1/3$ while deviating from the \num{2}:\num{1} line toward Mn-poor ($\XMn = 0.58$) and Mn-rich ($\XMn = 0.75$) limits, decoupling A-site and B-site effects.
Together, these six compositions enable a systematic assessment of how bulk stoichiometry shapes both thermodynamic stability (\Ehull) and the distribution of local vacancy formation energies.

\begin{table*}
\small
  \caption{
CCTM compositions examined in this work.
\XCe and \XMn denote A-site Ce and B-site Mn mole fractions, respectively.
\Ehull is the zero-kelvin energy above the convex hull (\unit{\meV\per\atom}).
Calculated oxidation states for Ce and Mn are assignments from SCAN+$U$ local magnetic moments (ESI \cref{S-sec:esi:magmoments}), not measurements; the integer labels classify calculated moments, and Ce sites labeled 3 have fractional $4f$ occupation (\CeIII-like).
}
  \label{tbl:cctm_compositions}
  \begin{tabular*}{\textwidth}{@{\extracolsep{\fill}}lllllll}
    \hline
    & \multicolumn{2}{l}{Mole fractions} & & & \multicolumn{2}{l}{Calculated oxidation states} \\
    \cline{2-3} \cline{6-7}
    CCTM chemical formula & \XCe & \XMn & $\XCe/\XMn$ & \Ehull (\unit{\meV\per\atom}) & Ce & Mn \\
    \hline
    \ce{Ca_{0.75}Ce_{0.25}Ti_{0.50}Mn_{0.50}O3} & 0.25 & 0.50 & 1/2 & 18 & 3 & 4, 3 \\
    \ce{Ca_{0.71}Ce_{0.29}Ti_{0.42}Mn_{0.58}O3} & 0.29 & 0.58 & 1/2 & 26 & 3 & 4, 3 \\
    \ce{Ca_{0.67}Ce_{0.33}Ti_{0.33}Mn_{0.67}O3} & 0.33 & 0.67 & 1/2 & 33 & 4, 3 & 4, 3 \\
    \ce{Ca_{0.63}Ce_{0.37}Ti_{0.25}Mn_{0.75}O3} & 0.37 & 0.75 & 1/2 & 39 & 3 & 4, 3 \\
    \ce{Ca_{0.67}Ce_{0.33}Ti_{0.42}Mn_{0.58}O3} & 0.33 & 0.58 & $>$1/2 & 31 & 3 & 4, 3 \\
    \ce{Ca_{0.67}Ce_{0.33}Ti_{0.25}Mn_{0.75}O3} & 0.33 & 0.75 & $<$1/2 & 35 & 3 & 4, 3, 2 \\
    \hline
  \end{tabular*}
\end{table*}

\subsection*{Phase stability and vacancy energetics}

\begin{figure}[ht]
    \centering
    \includegraphics[width=8.3cm]{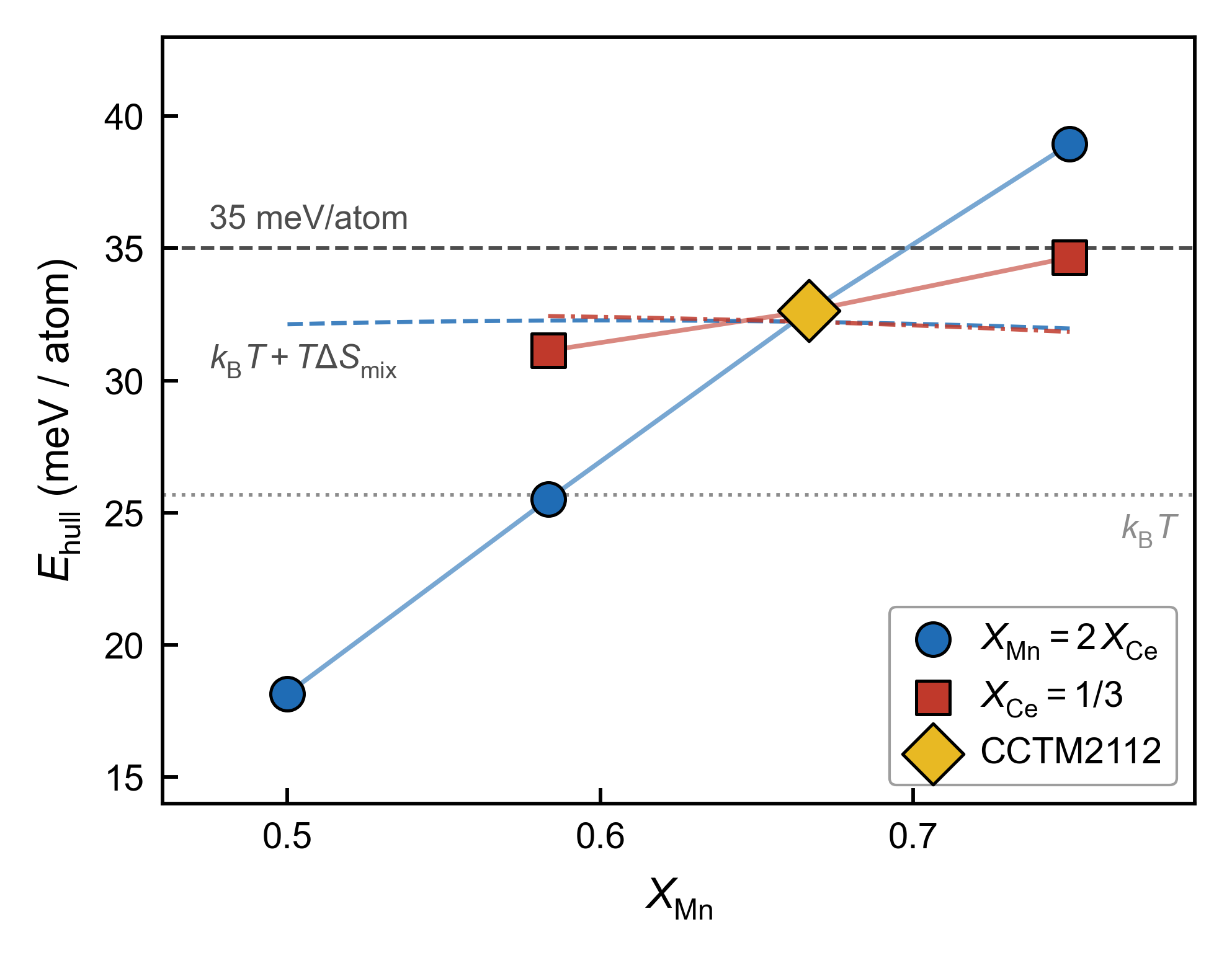}
    \caption{
Zero-kelvin energies above the convex hull (\Ehull) for two CCTM compositional series: $\XMn = 2\XCe$ (charge-compensation line, blue circles) and $\XCe = 1/3$ (constant-Ce line, red squares); the gold diamond marks CCTM2112.
Markers indicate \qty{0}{\kelvin} \Ehull in \unit{\meV\per\atom}.
The dotted horizontal line marks the vibrational-entropy scale ($\kB T$) at \qty{298}{\kelvin}, the dashed horizontal line the empirical \qty{35}{\meV\per\atom} synthesizability threshold, and the blue dashed and red dash-dotted curves the combined vibrational and configurational-mixing offset ($\kB T + T\Delta S_{\mathrm{mix}}$) at \qty{298}{\kelvin} for each series (the two offset curves nearly coincide over the plotted range). These room-temperature reference scales are exceeded by several of the \Ehull values; at the much higher synthesis temperature, the combined entropic offset is substantially larger and exceeds the \Ehull values of all six compositions, indicating thermal accessibility.
    }
    \label{fgr:2}
\end{figure}

Phase stability under repeated redox cycling is essential for practical thermochemical hydrogen production.
We assess intrinsic stability using the zero-kelvin energy above the convex hull, \Ehull, for each composition (\cref{tbl:cctm_compositions}).
All six values lie within \qtyrange[range-phrase = --, range-units = single]{18}{39}{\meV\per\atom}, comparable to the metastability range typical of experimentally observed complex oxides;\cite{sun_thermodynamic_2016} CCTM2112, in the upper portion of this range at \qty{33}{\meV\per\atom}, has been synthesized in prior work.\cite{wexler_multiple_2023}
Along the charge-compensation line $\XMn = 2\XCe$ (\cref{fgr:2}, blue circles), \Ehull increases monotonically with \XCe, consistent with progressive deviation from the stable \ce{CaTiO3} end member.
At fixed $\XCe = 1/3$ (\cref{fgr:2}, red squares), increasing \XMn from \num{0.58} to \num{0.75} raises \Ehull more modestly (\qtyrange[range-phrase = --, range-units = single]{31}{35}{\meV\per\atom}), indicating that B-site variation alone is less destabilizing than simultaneous A- and B-site changes.

Having established that all six compositions are thermodynamically accessible, we next assess their oxygen-exchange activity through the formation energy of a neutral oxygen vacancy, \Ev.
In these metallic hosts (ESI \cref{S-fgr:esi:dos}), the neutral vacancy is the relevant charge state (see \hyperref[sec:methods]{Methods}).
\Ev accounts for both the cost of removing a neutral oxygen atom to form \ce{ABO_{3-\delta}} and the stabilization gained when the two electrons left on the lattice are accommodated by reducible neighboring cations (equivalently, \ce{O^2-} departs as \ce{1/2 O2}, leaving its two electrons behind).
Reducible couples such as \MnIV/\MnIII lower \Ev by providing favorable electron-accepting states, whereas the less reducible \TiIV/\TiIII raises it; the role of Ce is intermediate and composition-dependent, as discussed below.

\begin{figure}[ht]
    \centering
    \includegraphics[width=8.3cm]{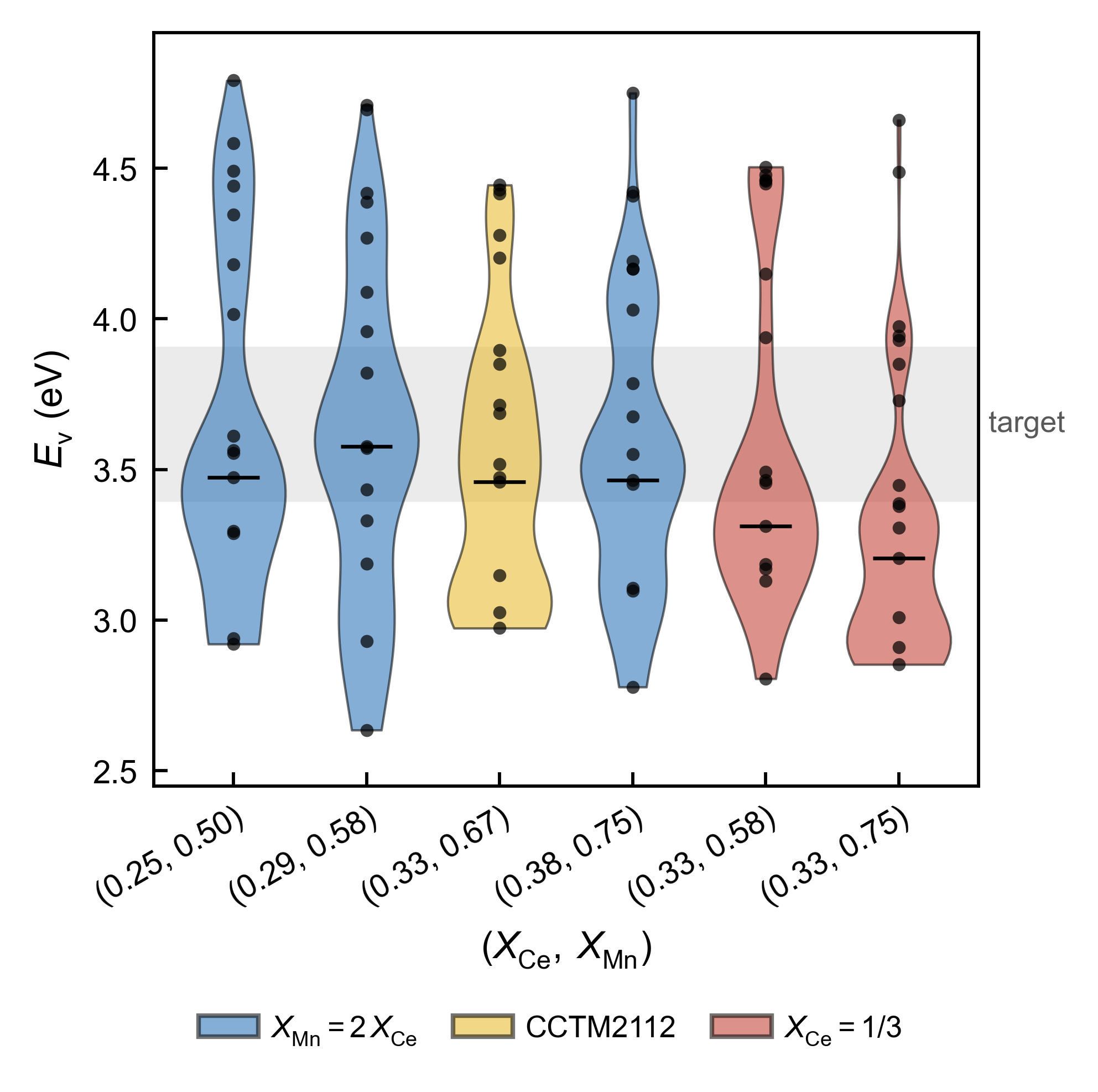}
    \caption{
Count-weighted distributions of oxygen-vacancy formation energies (\Ev) for the six CCTM compositions, shown as violins at each $(\XCe, \XMn)$.
Each violin is a kernel-density estimate over the symmetry-distinct vacancy environments weighted by their multiplicity; black bars mark the medians and black dots overlay the individual cell-mean \Ev values.
Violin fill color denotes the series: charge-compensation ($\XMn = 2\XCe$, blue), constant-Ce ($\XCe = 1/3$, red), and their intersection, CCTM2112 (gold).
The gray shaded band denotes the target \qtyrange[range-phrase = --, range-units = single]{3.4}{3.9}{\eV} window for balanced reduction--reoxidation cycling.
    }
    \label{fgr:3}
\end{figure}

Distributions centered in the \qtyrange[range-phrase = --, range-units = single]{3.4}{3.9}{\eV} range are expected to balance high-temperature reduction with steam reoxidation: the upper bound of the window reflects the requirement that \Ev be low enough for thermal reduction to be accessible, and the lower bound that it remain high enough to retain a thermodynamic driving force for steam reoxidation, the same competition formalized at finite temperature through the reduction and water-splitting oxygen chemical potentials (\cref{eqn:thermal-reduction-muO-split,eqn:water-splitting-muO-split} in \hyperref[sec:methods]{Methods}).\cite{wexler_materials_2023} This \qty{0}{\kelvin} \Ev target window is distinct from the explicit finite-temperature cycle-capacity evaluation, which uses the specific $T$--\POtwo values in \hyperref[sec:methods]{Methods}.
All six CCTM compositions populate this target window to varying extents (\cref{tbl:cctm_ev_stats} and \cref{fgr:3}).
Along the charge-compensation series $\XMn = 2\XCe$, the median \Ev values cluster tightly between \qtylist[list-units = single]{3.46;3.58}{\eV}, and the fraction of sites within the target range peaks at $\XCe = 0.33$, where \qty{49}{\percent} of vacancies fall between \qtylist[list-units = single]{3.4;3.9}{\eV}.
The mean--median difference is at most \qty{0.20}{\eV} across all compositions, indicating only mild skewness.

At fixed $\XCe = 1/3$, varying \XMn shifts both the central tendency and the spread of the distribution (\cref{fgr:3}).
The median rises from \qty{3.31}{\eV} at $\XMn = 0.58$ to \qty{3.46}{\eV} at $\XMn = 0.67$ (maximizing the in-window fraction at \qty{49}{\percent}), then drops to \qty{3.21}{\eV} at $\XMn = 0.75$.
In this Mn-rich limit, \qty{87}{\percent} of vacancies fall below \qty{3.4}{\eV}, indicating overly facile reduction and a diminished thermodynamic driving force for reoxidation.
These trends reflect the competing influences of cation reducibility: higher local Mn content lowers \Ev through the favorable \MnIV/\MnIII couple, while Ti raises \Ev.
An intermediate Mn fraction balances these effects, placing the largest fraction of sites in the desired window.
From a materials-selection perspective, \ce{Ca_{0.67}Ce_{0.33}Ti_{0.33}Mn_{0.67}O3} (\qty{49}{\percent} in-window; $\Ehull = \qty{33}{\meV\per\atom}$) and \ce{Ca_{0.71}Ce_{0.29}Ti_{0.42}Mn_{0.58}O3} (\qty{44}{\percent} in-window; $\Ehull = \qty{26}{\meV\per\atom}$) are the candidates that best combine a high in-window fraction with phase stability.

\begin{table*}
\small
  \caption{
Summary statistics for oxygen-vacancy formation energies (\Ev) across the six CCTM compositions.
Means and medians are computed over all symmetry-distinct vacancy sites.
``$\Ev \leq \qty{3.4}{\eV}$'' and ``$\Ev \leq \qty{3.9}{\eV}$'' columns give the percentage of sites at or below each threshold; ``In target range'' is the percentage with $\num{3.4} \leq \Ev \leq \qty{3.9}{\eV}$.
Percentages rounded to nearest integer.
}
  \label{tbl:cctm_ev_stats}
  \begin{tabular*}{\textwidth}{@{\extracolsep{\fill}}llllll}
    \hline
    & & & \multicolumn{2}{l}{Percent of sites with} & Percent in \\
    \cline{4-5}
    CCTM chemical formula & Mean & Median & $\Ev \leq \qty{3.4}{\eV}$ & $\Ev \leq \qty{3.9}{\eV}$ & target range \\
    \hline
    \ce{Ca_{0.75}Ce_{0.25}Ti_{0.50}Mn_{0.50}O3} & 3.60 & 3.47 & 40 & 72 & 32 \\
    \ce{Ca_{0.71}Ce_{0.29}Ti_{0.42}Mn_{0.58}O3} & 3.55 & 3.58 & 34 & 78 & 44 \\
    \ce{Ca_{0.67}Ce_{0.33}Ti_{0.33}Mn_{0.67}O3} & 3.48 & 3.46 & 40 & 89 & 49 \\
    \ce{Ca_{0.63}Ce_{0.37}Ti_{0.25}Mn_{0.75}O3} & 3.52 & 3.46 & 31 & 75 & 44 \\
    \ce{Ca_{0.67}Ce_{0.33}Ti_{0.42}Mn_{0.58}O3} & 3.51 & 3.31 & 55 & 76 & 21 \\
    \ce{Ca_{0.67}Ce_{0.33}Ti_{0.25}Mn_{0.75}O3} & 3.21 & 3.21 & 87 & 88 & 1 \\
    \hline
  \end{tabular*}
\end{table*}

\subsection*{Local cation control of vacancy energetics}

\begin{figure}[ht]
    \centering
    \includegraphics[width=8.3cm]{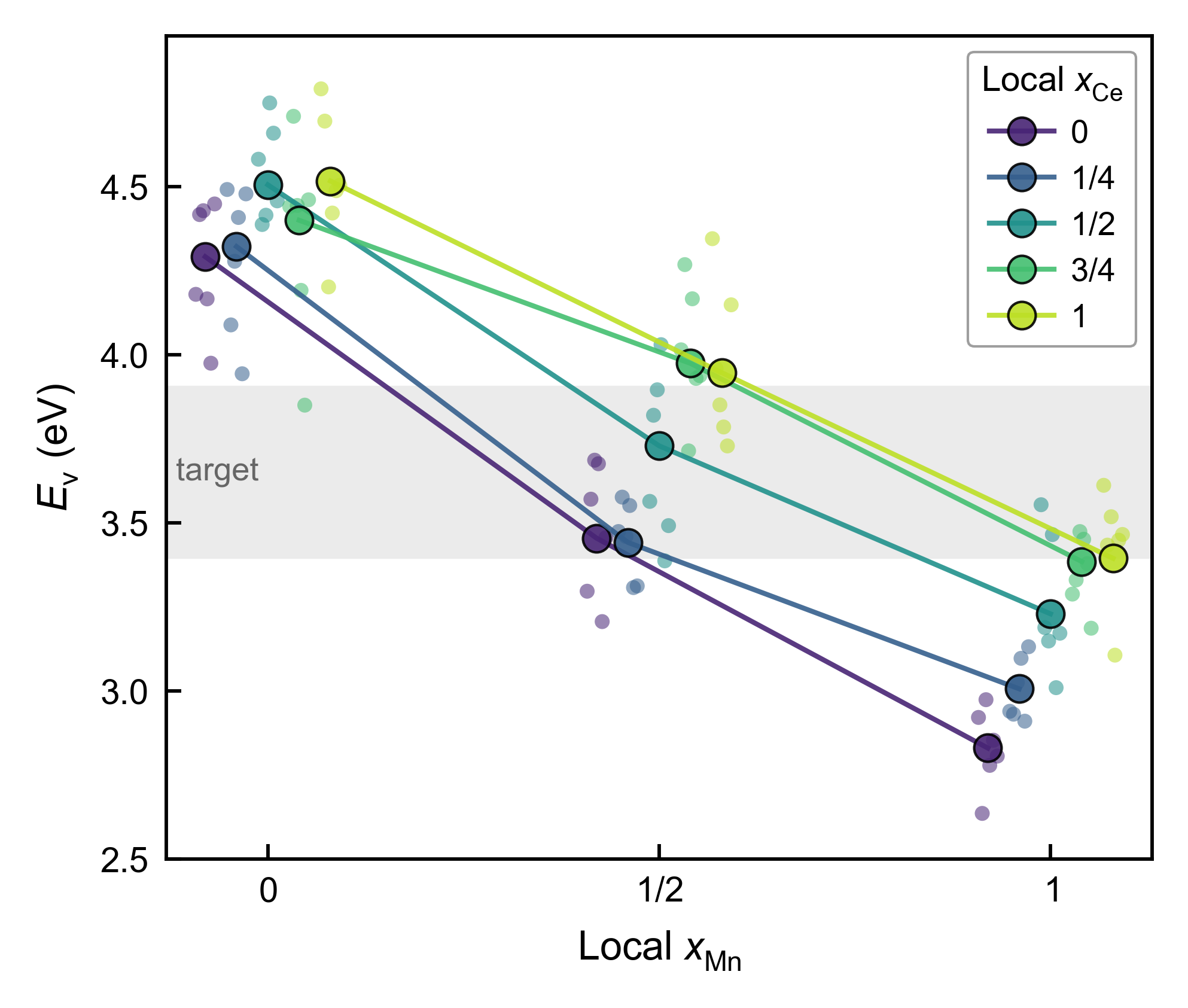}
    \caption{
Local-environment dependence of the oxygen-vacancy formation energy \Ev.
For each local nearest-neighbor composition (\xCe, \xMn), the \Ev values from the six bulk compositions are shown as small markers; larger markers connected by lines give the count-weighted mean within each local-\xCe series. The small markers are offset horizontally for visibility, both to separate the local-\xCe color groups within each \xMn position and to fan out the six bulk values within each cell; the underlying local \xMn is exactly $0$, $1/2$, or $1$, and the horizontal spread carries no information.
Increasing the local Mn fraction lowers \Ev (the dominant B-site effect), while the vertical separation between series at fixed \xMn reflects the weaker A-site modulation with local Ce.
The tight clustering of the six bulk markers within each cell indicates that the local-environment dependence is largely transferable across bulk compositions.
The shaded band marks the target \Ev window (\qtyrange[range-phrase = --, range-units = single]{3.4}{3.9}{\eV}) for solar thermochemical \ce{H2} production.
    }
    \label{fgr:4}
\end{figure}

To connect local cation chemistry with vacancy energetics, we define a vacancy's local composition (\cref{fgr:1}b) in terms of the Ce fraction among its four A-site nearest neighbors (\xCe) and the Mn fraction among its two B-site nearest neighbors (\xMn).
Uppercase symbols (\XCe, \XMn) denote bulk-averaged compositions (\cref{tbl:cctm_compositions}), while lowercase symbols denote local values around a specific vacancy site (\cref{fgr:4}).
Because only two cation types occupy each sublattice, the remaining fractions are fixed by stoichiometry: $\xCa = 1 - \xCe$ and $\xTi = 1 - \xMn$.
This mapping assigns each of the fifteen symmetry-distinct vacancy environments in the SQS supercells to a unique (\xCe, \xMn) pair, enabling direct correlation with the local vacancy formation energy \Ev.

The B-site chemistry is the primary control on \Ev.
Increasing \xMn from \num{0} to \num{1} lowers \Ev by \qty{\sim 1.5}{\eV} in \ce{Ca_{0.71}Ce_{0.29}Ti_{0.42}Mn_{0.58}O3} (\cref{fgr:4}).
The dependence on the local Mn fraction dominates, whereas the dependence on local Ce is weaker and nonmonotonic; varying \xCe shifts \Ev by \qtyrange[range-phrase = --, range-units = single]{0.2}{0.6}{\eV} depending on local Mn content.
The target \qtyrange[range-phrase = --, range-units = single]{3.4}{3.9}{\eV} window is thus accessible via two routes: high local Mn content with high Ce, or moderate Mn with low Ce.

These observations motivate a physics-informed linear model of the form
\begin{equation}
    \Ev = \bb \Eb + \br \Vr + \beta_0,
\end{equation}
where \Eb is the average M--O bond-dissociation energy and \Vr is the average crystal reduction potential (the reduction potential of a cation couple evaluated in the oxide crystal rather than in aqueous solution\cite{wexler_factors_2021}), both computed over all six nearest-neighbor cations (four A-site, two B-site).
The fitted coefficients encode the effective contributions of each term to vacancy formation.
Substituting Mn for Ti weakens M--O bonds and provides a more favorable reduction couple (\MnIV/\MnIII), both of which lower \Ev; replacing Ca with Ce modifies \Vr more modestly, consistent with the smaller and nonmonotonic influence of A-site composition.

A Huber regression (which down-weights outliers) fitted to \num{90} vacancy energies from the strongly constrained and appropriately normed functional with Hubbard $U$ corrections (SCAN+$U$) supports this picture (ESI \cref{S-fgr:cfm-parity}).
The model achieves a mean absolute error of \qty{0.175 \pm 0.015}{\eV} across \num{1000} random \qty{50}{\percent}/\qty{50}{\percent} train--test splits.
This accuracy is comparable to that of earlier linear descriptor models, which predict oxygen-vacancy formation energies in diverse oxides from intrinsic bulk properties (oxide formation enthalpy, midgap energy relative to the O~2p band center, and atomic electronegativity) to within \qty{\sim 0.2}{\eV};\cite{deml_intrinsic_2015} the crystal-feature model attains similar accuracy within the CCTM family using two descriptors chosen to expose the underlying Born--Haber thermochemistry.
The fitted coefficients have a direct physical interpretation.
Although \Eb and \Vr are each averaged over all six nearest neighbors, the coefficient $\bb \approx 2.0$ indicates that approximately two bonds' worth of dissociation energy contributes to \Ev, consistent with the two B-site cations forming direct M--O bonds to the removed oxygen, while the more distant A-site cations contribute primarily through electrostatics.
Similarly, $\br \approx -1.3$ reflects stabilization of the two electrons released upon vacancy formation into redox-active states on neighboring cations.
We evaluated \Ehull as a third descriptor but excluded it from the final model: with only six unique \Ehull values spanning a narrow range, its coefficient is not statistically resolved (the \qty{95}{\percent} confidence interval spans zero), and including it does not improve the cross-validated error (ESI \cref{S-tbl:cfm-subsets}).
This compact model directly encodes the underlying Born--Haber thermochemistry, enabling both rapid screening and mechanistic interpretation.

\subsection*{Cycle-level performance maps}

The preceding section established how local cation chemistry controls \Ev; here, we translate these site-level energetics into cycle-level performance. The local cation environments analyzed above produce a distribution of vacancy formation energies rather than a single characteristic value. For each local environment we compute the equilibrium probability that an oxygen site is vacant under representative reduction and oxidation conditions (\cref{eqn:occupation} in \hyperref[sec:methods]{Methods}); averaging this equilibrium vacancy population over the \Ev distribution gives the oxygen off-stoichiometry $\delta$ (\cref{eqn:delta-avg-main}), and the difference between the reduction and oxidation states yields the cycle capacity, $\Dd = \delta_{\mathrm{red}} - \delta_{\mathrm{ox}}$.
The thermodynamic formalism (\cref{eqn:occupation,eqn:thermal-reduction-muO-split,eqn:water-splitting-equilibrium,eqn:water-splitting-muO-split} in \hyperref[sec:methods]{Methods}) treats vacancies as neutral and noninteracting, an approximation supported by the modest electrostatic coupling estimated in ESI \cref{S-sec:esi:vv-electrostatic}.
We compare predictions from the crystal-feature model (CFM) and the defect graph neural network (dGNN) across the CCTM composition space.

\subsubsection*{Crystal feature model.~~}

\begin{figure*}
    \centering
    \includegraphics[width=\textwidth]{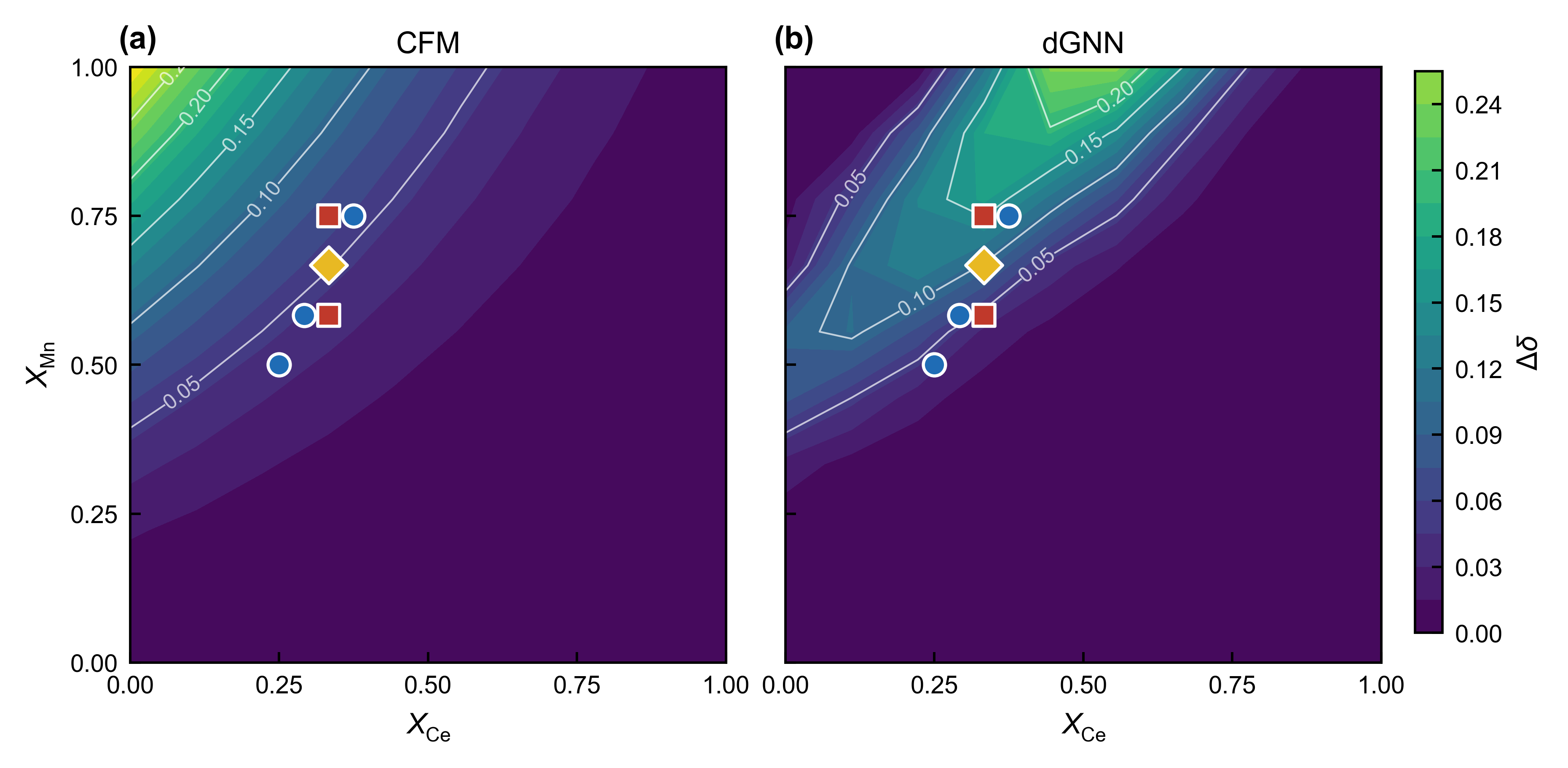}
    \caption{
Predicted oxygen-exchange capacity ($\Dd$) across CCTM composition space from (a) the crystal-feature model (CFM) and (b) the defect graph neural network (dGNN).
Axes denote bulk mole fractions \XCe and \XMn; the color scale indicates $\Dd$, the exchanged oxygen per perovskite formula unit.
White contours are spaced at intervals of \num{0.05} in $\Dd$.
Blue circles: charge-compensation series ($\XMn = 2\XCe$); red squares: constant-Ce series ($\XCe = 1/3$); gold diamond: CCTM2112.
The CFM surface increases monotonically toward the Mn-rich corner, whereas the dGNN surface exhibits a nonmonotonic ridge in mixed Ce--Mn regimes.
}
    \label{fgr:5}
\end{figure*}

The CFM gives a smooth, physically interpretable map of \Ev across composition space.
As shown in \cref{fgr:5}a, the predicted cycle capacity \Dd varies monotonically along the edges of the (\XCe, \XMn) composition square, increasing from the Ce-rich corner ($\XCe = 1$, $\XMn = 0$; \ce{CeTiO3} limit) toward the Mn-rich corner ($\XCe = 0$, $\XMn = 1$; \ce{CaMnO3} limit).
This trend reflects the correlation between local Mn content and the favorable \MnIV/\MnIII reduction couple.
For the cycle conditions employed here ($\Tred = \qty{1350}{\degreeCelsius}$, $\POtwo = \qty[exponent-mode = scientific]{0.0001}{\bar}$; $\Tox = \qty{850}{\degreeCelsius}$, $\PHtwo/\PHtwoO = \num[exponent-mode = scientific]{0.001}$), the predicted cycle capacity $\Dd$ spans \numrange[range-phrase = --]{0}{0.14} per formula unit across the accessible composition range.
For comparison, ceria achieves $\Dd/3 \approx \num{0.010}$ under more aggressive cycling conditions ($\Tred \approx \qty{1600}{\degreeCelsius}$, $\POtwo = \qty[exponent-mode = scientific]{1e-5}{\bar}$; $\Tox \approx \qty{900}{\degreeCelsius}$).\cite{chueh_highflux_2010}
On a per-atom basis, \Dd is normalized by the number of atoms per formula unit: \num{5} for the \ce{ABO3} CCTM perovskites and \num{3} for the \ce{CeO2} fluorite that defines the $\Dd/3$ ceria value above. On this basis, the synthesized CCTM compositions ($\Dd/5 \approx \numrange[range-phrase = \text{--}]{0.007}{0.010}$) approach or exceed this benchmark under our more moderate conditions, suggesting competitive performance with reduced thermal demands.

Practical synthesis constraints narrow the accessible region.
Prior reports indicate that Ce solubility in the Ca--Mn--O perovskite host is limited, with \ce{CeO2} segregating beyond the solubility limit,\cite{dukic_crystal_2009, naik_cationdeficient_2023} and single-phase CCTM has been demonstrated up to $\XCe = 1/3$ by powder X-ray diffraction;\cite{wexler_multiple_2023} our \qty{0}{\kelvin} stability analysis additionally shows that compositions with $\XMn \geq 0.75$ have \Ehull values exceeding that of CCTM2112.
These boundaries exclude a region at high Ce and high Mn on stability and solubility grounds.
Within the remaining domain, the CFM predicts that \Dd is maximized near $\XCe \approx 0$ and $\XMn \approx 2/3$.
Selecting this composition would forgo any potential kinetic or entropic benefits of A-site Ce, a tradeoff that merits experimental investigation.

\subsubsection*{Defect graph neural network.~~}

Predictions from the dGNN are broadly consistent with the CFM but display nonmonotonic behavior along the \ce{CaTiO3}--\ce{CaMnO3} and \ce{CaMnO3}--\ce{CeMnO3} edges (\cref{fgr:5}b).
Because the dGNN requires explicit structural input, we generated SQS supercells at \num{100} grid points across composition space and interpolated between them.
The nonmonotonic features may reflect composition--structure--defect coupling that the linear CFM cannot capture, although some deviations could also arise from sparse sampling in certain composition regions.
Without the stability constraint, the dGNN \Dd maximum lies near $(\XCe, \XMn) \approx (0.5, 1)$.
When the \Ehull and solubility constraints defined above are applied, compositions near $(\XCe, \XMn) \approx (0.22, 0.67)$ are favorable.

The CFM and dGNN give complementary descriptions of the CCTM redox behavior: the CFM represents mean thermodynamic trends with transparent physical interpretation, and the dGNN represents potential higher-order structure--composition--defect interactions.
Composition ranges where the two predictions overlap balance reducibility with phase stability and are prioritized for experiment.

\subsection*{Experimental validation\label{sec:results:experimental:validation}}

\begin{figure*}
    \centering
    \includegraphics[width=\textwidth]{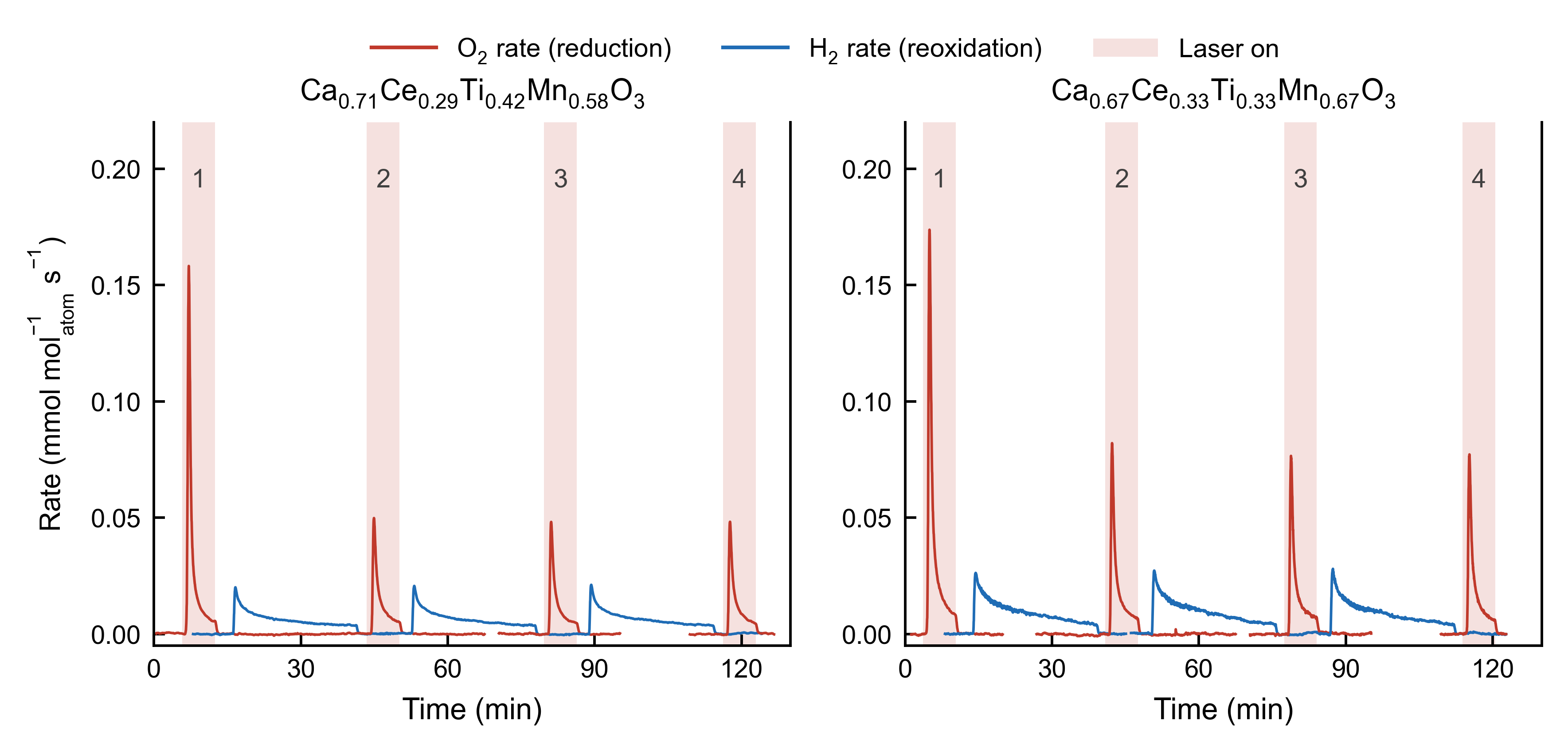}
    \caption{
Time-resolved oxygen and hydrogen release rates measured in the laser-heated stagnation flow reactor for \ce{Ca_{0.71}Ce_{0.29}Ti_{0.42}Mn_{0.58}O3} ($\XCe = 0.29$; left) and CCTM2112 ($\XCe = 0.33$; right) over four redox cycles: \ce{O2} release rate during reduction (red) and \ce{H2} production rate during reoxidation (blue), normalized to total atoms in the perovskite formula unit.
Shaded bands mark the intervals with the laser on, including its heating and cooling ramps; the laser-heated sample reached \qty{1350}{\degreeCelsius} during each \qty{5.5}{\minute} reduction under \ce{Ar}.
Reoxidation was carried out for \qty{25}{\minute} under \qty{40}{\percent}~\ce{H2O}/\ce{Ar} at \qty{850}{\degreeCelsius}, set by the furnace control thermocouple.
The cycle-\num{1} \ce{O2} peak exceeds the cycle-\numrange[range-phrase = --]{2}{4} peaks by a factor of \numrange[range-phrase = --]{2}{3.3} in both compositions, reflecting preconditioning of as-synthesized material; tabulated \Dd values (\cref{tbl:cctm_delta_delta}) are averaged over cycles \numrange[range-phrase = --]{2}{4}.
}
    \label{fgr:6}
\end{figure*}

We synthesized three CCTM compositions along the charge-compensation series at $\XCe = 0.25$, $0.29$, and $0.33$, the last corresponding to CCTM2112 (full chemical formulas in \cref{tbl:cctm_delta_delta}), and measured their oxygen-exchange capacities by thermogravimetric analysis (TGA) and laser-heated stagnation flow reactor (LSFR) cycling.
Only orthorhombic perovskite reflections were observed by powder X-ray diffraction for each of the three samples; in contrast to previous reports,\cite{wexler_multiple_2023} no \ce{CeO2} or other secondary phase was detected.
Laboratory XRD does not exclude nanocrystalline or amorphous \ce{CeO2}.
Powder X-ray diffraction of the $\XCe = 0.29$ sample after the four LSFR cycles of \cref{fgr:6} and of CCTM2112 after \num{20} TGA cycles shows no secondary phases, decomposition, or structural changes relative to the as-synthesized patterns (ESI \cref{S-sec:esi:xrd,S-fgr:esi:xrd-postcycling}).
We applied three TGA cycling protocols (A, B, and C; \hyperref[sec:experimental-methods]{Experimental methods}) and one LSFR protocol to test how reducing atmosphere and dwell time affect the measured \Dd.
Full TGA cycling traces and protocol details appear in ESI \cref{S-sec:esi:tga}.
Steady-state \Dd values are summarized in \cref{tbl:cctm_delta_delta}.

\begin{table*}
\small
  \caption{
Predicted and measured oxygen-exchange capacity (\Dd) for three CCTM perovskites across cycling protocols.
SCAN+$U$: site-resolved DFT vacancy formation energies for the \num{15} symmetry-distinct nearest-neighbor cation environments of each composition, weighted by their ideal-solution (binomial) probabilities at the bulk composition and evaluated through the thermodynamic formalism of ESI \cref{S-sec:esi:thermo} (ESI \cref{S-eqn:occupation,S-eqn:delta-avg}).
CFM: same binomial environment weighting applied to crystal-feature-model predictions of the per-environment \Ev.
dGNN: defect graph neural network ensemble-mean predictions on SQS supercells at \num{100} $(\XCe, \XMn)$ grid points; off-grid compositions obtained by linear interpolation between the four nearest grid points.
All three predictions use identical cycling conditions: $\Tred = \qty{1350}{\degreeCelsius}$, $\POtwo^\mathrm{red} = \qty[exponent-mode = scientific]{0.0001}{\bar}$; $\Tox = \qty{850}{\degreeCelsius}$, $\PHtwo/\PHtwoO = \num[exponent-mode = scientific]{0.001}$.
TGA Protocol~A: \qty{1350}{\degreeCelsius} (\qty{0.5}{\hour}) / \qty{1300}{\degreeCelsius} (\qty{1}{\hour}) reduction in \ce{Ar}\textsuperscript{\emph{a}}, \qty{900}{\degreeCelsius} (\qty{0.5}{\hour}) reoxidation in \qty{20}{\percent}~\ce{O2}/\ce{Ar}.
Protocol~B: \qty{1350}{\degreeCelsius} (\qty{1}{\hour}) / \qty{1300}{\degreeCelsius} (\qty{0.5}{\hour}) reduction in \ce{Ar}\textsuperscript{\emph{a}}; same reoxidation.
Protocol~C: \qty{1350}{\degreeCelsius} (\qty{1.5}{\hour}) reduction in \qty{0.11}{\percent}~\ce{O2}/\ce{Ar}, \qty{850}{\degreeCelsius} (\qty{1}{\hour}) reoxidation in the same atmosphere.
LSFR: \qty{1350}{\degreeCelsius} (\qty{5.5}{\minute}) reduction in \ce{Ar}\textsuperscript{\emph{a}}, \qty{850}{\degreeCelsius} (\qty{25}{\minute}) reoxidation in \qty{40}{\percent}~\ce{H2O}/\ce{Ar}; values are mean of cycles \numrange[range-phrase = --]{2}{4} from \ce{O2} integration.
Experimental uncertainties approximately \qty{\pm 10}{\percent}.
Both TGA and LSFR \Dd are lower bounds on the equilibrium oxygen-exchange capacity owing to incomplete equilibration during the reduction dwell.
}
  \label{tbl:cctm_delta_delta}
  \begin{tabular*}{\textwidth}{@{\extracolsep{\fill}}lcccccccccc}
    \hline
    & & & \multicolumn{3}{c}{Predicted} & \multicolumn{3}{c}{TGA Protocol} & & \\
    \cline{4-6} \cline{7-9}
    Composition & \XCe & \XMn & SCAN+$U$ & CFM & dGNN & A & B & C & LSFR & \\
    \hline
    \ce{Ca_{0.75}Ce_{0.25}Ti_{0.50}Mn_{0.50}O3} & 0.25 & 0.50 & 0.043 & 0.036 & 0.042 & 0.061 & 0.066 & 0.036 & ---  & \textsuperscript{\emph{b}} \\
    \ce{Ca_{0.71}Ce_{0.29}Ti_{0.42}Mn_{0.58}O3} & 0.29 & 0.58 & 0.069 & 0.044 & 0.058 & 0.066 & 0.071 & 0.044 & 0.047 & \\
    \ce{Ca_{0.67}Ce_{0.33}Ti_{0.33}Mn_{0.67}O3} & 0.33 & 0.67 & 0.044 & 0.051 & 0.098 & 0.067 & 0.072 & 0.048 & 0.067 & \\
    \hline
  \end{tabular*}
  \begin{flushleft}
    \textsuperscript{\emph{a}} The \ce{Ar} reduction atmosphere contains residual oxygen at $\POtwo \approx \qtyrange[range-phrase = \text{--}, range-units = single]{0.1}{1}{\Pa}$ (\qtyrange[range-phrase = --, range-units = single, exponent-mode = scientific]{\approx 1e-6}{1e-5}{\bar}). \\
    \textsuperscript{\emph{b}} LSFR not performed; the $\XCe = 0.25$ composition is the least reducible in the experimental set.
  \end{flushleft}
\end{table*}

Under TGA Protocol~C, in which the reduction \POtwo (\qty[exponent-mode = scientific]{\approx 1.1e-3}{\bar}) is closest to the model conditions, \Dd increases monotonically with \XCe from \num{0.036} at $\XCe = 0.25$ to \num{0.048} at $\XCe = 0.33$.
Under Protocols~A and B, in which residual oxygen in the \ce{Ar} carrier gives an effective $\POtwo \approx \qty[exponent-mode = scientific]{1e-6}{\bar}$ (substantially below the model conditions), measured \Dd is systematically higher (\numrange[range-phrase = --]{0.061}{0.072}) and the \XCe-dependent ordering is preserved but flatter.
Measured \Dd values are lower bounds on the equilibrium oxygen-exchange capacity in all four protocols, evidenced by the nonzero \ce{O2} release rate at the end of the reduction step in the LSFR (\cref{fgr:6}).
The specific rate-limiting step (surface exchange, bulk diffusion, or gas-phase transport) is not separately identified from the present data and likely differs across protocols owing to the different temperature ramps, atmospheres, and reactor geometries involved.

LSFR cycling of the $\XCe = 0.29$ and $0.33$ compositions yields steady-state \Dd values of \num{0.047} and \num{0.067}, respectively, comparable to the corresponding TGA Protocol~C values (\num{0.044} and \num{0.048}) despite the substantially shorter reduction dwell.
This is consistent with laser heating reaching the reduction temperature faster than the TGA furnace ramp.
Two features are apparent in the time-resolved \ce{O2} and \ce{H2} release rates (\cref{fgr:6}).
First, the cycle-\num{1} \ce{O2} peak exceeds the cycle-\numrange[range-phrase = --]{2}{4} peaks in both compositions; this behavior is commonly observed and motivates a redox break-in cycle, so tabulated \Dd values are averaged over cycles \numrange[range-phrase = --]{2}{4}.
The integrated gas amounts account for the larger first release.
For CCTM2112, the \ce{O2} released in cycle~\num{1} corresponds to \num{0.120} O per formula unit, the subsequent steam reoxidation restored \num{0.072} O per formula unit (from the integrated \ce{H2}), and cycles \numrange[range-phrase = --]{2}{4} then released \numrange[range-phrase = --]{0.066}{0.069}; for $\XCe = 0.29$, the corresponding values are \num{0.092}, \num{0.048}, and \numrange[range-phrase = --]{0.046}{0.049}.
The as-synthesized powder was equilibrated with air, whereas in subsequent cycles the material is reoxidized in \qty{40}{\percent}~\ce{H2O}/\ce{Ar} at \qty{850}{\degreeCelsius}, a much weaker oxidant than air, so the first reduction also removes oxygen that steam does not restore.
Second, the \ce{O2} peak heights decrease slightly from cycle~\num{2} to cycle~\num{4}, and the measured \ce{H2}/\ce{O2} material balance shows a small (\qtyrange[range-phrase = --, range-units = single]{5}{7}{\percent}) mismatch over time.
Reoxidation of CCTM and related materials is much slower in water vapor (here \qty{40}{\percent}~\ce{H2O}) than in \ce{O2}/\ce{Ar} mixtures such as those used in the TGA, so this trend is consistent with incomplete reoxidation to the starting state of the material in each cycle; the \ce{O2} evolution is also truncated by the \qty{5.5}{\minute} reduction window in each cycle.
Both the peak-height decline and the material-balance mismatch therefore reflect an experimental artifact of reoxidation kinetics and cycle times rather than deactivation over consecutive cycles; a few-percent mass-spectrometer calibration uncertainty (certified gas standards; \hyperref[sec:experimental-methods]{Experimental methods}) may also contribute.
This supports a kinetic rather than thermodynamic origin for the lower-bound character of the LSFR \Dd described above.
The $\XCe = 0.25$ composition was not measured in the LSFR because its predicted and TGA-measured \Dd are the lowest of the three compositions and below the productivity threshold of practical interest for subsequent characterization.

At cycling conditions close to the model conditions (TGA Protocol~C, LSFR), measured \Dd at $\XCe = 0.25$ (\num{0.036}) matches the CFM prediction (\num{0.036}) and lies below the SCAN+$U$ and dGNN predictions (\num{0.043} and \num{0.042}, respectively).
At CCTM2112 ($\XCe = 0.33$), measured \Dd (\numrange[range-phrase = --]{0.048}{0.067}) brackets the CFM prediction (\num{0.051}); the SCAN+$U$ value (\num{0.044}) lies just below the measured range, and the dGNN prediction (\num{0.098}) well above it.
At $\XCe = 0.29$, the SCAN+$U$ (\num{0.069}) and dGNN predictions exceed the measured range (\numrange[range-phrase = --]{0.044}{0.047}) while the CFM (\num{0.044}) falls at the lower edge of the measured range, the closest of the three; the ordering is consistent with experimental \Dd being a lower bound on the equilibrium value, but the magnitude of the SCAN+$U$--experiment gap is larger here than at the other two compositions.
Under Protocols~A and B, measured \Dd exceeds all three predictions for all three compositions, consistent with the more strongly reducing residual-\POtwo atmospheres of these protocols falling outside the model conditions.
Across the three compositions, the two fitted models deviate from SCAN+$U$ in different ways: the CFM yields a narrower predicted \Dd range, while the dGNN yields a substantially larger \Dd at CCTM2112 (ESI \cref{S-sec:esi:dgnn:validation}).
The implications of these complementary biases for fitted-model selection in composition-space screening are addressed in the Discussion.

\section*{Discussion}

\subsection*{B-site chemistry as the primary control over vacancy energetics}

Local B-site chemistry, not bulk stoichiometry alone, dominates oxygen-vacancy formation energetics in CCTM perovskites.
Varying the Mn fraction among nearest-neighbor B sites shifts \Ev by \qtyrange[range-phrase = --, range-units = single]{1.0}{1.5}{\eV} depending on local Ce content, whereas A-site Ce variation produces a smaller, Mn-dependent shift of \qtyrange[range-phrase = --, range-units = single]{0.2}{0.6}{\eV} (\cref{fgr:4}). This B-site shift is large relative to the targeted energetics: it spans \numrange[range-phrase = --]{2}{3} times the \qty{0.5}{\eV} width of the thermochemically favorable \qtyrange[range-phrase = --, range-units = single]{3.4}{3.9}{\eV} window, and its upper end of \qty{1.5}{\eV} is \qtyrange[range-phrase = --, range-units = single]{38}{44}{\percent} of the target \Ev itself.
The larger B-site contribution suggests that short-range B-site cation order, rather than bulk composition, is the more effective target for tuning the \Ev distribution: bulk optimization averages over local environments, whereas targeted B-site ordering could narrow the distribution and increase the fraction of sites within the thermochemically favorable \qtyrange[range-phrase = --, range-units = single]{3.4}{3.9}{\eV} window.
Adopting a uniformly narrow distribution as the optimization goal assumes that per-vacancy electronic entropy contributions are approximately uniform across local environments; differences between Ce- and Mn-localized vacancies could shift the optimum from a uniformly narrow distribution to one weighted toward specific local environments.

Several processing routes might achieve such control.
Layer-by-layer thin-film growth could bias toward specific B-site ordering motifs; epitaxial strain, which couples to octahedral tilting and B--O bond lengths,\cite{rondinelli_structure_2011, rondinelli_control_2012} could indirectly bias cation arrangements through these structural distortions; thermal histories that permit partial B-site equilibration before quenching offer a bulk-compatible alternative; and aliovalent or size-mismatched dopants may promote B-site cation ordering directly in bulk material.\cite{bai_asite_2012, rautama_electrondoping_2005} In related double perovskites, such substitution enhances B-site cation order, although the effect is system-dependent and the smaller B-site charge contrast in CCTM makes the outcome less certain. Any such order would be metastable, requiring kinetic trapping during synthesis, and could re-equilibrate toward the thermodynamically preferred arrangement at the high reduction temperatures of cycling; retention of the targeted ordering is therefore a prerequisite for this approach.
The present work treats disordered configurations via SQS supercells; future first-principles calculations comparing representative ordered motifs to this baseline would quantify how much ordering can shift the \Ev distribution and whether such shifts improve \Dd without compromising phase stability.
Such ordering can be quantified by short-range order parameters (e.g., Warren--Cowley pair correlations\cite{cowley_approximate_1950,cowley_shortrange_1965} on the B-site Ti--Mn sublattice, with secondary parameters describing A-site Ca--Ce mixing and the cross-sublattice Ce--Mn correlation, extended to multiple sublattices in standard fashion\cite{gehringer_models_2023}), which are measurable by high-resolution electron microscopy and can be computed from the same supercell configurations used to evaluate the \Ev distribution.

The larger B-site contribution to \Ev is reflected in the CFM coefficients, whose fitted values ($\bb \approx 2.0$, $\br \approx -1.3$) recover the direct M--O bonding of the two B-site cations to the removed oxygen and the stabilization of the two released electrons, as detailed above.
Substituting Mn for Ti weakens the M--O bonds and provides a more favorable reduction couple (\MnIV/\MnIII~vs.~\TiIV/\TiIII), both of which lower \Ev.\cite{qian_favorable_2020, qian_outstanding_2021}
The smaller influence of A-site Ce likely reflects the interplay between Ce's reduction potential, which is intermediate between the more reducible \MnIV/\MnIII and the less reducible \TiIV/\TiIII couples, and local structural distortions, though a full mechanistic decomposition remains an open question; the sense in which Ce can be both \CeIII-like and electronically active in reduction is addressed below.

\subsection*{Cerium oxidation states and delocalized reduction}

SCAN+$U$ predicts predominantly \CeIII-like character in pristine CCTM; this assignment has not been verified experimentally for the present compositions.
In the CCTM2112 supercell, \num{23} of \num{24} Ce sites have calculated moments of \qtyrange[range-phrase = --, range-units = single]{0.45}{0.85}{\bohrmagneton}, below the \qty{0.94}{\bohrmagneton} of fully localized \CeIII in \ce{Ce2O3} in the same framework, which indicates fractional $4f$ occupation closer to $4f^1$ than to $4f^0$ (ESI \cref{S-fgr:esi:magmoments,S-tbl:esi:ce-counts}). Although the pristine compositions thus exhibit predominantly \CeIII-like character within the SCAN+$U$ description, the changes in Ce $4f$ occupation that accompany vacancy formation (below) show that Ce remains electronically active during reduction and reoxidation, participating in charge compensation through redistribution of electronic density rather than through a discrete $\CeIV \rightarrow \CeIII$ transition.
This finding contrasts with earlier work emphasizing $\CeIV \rightarrow \CeIII$ reduction as a primary contributor to redox activity.\cite{wexler_multiple_2023}
Both studies used \num{360}-atom SQS supercells, and starting from the initial magnetic moments of the earlier work recovers the \CeIII-like state at fixed geometry (ESI \cref{S-sec:esi:magmoments}), so the origin of this difference is not established; it may reflect the different cation configurations of the two SQS realizations or their relaxed structures.
The experimental evidence on Ce and Mn oxidation states in CCTM is limited to soft X-ray absorption spectroscopy (XAS) of CCTM2112 prepared by solid-state reaction,\cite{wexler_multiple_2023} which found predominantly \CeIV and \MnIII in the oxidized material and, on reduction, a reversible decrease of the average Ce oxidation state to \num{3.90 \pm 0.03} with a small net change in Mn.
The delocalized-reduction picture described below is therefore a result of the SCAN+$U$ description that remains to be tested experimentally for the present compositions.
Reduction can still occur at \CeIII-like sites: the partially occupied Ce $4f$ states are near the Fermi level, providing a manifold that accommodates the electrons released on vacancy formation, distributing a small additional reduction over many \CeIII-like sites rather than driving a discrete change in the formal oxidation state of any single Ce ion.\cite{naghavi_giant_2017}
Nonlocal accommodation of the electrons released upon oxygen-vacancy formation has precedent in related perovskites: in \ce{SrFeO_{3-\delta}}, these electrons localize on second-nearest-neighbor Fe rather than the directly coordinated cation.\cite{das_longrange_2017}

In all six compositions, the vacancy-induced changes in the calculated local magnetic moments occur on Ce and Mn, with smaller induced moments on O, whereas individual Ti and Ca moments change by less than \qty{0.05}{\bohrmagneton} (ESI \cref{S-tbl:esi:dmu}).
Averaged over the \num{15} vacancies of each composition, the net change in the Ce moment magnitudes, summed over the Ce sites, is \qtyrange[range-phrase = --, range-units = single]{0.55}{0.94}{\bohrmagneton}, and that of Mn is \qtyrange[range-phrase = --, range-units = single]{0.52}{0.70}{\bohrmagneton} in five of the six compositions, rising to \qty{1.13}{\bohrmagneton} in the most Ce-rich composition ($\XCe = 0.37$).
For CCTM2112, the net Ce and Mn changes are similar (\qtylist[list-units = single]{0.70;0.65}{\bohrmagneton}), whereas the XAS placed most of the net reduction on Ce.\cite{wexler_multiple_2023}
These moment changes are computational indicators of the redox response, not a measure of electron transfer.

Resolving the precise partitioning of reduction between Ce and Mn will require further investigation.
Local compositional fluctuations, combined with the sensitivity of oxidation-state assignments to the choice of DFT functional and Hubbard $U$ parameters, complicate definitive mechanistic attribution.
For thermochemical screening purposes, however, the key point is that the CFM and dGNN capture the net effect of these contributions on \Ev without requiring a complete mechanistic decomposition, an advantage for high-throughput materials discovery.

\subsection*{Model validation and limitations}

The CFM and dGNN deviate from SCAN+$U$ (the DFT reference in \cref{tbl:cctm_delta_delta}) in different ways.
The CFM's predicted \Dd range across the three measured compositions (\numrange[range-phrase = --]{0.036}{0.051}) is narrower than that of SCAN+$U$ (\numrange[range-phrase = --]{0.043}{0.069}); this is consistent with the linear functional form of the CFM, which cannot represent composition-specific changes in the shape of the underlying \Ev distribution, whose low-\Ev tail governs the Boltzmann average.
The dGNN agrees with SCAN+$U$ within \qty{16}{\percent} at $\XCe = 0.25$ and $0.29$ but predicts substantially larger \Dd at CCTM2112 (\num{0.098}~vs.~SCAN+$U$ \num{0.044}); we attribute this to a broader, lower-\Ev tail of the dGNN \Ev distribution at this composition (ESI \cref{S-sec:esi:dgnn:validation}).
At $\XCe = 0.29$, the SCAN+$U$ value (\num{0.069}) exceeds both fitted-model predictions and the measured range (\numrange[range-phrase = --]{0.044}{0.047}); the present data do not distinguish whether this reflects composition-specific incomplete equilibration in the experiment, a nonrepresentative low-\Ev tail in the \num{15}-site DFT sample, or both.
The spread among the three predictions defines an effective uncertainty band for composition-space screening: within its training composition window the CFM tends to give a conservative (typically lower) \Dd estimate, lying below SCAN+$U$ at two of the three measured compositions (the exception being CCTM2112). Compositions with large dGNN--SCAN+$U$ disagreement warrant cross-checking with SCAN+$U$ before committing to synthesis.

Two further caveats qualify this comparison.
First, the noninteracting vacancy approximation does not include possible vacancy--vacancy clustering at high $\delta$; electrostatic repulsion remains modest (a few tens of \unit{\meV}) over the relevant $\delta$ range, per the screened-monopole analysis in ESI \cref{S-fgr:esi:vv-electrostatic}, but short-range chemical interactions could modify the effective \Ev distribution at elevated defect concentrations.
Second, the dGNN screening surface (\cref{fgr:5}b) exhibits nonmonotonic features in composition regions outside the measured set; the present validation cannot distinguish whether these reflect genuine composition--structure coupling captured by the dGNN's structural input or sparse-sampling artifacts of the underlying training data.
Expanding the SCAN+$U$ data set or the dGNN training set with additional CCTM compositions, particularly in the mixed Ce--Mn regime at intermediate \XCe, would help discriminate between these possibilities.

The thermodynamic model includes the gas-phase entropy, through the oxygen chemical potential, and the configurational entropy of the vacancies, through the lattice-gas occupations; the latter depends on composition through the distribution of \Ev over the fifteen local environments.
The model omits the vibrational entropy of the solid, which harmonic-phonon calculations for related Mn-containing perovskites show to be smaller than the gas-phase and configurational contributions,\cite{baldassarri_vibrational_2024} and the electronic entropy, whose magnitude depends on how the released electrons are distributed over the Ce $4f$ and Mn $3d$ states.
A composition-dependent electronic entropy would shift the predicted \Dd and could favor \Ev distributions weighted toward particular local environments.

\subsection*{Outlook and transferability}

The interpretable modeling framework developed here provides design principles that should transfer to related perovskite families.
The finding that B-site chemistry dominates \Ev suggests that other systems with mixed B-site occupancy, such as Sr--Ce--Mn--O or Ba--Ce--Mn--O analogs, may exhibit similar tunability through short-range cation order.\cite{barcellos_bace_2018, kumar_firstprinciples_2025, naik_cationdeficient_2023}
The Born--Haber decomposition represented by the CFM coefficients (bond-breaking~vs.~reduction stabilization) provides a physically grounded basis for predicting how alternative dopants will shift the \Ev distribution: species with weaker M--O bonds or more favorable reduction couples should lower \Ev, while the opposite perturbations should raise it.

Extending this approach to new compositions requires generating appropriate SQS supercells, computing \Ev for representative local environments, and optionally fine-tuning the dGNN on the expanded dataset.
The computational cost is modest, roughly \num{16} DFT calculations per composition for complete coverage of the \num{15} nearest-neighbor environments, making systematic screening across multi-cation perovskite spaces tractable.
We note that the present analysis addresses thermodynamic capacity; practical performance also depends on oxygen-transport kinetics and surface-exchange rates, which merit complementary investigation. The reoxidation measurements reported here were performed in steam (\qty{40}{\percent}~\ce{H2O}/\ce{Ar}, with no added \ce{H2}); CCTM performance under high \ce{H2}/\ce{H2O} ratios, the stricter and more commercially relevant reoxidation test, is left for future work.
Element-specific measurements on the synthesized compositions, such as Ce and Mn X-ray absorption spectroscopy before and after reduction, would test the calculated partition of reduction between Ce and Mn, which has not been verified experimentally for these compositions.

\section*{Conclusions}

This work establishes an interpretable computational framework for mapping oxygen-vacancy formation energetics across compositionally complex perovskites.
Using a coverage-constrained special quasirandom structure methodology that samples all fifteen symmetry-distinct oxygen nearest-neighbor environments, we computed vacancy formation energies for six Ca--Ce--Ti--Mn compositions and developed two complementary fitted models: a crystal-feature model (CFM) whose fitted coefficients directly encode the underlying Born--Haber thermochemistry, and a fine-tuned defect graph neural network (dGNN) that models potential nonlinear structure--composition--defect coupling beyond the linear form of the CFM.
Vacancy formation energetics in CCTM are dominated by local B-site chemistry: changing the nearest-neighbor Mn fraction shifts \Ev by \qtyrange[range-phrase = --, range-units = single]{1.0}{1.5}{\eV} depending on local Ce content, whereas A-site Ce variation produces a smaller, Mn-dependent shift of \qtyrange[range-phrase = --, range-units = single]{0.2}{0.6}{\eV}.
This hierarchy indicates that short-range B-site cation order, if it can be established and kinetically retained through processing, is a candidate means of tuning redox performance.
Composition-space maps show a Ce/Mn-balanced region ($\XCe \approx \numrange[range-phrase = \text{--}]{0.29}{0.33}$, $\XMn \approx \numrange[range-phrase = \text{--}]{0.58}{0.67}$) that combines a high in-window \Ev fraction with phase stability and solubility, where the predicted cycle capacity \Dd is comparable to or exceeds the ceria benchmark ($\Dd/3 \approx \num{0.010}$) at substantially lower reduction temperatures ($\Tred = \qty{1350}{\degreeCelsius}$~vs.~\qty{\sim 1600}{\degreeCelsius} for ceria).\cite{chueh_highflux_2010}
Experimental measurements on three compositions along the charge-compensation line $\XMn = 2 \XCe$ show \Dd increasing monotonically with \XCe under cycling protocols close to the model conditions, consistent with the qualitative trend from the fitted models; the measured \Dd values exceed the model predictions under more strongly reducing protocols.
The interpretable framework provides compositional screening capability and mechanistic design rules expected to transfer to related perovskite families for solar thermochemical hydrogen production.

\section*{Methods\label{sec:methods}}

\subsection*{Computational and theoretical methods}

\subsubsection*{Supercell construction and vacancy site selection.~~}

To model configurational disorder in the CCTM solid solution, we generated special quasirandom structures (SQSs)\cite{zunger_special_1990a, vandewalle_efficient_2013} using the \textsc{icet} package.\cite{angqvist_icet_2019a}
A \num{360}-atom orthorhombic $3\times3\times2$ supercell was constructed with initial lattice parameters $a = \qty{16.12}{\angstrom}$, $b = \qty{16.39}{\angstrom}$, and $c = \qty{15.27}{\angstrom}$; lattice vectors and internal coordinates were subsequently relaxed by DFT for each composition.
Additional SQS methodology details and prior validation for multication perovskites are provided in ref.~\citenum{wexler_multiple_2023}.
The cluster space was defined with pair and triplet cutoffs of \qty{7.0}{\angstrom} and \qty{4.0}{\angstrom} (\textsc{icet} v2.2; symmetry tolerance \num{e-4}), and the SQS were optimized by \num{100001} Monte Carlo trial steps (optimality weight \num{1}) from a fixed random seed; the realized supercells are provided in the data release.

The default SQS objective in \textsc{icet} does not ensure that every symmetry-distinct oxygen first-nearest-neighbor (1NN) environment is realized.
In CCTM, the four A-site and two B-site neighbors surrounding each oxygen yield up to fifteen symmetry-distinct 1NN environments.
To promote complete coverage, we augmented the SQS objective with a penalty term:
\begin{equation}
    Q' = Q + k \left| 15 - N_{\mathrm{env}} \right|,
\end{equation}
where $Q$ and $Q'$ are the original and modified objectives, $N_{\mathrm{env}}$ is the number of distinct O-site 1NN environments realized in the supercell, and $k$ is a penalty factor.
Setting $k = 2000$ reliably produced SQS cells containing all fifteen environments.
The coverage constraint also removes a source of uncontrolled variance in computed bulk energetics, since standard SQS realizations at the same composition can sample different 1NN environment subsets; ESI \cref{S-sec:esi:sqs-sensitivity} estimates the magnitude of this variance for CCTM2112 by comparison with the standard-SQS realization of ref.~\citenum{wexler_multiple_2023}.

Having generated supercells with complete 1NN coverage, we next selected representative vacancy sites for DFT calculations.
Each \num{360}-atom supercell contains \num{216} oxygen sites, of which \num{15} are symmetry-distinct by 1NN environment.
To avoid redundant calculations, we created one neutral oxygen vacancy per environment class.
The representative site within each class was chosen from a two-dimensional Gaussian kernel density estimate (\textsc{scipy} \texttt{gaussian\_kde},\cite{virtanen_scipy_2020} bandwidth \num{0.5}) of the second-shell Ce and Mn cation fractions among all O sites sharing that 1NN environment; the site nearest the density maximum was designated as the representative vacancy, so that each selected site is statistically typical of its class. For classes containing two or fewer sites, the first such site was used.

\subsubsection*{Density functional theory calculations.~~}

All calculations were performed with spin-polarized DFT using the Vienna ab initio simulation package (\textsc{VASP}) v5.4.4 and the projector-augmented-wave (PAW) formalism.
The PAW potentials, from the standard \texttt{PAW\_PBE} dataset, treated the valence configurations \ce{Ca} $3s^2 3p^6 4s^2$ (\texttt{Ca\_sv}), \ce{Ce} $5s^2 5p^6 6s^2 4f^1 5d^1$ (\texttt{Ce}, with the $4f$ states in the valence), \ce{Ti} $3p^6 3d^2 4s^2$ (\texttt{Ti\_pv}), \ce{Mn} $3p^6 3d^5 4s^2$ (\texttt{Mn\_pv}), and \ce{O} $2s^2 2p^4$ (\texttt{O}).
We employed the SCAN exchange--correlation functional with Hubbard $U$ corrections (SCAN+$U$) to account for onsite correlation in transition-metal and $f$-electron states, adopting $U_{\ce{Ce}} = \qty{2.0}{\eV}$, $U_{\ce{Ti}} = \qty{2.5}{\eV}$, and $U_{\ce{Mn}} = \qty{2.7}{\eV}$ based on prior benchmarks against experimental oxidation enthalpies.\cite{saigautam_evaluating_2018,long_evaluating_2020}
The corrections used the rotationally invariant (Dudarev) scheme (\texttt{LDAUTYPE = 2}) with $J = 0$, so the quoted values are effective $U_{\mathrm{eff}} = U - J$ applied to the \ce{Ce} $4f$, \ce{Ti} $3d$, and \ce{Mn} $3d$ states, with \texttt{LMAXMIX = 6}.
The plane-wave energy cutoff was \qty{520}{\eV}, and Brillouin-zone sampling was restricted to the $\Gamma$ point, given the large supercell size.
All calculations used the \texttt{Accurate} precision setting and the aspherical-gradient correction (\texttt{LASPH = True}) required for the meta-generalized gradient approximation.
Total energies were minimized with the all-bands algorithm (\texttt{ALGO = All}) to an electronic self-consistency threshold of \qty{e-5}{\eV} (\texttt{EDIFF}) using Gaussian smearing (\texttt{ISMEAR = 0}) with a width of \qty{0.05}{\eV}, with symmetry disabled (\texttt{ISYM = 0}).
The pristine supercells were relaxed with respect to both cell and ionic degrees of freedom (\texttt{ISIF = 3}); each defective supercell was then relaxed at the fixed pristine cell shape and volume, allowing only the ions to move (\texttt{ISIF = 2}). Both used a quasi-Newton algorithm (\texttt{IBRION = 1}) with an ionic force threshold of \qty{0.03}{\eV\per\angstrom} (\texttt{EDIFFG = -0.03}).
Collinear spin polarization was applied with ferromagnetic initial states; initial magnetic moments were set to \qty{0.6}{\bohrmagneton} on Ca and Ti and to \qty{1.0}{\bohrmagneton} and \qty{5.0}{\bohrmagneton} on Ce and Mn, respectively, then allowed to relax self-consistently.

The chemical and magnetic disorder in these \num{360}-atom cells precludes exhaustive enumeration of magnetic orderings or systematic $U$-parameter sweeps.
Prior work indicates that compositional rankings by \Ev are robust to moderate $U$ variations despite $U$-dependent absolute values,\cite{baldassarri_accuracy_2023} and that magnetic energy scales are typically smaller than redox energetics in transition-metal perovskite oxides.\cite{lee_initio_2009}
A more comprehensive treatment of magnetic disorder, such as disordered-local-moment averaging, is left for future work.

Site-resolved magnetic moments were obtained by integrating the spin density ($\rho_{\uparrow} - \rho_{\downarrow}$) within PAW spheres; Bader moments\cite{henkelman_fast_2006, tang_gridbased_2009} for representative structures confirm robustness to the partition choice (ESI \cref{S-sec:esi:magmoments}).
Thermodynamic stability was assessed via the \qty{0}{\kelvin} energy above the convex hull (\Ehull), computed with the \textsc{pymatgen} \texttt{PhaseDiagram} module.\cite{ong_lifepo2_2008, ong_thermal_2010, ong_python_2013}
The hull was constructed from a reference set of elemental, binary, and ternary phases spanning the \ce{Ca-Ce-Ti-Mn-O} system, with total energies taken from our prior work\cite{saigautam_evaluating_2018, long_evaluating_2020, saigautam_exploring_2020, wexler_factors_2021} at the same SCAN+$U$ level as the CCTM supercells (elemental metals at the SCAN level), so that no cross-functional energy correction is applied; \Ehull is the energy of each composition above this hull.
Density of states calculations used Gaussian smearing (\qty{0.05}{\eV} width) at the $\Gamma$ point; a $2\times2\times2$ $k$-mesh produced no qualitative differences (ESI \cref{S-fgr:esi:bz-convergence}).

The partially occupied Ce $4f$ states cross the Fermi level in all six compositions, resulting in metallic electronic structure (ESI \cref{S-fgr:esi:dos}).
In such metallic hosts, positively charged oxygen vacancies are disfavored because excess electrons are readily accommodated in the conduction manifold;\cite{freysoldt_firstprinciples_2014} we therefore treat neutral vacancies as the relevant charge state, consistent with the computed metallic electronic structure of all compositions studied.

Neutral oxygen-vacancy formation energies were computed as
\begin{equation}
    \Ev = E(\mathrm{defective}) - E(\mathrm{pristine}) + \frac{1}{2} E_{\ce{O2}},
\end{equation}
where $E(\mathrm{defective})$ and $E(\mathrm{pristine})$ are the total energies of the vacancy-containing and stoichiometric supercells, respectively.

\subsubsection*{Thermodynamic modeling.~~}

To translate oxygen-vacancy formation energies into cycle-level performance metrics, we modeled the equilibrium oxygen off-stoichiometry $\delta(T, \muO)$ under representative thermochemical cycling conditions.
The lattice is treated as a set of oxygen sites that are either occupied or vacant.
For neutral, noninteracting vacancies with temperature-independent formation energy $\Ev(\mathbf{x}, \mathbf{X})$, where $\mathbf{x} = (\xCe, \xMn)$ denotes the local 1NN environment and $\mathbf{X}$ the bulk composition, the probability that a site is vacant at temperature $T$ and oxygen chemical potential $\muO$ is
\begin{equation}
    n(\mathbf{x}, \mathbf{X}) = \frac{1}{1 + \exp\left[ \frac{\Ev(\mathbf{x}, \mathbf{X}) + \muO(T, \{ P_i \})}{\kB T} \right]}.
    \label{eqn:occupation}
\end{equation}
The overall oxygen deficiency follows from averaging over all local environments,
\begin{equation}
    \delta(T, \muO) = 3 \langle n(\mathbf{x}, \mathbf{X}) \rangle,
    \label{eqn:delta-avg-main}
\end{equation}
where the factor of \num{3} reflects the three oxygen atoms per \ce{ABO3} formula unit.

During thermal reduction, the oxide equilibrates with \ce{O2(g)}, and the oxygen chemical potential per atom is
\begin{equation}
    \muO(T, \POtwo) = \frac{1}{2} \left[ \mu_{\ce{O2}}^{\circ}(T) + \kB T \ln\left(\frac{\POtwo}{P^{\circ}}\right) \right],
    \label{eqn:thermal-reduction-muO-split}
\end{equation}
where $\mu_{\ce{O2}}^{\circ}(T)$ is the standard chemical potential of \ce{O2} from NIST-JANAF thermochemical tables\cite{chase_nistjanaf_1998} and $P^{\circ} = \qty{1}{\bar}$.
All oxygen chemical potentials are referenced to the SCAN total energy of \ce{O2}, computed for a spin-triplet molecule in a \qty{15}{\angstrom} cubic cell at the $\Gamma$ point with the same functional, cutoff, and smearing as the solids.
No empirical \ce{O2} energy correction is applied: SCAN reproduces the \ce{O2} equilibrium bond length (\qty{1.22}{\angstrom}~vs.~the experimental \qty{1.208}{\angstrom}) and yields a dissociation energy of \qty{5.28}{\eV}, consistent with its benchmarked molecular accuracy,\cite{sun_strongly_2015} and in any case the \ce{O2} reference energy cancels between \Ev and the oxygen chemical potential (ESI \cref{S-sec:esi:thermo}), so its absolute value does not affect the computed \Dd.
The connection between \qty{0}{\kelvin} DFT \Ev and the finite-temperature formulation is derived in ESI \cref{S-sec:esi:thermo}.

During water splitting, the relevant equilibrium is
\begin{equation}
    \ce{H2(g) + O_{lattice} <=> H2O(g)},
    \label{eqn:water-splitting-equilibrium}
\end{equation}
with corresponding oxygen chemical potential
\begin{equation}
    \muO(T, \PHtwoO, \PHtwo) = \mu_{\ce{H2O}}^{\circ}(T) - \mu_{\ce{H2}}^{\circ}(T) + \kB T \ln\left( \frac{\PHtwoO}{\PHtwo} \right).
    \label{eqn:water-splitting-muO-split}
\end{equation}
Substituting \cref{eqn:thermal-reduction-muO-split} into \cref{eqn:occupation} and averaging via \cref{eqn:delta-avg-main} yields the equilibrium deficiency during reduction, $\delta_{\mathrm{red}}(\Tred, \POtwo)$; substituting \cref{eqn:water-splitting-muO-split} yields the oxidation deficiency, $\delta_{\mathrm{ox}}(\Tox, \PHtwoO/\PHtwo)$.
The cycle capacity is $\Dd = \delta_{\mathrm{red}} - \delta_{\mathrm{ox}}$.
For the screening calculations, we used $\Tred = \qty{1350}{\degreeCelsius}$, $\POtwo = \qty[exponent-mode = scientific]{0.0001}{\bar}$; $\Tox = \qty{850}{\degreeCelsius}$, $\PHtwoO/\PHtwo = \num{1000}$ (i.e., $\PHtwo/\PHtwoO = \num[exponent-mode = scientific]{0.001}$).
These conditions are representative of those used in prior solar thermochemical measurements on Sr--La--Mn--Al (SLMA) and Ba--Ce--Mn (BCM) perovskites, ceria, and CCTM2112,\cite{barcellos_bace_2018, wexler_multiple_2023} and are adopted here for comparability rather than as a CCTM-specific optimum.

To map \Dd across the full CCTM composition space, we replaced DFT \Ev values with CFM or dGNN predictions.
For the dGNN, which requires explicit structural input, we generated SQS supercells at \num{100} grid points and linearly interpolated for off-grid compositions.

\subsubsection*{Model construction.~~}

\paragraph*{Crystal feature model.~~}

The crystal-feature model (CFM) uses two physically motivated descriptors: the average metal--oxygen bond-dissociation energy \Eb and the average crystal reduction potential \Vr.
Both \Eb and \Vr are computed by averaging tabulated atomic values (ESI \cref{S-tbl:esi:bond-energies,S-tbl:esi:reduction-potentials}) over all six nearest-neighbor cations surrounding the vacancy (four A-site, two B-site), reflecting the local chemical environment.
Because all CCTM compositions exhibit metallic electronic structure, band gap is not included as a feature.

Model coefficients were obtained by Huber regression,\cite{huber_robust_1964, huber_robust_2009} which provides robust estimates for datasets with heterogeneous variance.
The reported coefficients are obtained from a single Huber fit to the full \num{90}-site dataset, with \qty{95}{\percent} confidence intervals from bootstrap resampling of sites with replacement (\num{2000} resamples).
Prediction accuracy is assessed separately by \num{1000} random \qty{50}{\percent}/\qty{50}{\percent} train--test splits and by leave-one-composition-out cross-validation, both reported in ESI \cref{S-sec:esi:cfm-parity}.
The fitted coefficients ($\bb \approx 2.0$, $\br \approx -1.3$, $\beta_0 \approx -4.26$) and their physical interpretation are discussed in the \hyperref[sec:results]{Results}.

\paragraph*{Defect graph neural network model.~~}

We validated and extended the defect graph neural network (dGNN) approach previously developed by Witman et al.\cite{witman_defect_2023} by testing it on CCTM oxygen-vacancy formation energies.
The original dGNN training data (``Base'' dataset) consisted primarily of binary and ternary oxides, with limited quaternary compounds; it has since been augmented with additional quaternary oxides and \ce{ABO3} perovskite data from ref.~\citenum{wexler_factors_2021}.
The CCTM system lies outside the chemical space of the Base training data and contains O nearest-neighbor environments absent from that data.
To improve predictions, we constructed a fine-tuning (FT) dataset by augmenting the Base data with \qty{20}{\percent} of the CCTM vacancy energies computed in this work.

Model architecture and training hyperparameters are described in ESI \cref{S-sec:esi:dgnn}.
Using $K$-fold cross-validation, each \Ev is predicted by an ensemble of $K$ models; the ensemble mean provides the final prediction, and the standard deviation serves as an uncertainty estimate.
ESI \cref{S-fgr:esi:dgnn:parity} compares parity plots for the Base and FT models: the Base model performs reasonably on the CCTM hold-out set, but fine-tuning substantially reduces both error and uncertainty.

To screen across composition space, we generated SQS supercells at \num{100} grid points spanning the $(\XCe, \XMn)$ domain and predicted \Ev distributions for each.
The resulting distributions (ESI \cref{S-fgr:esi:dgnn:vs:cfm:distributions,S-fgr:esi:dgnn:summary}) indicate composition-dependent nonlinearity in the dGNN predictions and notable variation in \Ev spread.
These distributions were used to compute $\delta(T, \muO)$ via \cref{eqn:delta-avg-main}.

\subsection*{Experimental methods\label{sec:experimental-methods}}

\subsubsection*{Synthesis.~~}

We synthesized CCTM samples by a modified Pechini route.\cite{pechini_method_1967}
We dissolved stoichiometric amounts of titanium bis(ammonium lactato)dihydroxide (TALH; \qty{50}{\percent} (w/w) in \ce{H2O}) and the acetates of calcium, cerium, and manganese in deionized water with citric acid as a chelating agent (\num{1}:\num{3} metal-cation to citric-acid molar ratio); all reagents were obtained from Sigma-Aldrich (cerium acetate, \qty{99.9}{\percent} trace-metals basis; calcium acetate and citric acid, \qty{>=99.0}{\percent}; manganese acetate, \qty{>=99}{\percent}).
We evaporated the solution to dryness in an oven at \qty{150}{\degreeCelsius} for \qty{15}{\hour}, ground the resulting gel to a coarse powder, and calcined this powder at \qty{1000}{\degreeCelsius} for \qty{6}{\hour} in static air (heating rate \qty{3}{\degreeCelsius\per\minute}).
We pressed the calcined powder in a \qty{6.35}{\mm} die at \qty{2.5}{\mega\pascal}, pre-fired the resulting pellets at \qty{1000}{\degreeCelsius} for \qty{6}{\hour} in air, and sintered them at \qty{1425}{\degreeCelsius} (\qty{1450}{\degreeCelsius} for CCTM2112) for \qty{6}{\hour} in static air on \ce{CeO2} coasters (to prevent reaction with the crucible) using a three-stage ramp (\qty{2.5}{\degreeCelsius\per\minute} to \qty{850}{\degreeCelsius}, \qty{2.25}{\degreeCelsius\per\minute} to \qty{1300}{\degreeCelsius}, \qty{1.25}{\degreeCelsius\per\minute} to the sintering temperature), and ground the sintered pellets to a fine powder using a mortar and pestle for subsequent characterization.
The reported compositions are nominal cation stoichiometries; analyzed compositions will be reported separately.\cite{ali_synthesis_inprep}
Within the CFM, a deviation of \num{0.01} in \XCe or \XMn from the nominal composition changes the predicted \Dd of the three synthesized compositions by at most \num{0.002}, less than the experimental uncertainty.

\subsubsection*{X-ray diffraction.~~}

We characterized phase purity by powder X-ray diffraction (XRD) on a Malvern Panalytical Aeris powder diffractometer with \ce{Cu K\alpha} radiation (\qty{40}{\kV}, \qty{15}{\mA}; \ce{K\alpha_2} not stripped, $I_{\ce{K\alpha_2}} / I_{\ce{K\alpha_1}} = \num{0.5}$) in Bragg--Brentano geometry, without a secondary monochromator and with a fixed \qty{0.76}{\mm} divergence slit.
Powder samples were mounted on a zero-background silicon holder and scanned in continuous mode over $2\theta = $ \qtyrange[range-phrase = --, range-units = single]{5}{90}{\degree} with a step size of \qty{0.0217}{\degree} and an effective counting time of \qty{25.2}{\second} per step (\num{3911} points per scan).
The calculated reference pattern shown in ESI \cref{S-sec:esi:xrd} was generated from a \num{360}-atom $3 \times 3 \times 2$ SQS supercell of CCTM2112 using the \texttt{XRDCalculator} module of \textsc{pymatgen}.\cite{ong_python_2013}

\subsubsection*{TGA measurements.~~}

We measured oxygen uptake and release on a NETZSCH STA 449 F3 Jupiter thermogravimetric analyzer equipped with a graphite furnace.
Carrier gases were ultra-high-purity \ce{Ar} (\qty{99.99}{\percent}) for Protocols A and B and a certified \qty{1102}{\ppm}~\ce{O2}-in-\ce{Ar} standard (nominal \qty{0.11}{\percent}~\ce{O2}/\ce{Ar}, $\POtwo = \qty[exponent-mode = scientific]{1.1e-3}{\bar}$) for Protocol C.
For each measurement, we loaded \qty{80}{\mg} of CCTM powder into a calcia-stabilized zirconia crucible and heated the sample at \qty{10}{\degreeCelsius\per\minute} to the reduction temperature under flowing carrier gas at a total flow rate of \qty{50}{\milli\liter\per\minute}.
The three TGA protocols differed in reduction atmosphere, reduction dwell-time partition, and reoxidation conditions.
In Protocol~A, samples were reduced at \qty{1350}{\degreeCelsius} (\qty{0.5}{\hour}) followed by \qty{1300}{\degreeCelsius} (\qty{1}{\hour}) under flowing \ce{Ar} and reoxidized at \qty{900}{\degreeCelsius} (\qty{0.5}{\hour}) under \qty{20}{\percent}~\ce{O2}/\ce{Ar}.
Protocol~B used the same atmospheres and reoxidation conditions as Protocol~A but exchanged the two reduction-step durations to \qty{1350}{\degreeCelsius} (\qty{1}{\hour}) followed by \qty{1300}{\degreeCelsius} (\qty{0.5}{\hour}). In both two-step protocols, the initial step at the higher \qty{1350}{\degreeCelsius} accelerates the approach to the reduction equilibrium at \qty{1300}{\degreeCelsius}, which is otherwise slow to reach at that temperature alone.
In Protocol~C, samples were reduced at \qty{1350}{\degreeCelsius} (\qty{1.5}{\hour}) under \qty{0.11}{\percent}~\ce{O2}/\ce{Ar} and reoxidized at \qty{850}{\degreeCelsius} (\qty{1}{\hour}) under the same atmosphere.
Each protocol comprised five reduction--reoxidation cycles; reported \Dd values are means over cycles \numrange[range-phrase = --]{2}{5}, with cycle \num{1} excluded to avoid preconditioning effects (see \hyperref[sec:results:experimental:validation]{Experimental validation}).
Apparent sample-mass changes were corrected for buoyancy by subtracting an empty-crucible blank acquired under the same temperature and gas program before conversion.
We converted the corrected mass change to oxygen off-stoichiometry $\delta$ assuming
\begin{equation}
    \delta = \left( \frac{\Delta m}{m_{\mathrm{initial}}} \right) \left( \frac{\mathrm{MW}_{\mathrm{CCTM}}}{\mathrm{MW}_{\ce{O}}} \right),
\end{equation}
where $\Delta m$ is the measured mass change, $m_{\mathrm{initial}}$ is the initial sample mass, and $\mathrm{MW}_{\mathrm{CCTM}}$ and $\mathrm{MW}_{\ce{O}}$ are the molecular weights of the oxide and atomic oxygen, respectively.
Experimental uncertainty in \Dd is approximately \qty{\pm 10}{\percent}, dominated by uncertainty in the residual \POtwo (\qty{\sim 1}{\Pa}) in the \ce{Ar} carrier supply for Protocols~A and B.

\subsubsection*{LSFR measurements.~~}

We measured oxygen-release and hydrogen-production rates on the Sandia National Laboratories laser-heated stagnation flow reactor (LSFR),\cite{scheffeHydrogenProductionChemical2011, arifinCoFe2O4PorousAl2O32012, scheffeKineticsMechanismSolarthermochemical2013, mcdanielSrMndopedLaAlO3d2013} in which samples are heated rapidly by focused continuous-wave laser irradiation.
For each measurement, we loaded \qtyrange[range-phrase = --, range-units = single]{80}{100}{\mg} of CCTM powder onto a platinum foil supported on a zirconia sample holder; sample-surface temperature was measured by a two-color pyrometer (Process Sensors PYROSPOT DGR~55N) positioned directly above the sample. The pyrometer was calibrated against a tube-furnace control thermocouple as the temperature standard: at \qty{1400}{\degreeCelsius} its reading was matched to the thermocouple, and the agreement was verified at other temperatures; the emissivity was set to \num{1} with a two-color correction factor of \num{1.004}.
Before cycle~\num{1}, the furnace and sample were preheated to \qty{850}{\degreeCelsius} under \ce{Ar}; every reduction, including the first, was driven from \qty{850}{\degreeCelsius} to \qty{1350}{\degreeCelsius} by the laser.
We reduced samples at \qty{1350}{\degreeCelsius} for \qty{5.5}{\minute} under flowing \ce{Ar} at a total flow rate of \qty{300}{\milli\liter\per\minute}.
We then cooled to \qty{850}{\degreeCelsius} and reoxidized for \qty{25}{\minute} under \qty{40}{\percent}~\ce{H2O}/\ce{Ar} at a total flow rate of \qty{500}{\milli\liter\per\minute} (\qty{300}{\milli\liter\per\minute} \ce{Ar} carrier + \qty{200}{\milli\liter\per\minute} \ce{H2O}).
The furnace control thermocouple, which sets the reoxidation temperature, read \qty{850.0 \pm 0.5}{\degreeCelsius} throughout every reoxidation step; at this set point, the pyrometer read \qtyrange[range-phrase = --, range-units = single]{852}{855}{\degreeCelsius} under \ce{Ar} but \qtyrange[range-phrase = --, range-units = single]{873}{877}{\degreeCelsius} while the \ce{H2O}/\ce{Ar} reoxidation gas was flowing, an offset of approximately \qty{20}{\degreeCelsius} that appears in every cycle as a step when the steam flow starts.
Reactor effluent was sampled continuously by a downstream Extrel C50 modulated effusive-beam quadrupole mass spectrometer.
The \ce{O2} and \ce{H2} signals were calibrated using \qty{25}{\percent}~\ce{O2}/\ce{Ar} and \qty{5}{\percent}~\ce{H2}/\ce{Ar} certified gas standards introduced at known flow rates.
Raw mass-spectrometer signals were processed with a custom Mathematica routine provided by the Sandia LSFR team: a stationary wavelet transform for denoising, subtraction of a polynomial fit to a user-selected baseline region, and a linear-regression calibration factor converting the ion signals to mole fractions and then, following the procedure of ref.~\citenum{mcdanielSrMndopedLaAlO3d2013}, to molar release rates normalized per mole of perovskite atoms; per-cycle \ce{O2} and \ce{H2} amounts were obtained by integrating over user-selected cycle bounds.
Each composition was cycled four times; reported \Dd values are means over cycles \numrange[range-phrase = --]{2}{4} from integrated \ce{O2} release, with cycle \num{1} excluded to avoid preconditioning effects (see \hyperref[sec:results:experimental:validation]{Experimental validation}).

\section*{Author contributions}

Manish Kumar: Data curation, Formal analysis, Investigation, Methodology, Software, Validation, Visualization, Writing -- original draft, Writing -- review \& editing; Natalia Ali: Data curation, Formal analysis, Investigation, Methodology, Validation, Visualization, Writing -- original draft, Writing -- review \& editing; Matthew D.\ Witman: Data curation, Formal analysis, Funding acquisition, Investigation, Resources, Software, Validation, Visualization, Writing -- original draft, Writing -- review \& editing; Shang Zhai: Funding acquisition, Writing -- review \& editing; James E.\ Miller: Formal analysis, Methodology, Supervision, Validation, Writing -- original draft, Writing -- review \& editing; Ivan Ermanoski: Formal analysis, Methodology, Supervision, Validation, Writing -- original draft, Writing -- review \& editing; Ellen B.\ Stechel: Conceptualization, Formal analysis, Funding acquisition, Methodology, Project administration, Resources, Supervision, Validation, Writing -- original draft, Writing -- review \& editing; Robert B.\ Wexler: Conceptualization, Formal analysis, Funding acquisition, Methodology, Project administration, Resources, Software, Supervision, Validation, Visualization, Writing -- original draft, Writing -- review \& editing.

\section*{Conflicts of interest}

There are no conflicts to declare.

\section*{Data availability}

The processed data supporting this article, including the first-principles oxygen-vacancy formation energies, the crystal-feature-model and defect graph neural network predictions, and the experimental thermogravimetric, stagnation-flow-reactor, and powder X-ray diffraction measurements, together with the scripts that generate all figures, are available at \url{https://github.com/wexlergroup/cctm-screening}; that repository will be archived at Zenodo, with the DOI provided prior to publication. The underlying VASP density-functional-theory inputs and outputs are deposited in the NOMAD repository (DOI: \href{https://doi.org/10.17172/nomad.6d7e-hvb1}{\texttt{10.17172/nomad.6d7e-hvb1}}).

\section*{Acknowledgements}

This material is based upon work supported by the U.S.\ Department of Energy's Office of Energy Efficiency and Renewable Energy (EERE) under the Fuel Cell Technologies Office (FCTO) under Award Number DE-EE0010733.
The authors gratefully acknowledge research support from the HydroGEN Advanced Water Splitting Materials Consortium, established as part of the Energy Materials Network under the U.S.\ Department of Energy, Office of Energy Efficiency and Renewable Energy, Fuel Cell Technologies Office, under Award Number DE-EE0010733.
Specifically, the authors would like to recognize the following HydroGEN experts and capabilities: Anthony McDaniel, Ethan Hecht, Keith King, and Maria Syrigou at the Sandia National Laboratories Laser Heated Stagnation Flow Reactor for Characterizing Redox Chemistry, for their assistance with LSFR access, training, and loading of samples into the reactor for remote experimentation.
The authors gratefully acknowledge the use of XRD facilities within the Eyring Materials Center at Arizona State University, supported in part by NNCI-ECCS-1542160.
A portion of the research was performed using computational resources sponsored by the Department of Energy's Office of Energy Efficiency and Renewable Energy and located at the National Renewable Energy Laboratory.
The views and opinions of the authors expressed herein do not necessarily state or reflect those of the United States Government or any agency thereof.
Neither the United States Government nor any agency thereof, nor any of their employees, makes any warranty, expressed or implied, or assumes any legal liability or responsibility for the accuracy, completeness, or usefulness of any information, apparatus, product, or process disclosed, or represents that its use would not infringe privately owned rights.

Sandia National Laboratories is a multi-mission laboratory managed and operated by National Technology \& Engineering Solutions of Sandia, LLC (NTESS), a wholly owned subsidiary of Honeywell International Inc., for the U.S.\ Department of Energy's National Nuclear Security Administration (DOE/NNSA) under contract DE-NA0003525. This written work is authored by an employee of NTESS. The employee, not NTESS, owns the right, title and interest in and to the written work and is responsible for its contents. Any subjective views or opinions that might be expressed in the written work do not necessarily represent the views of the U.S.\ Government. The publisher acknowledges that the U.S.\ Government retains a non-exclusive, paid-up, irrevocable, world-wide license to publish or reproduce the published form of this written work or allow others to do so, for U.S.\ Government purposes. The DOE will provide public access to results of federally sponsored research in accordance with the DOE Public Access Plan.


\balance

\bibliography{references}
\bibliographystyle{rsc}

\end{document}


\maketitle
{\renewcommand{\thefootnote}{\fnsymbol{footnote}}%
\footnotetext[3]{These authors contributed equally to this work.}}

\section{Magnetic moments and oxidation-state assignments}\label{sec:esi:magmoments}

\begin{figure}[ht]
    \centering
    \includegraphics[width=0.6\columnwidth]{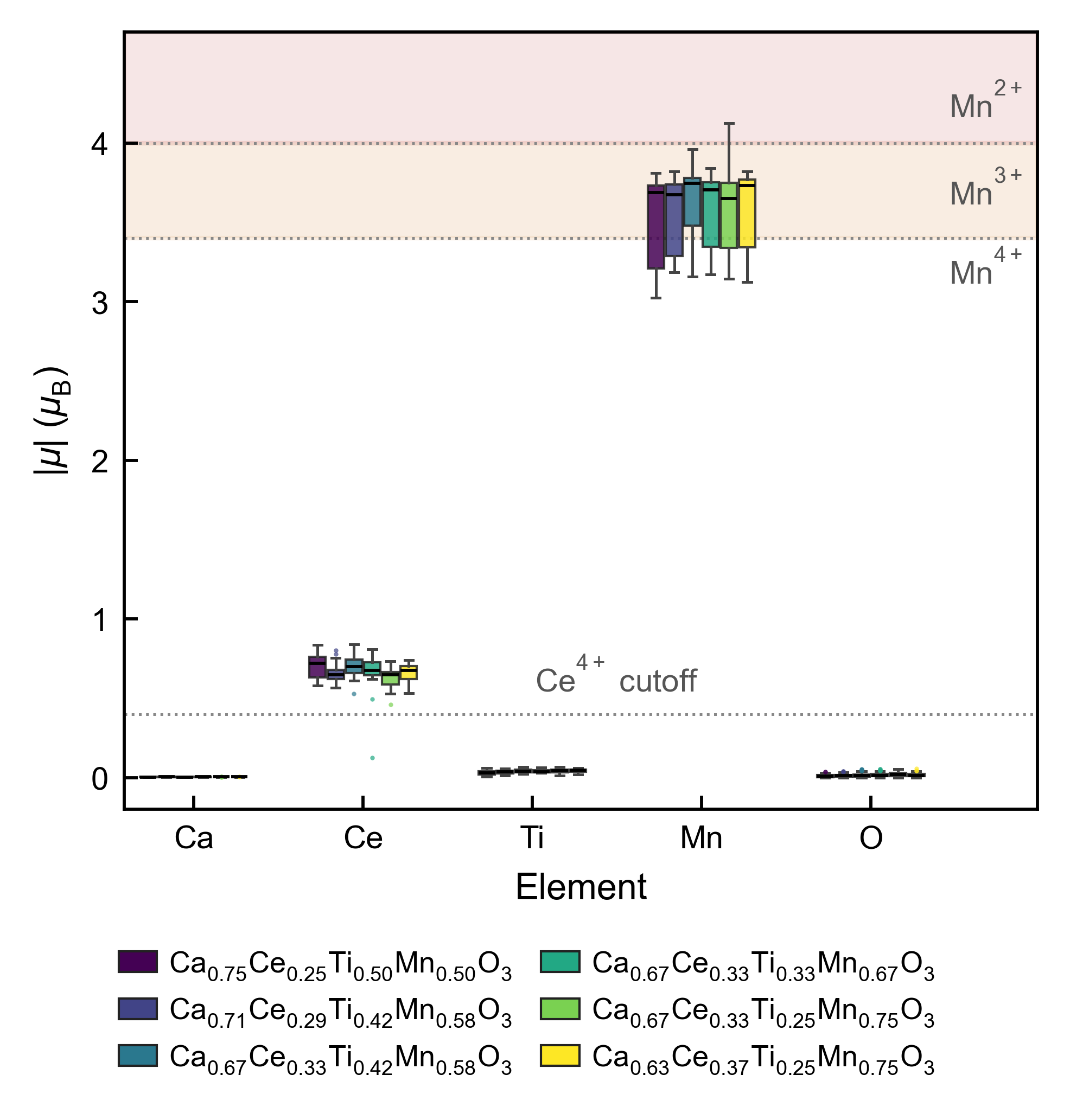}
    \caption{
Distribution of site-resolved $|\mu|$ from PAW-sphere integration, by element, across all six bulk \num{360}-atom special quasirandom structure (SQS) supercells.
Boxes span the interquartile range (IQR), and whiskers extend to $1.5 \times$ IQR.
Shaded bands indicate the Mn assignment ranges (\MnIII, \MnII); the dotted lines mark the \MnIV/\MnIII/\MnII cutoffs at \qty{3.4}{\bohrmagneton} and \qty{4.0}{\bohrmagneton} and the \CeIV/\CeIII cutoff at \qty{0.4}{\bohrmagneton}.
    }
    \label{fgr:esi:magmoments}
\end{figure}

Calculated oxidation-state assignments in CCTM perovskites were made from site-resolved magnetic moments obtained by integrating the spin density within projector augmented wave (PAW) spheres, as reported in the Vienna ab initio simulation package (VASP) \texttt{OUTCAR} file.
\Cref{fgr:esi:magmoments} shows the distribution of site-resolved $|\mu|$ by element across all six CCTM compositions.
Ca, Ti, and O sites carry negligible moments (median $|\mu| < \qty{0.05}{\bohrmagneton}$), consistent with \CaII, \TiIV, and \ce{O^2-}.
Ce moments fall almost entirely within a narrow window of \qtyrange[range-phrase = --, range-units = single]{0.45}{0.85}{\bohrmagneton} (\CeIII-like), with a single site at \qty{0.13}{\bohrmagneton} assigned to \CeIV-like $4f^0$ occupation.
Mn moments span \qtyrange[range-phrase = --, range-units = single]{3.0}{4.1}{\bohrmagneton} and are assigned to oxidation states using ionic crystal-field expectations:\cite{wexler_multiple_2023, xu_local_2024} $|\mu_{\ce{Mn}}| < \qty{3.4}{\bohrmagneton}$ for \MnIV ($3d^3$, $S = 3/2$), \qtyrange[range-phrase = --, range-units = single]{3.4}{4.0}{\bohrmagneton} for \MnIII ($3d^4$, $S = 2$), and \qty{> 4.0}{\bohrmagneton} for \MnII ($3d^5$, $S = 5/2$).
Some overlap between adjacent ranges is expected from local distortions and \ce{Mn-O} covalency, which modulate the spin density within PAW spheres.
These assignments are consistent with the projected density of states reported in \cref{sec:esi:dos}.

Of the \num{138} Ce sites across the six bulk supercells, \num{137} are labeled \CeIII and \num{1} \CeIV using a cutoff at $|\mu_{\ce{Ce}}| = \qty{0.4}{\bohrmagneton}$, placed in the empty interval between the lone low-moment site and the main cluster.
This binary labeling is coarse-grained, and the \CeIII/\CeIV labels classify calculated moments rather than measured oxidation states.
Ce sites classified as \CeIII-like have calculated PAW-sphere spin moments of \qtyrange[range-phrase = --, range-units = single]{0.45}{0.85}{\bohrmagneton}, compared with \qty{0.94}{\bohrmagneton} for fully localized \CeIII in hexagonal \ce{Ce2O3} within the same strongly constrained and appropriately normed functional with Hubbard $U$ corrections (SCAN+$U$) framework;\cite{saigautam_evaluating_2018} these moments indicate fractional $4f$ occupation, consistent with partial \ce{Ce}~$4f$--\ce{O}~$2p$ hybridization on the A sublattice.
No oxidation state was measured for the compositions synthesized in this work.
The single low-moment site occurs in CCTM2112 (\ce{Ca_{0.67}Ce_{0.33}Ti_{0.33}Mn_{0.67}O3}), the supercell referenced in the main text.
\Cref{tbl:esi:ce-counts} gives the per-composition breakdown.

We checked that these assignments are insensitive to the choice of real-space partition by comparing PAW-sphere moments to Bader moments, obtained with the Henkelman-group grid-based \texttt{bader} code\cite{henkelman_fast_2006,tang_gridbased_2009} using the self-consistent valence charge density as the charge reference for basin assignment and integrating the spin density over the resulting atomic basins.
Across the \num{2160} sites in the six bulk supercells, the mean absolute deviation (MAD) is \qty{0.014}{\bohrmagneton}; across the \num{32310} matched sites in the \num{90} defective supercells used in \cref{sec:esi:cfm-parity} the MAD for vacancy-induced moment changes $\Delta|\mu|$ is \qty{0.005}{\bohrmagneton}.
The Ce classification is unchanged, so it does not depend on the partition, although neither partition measures a formal oxidation state; \num{7} of the \num{276} Mn sites, all within \qty{0.04}{\bohrmagneton} of the \qty{3.4}{\bohrmagneton} \MnIV/\MnIII cutoff, reclassify from \MnIII to \MnIV under Bader partitioning (shifting the bulk counts from \numlist{84;191;1} to \numlist{91;184;1}), consistent with the near-boundary overlap noted above.

The \CeIII-like state does not depend on the initial magnetic moments.
Our earlier calculations on CCTM2112\cite{wexler_multiple_2023} initialized the Ce and Mn moments at \qtylist[list-units = single]{0.6;4.0}{\bohrmagneton}, whereas the present calculations used \qtylist[list-units = single]{1.0;5.0}{\bohrmagneton}.
Two single-point SCAN+$U$ calculations of the relaxed CCTM2112 supercell, started from the earlier and the present initial moments, converge to the same electronic state: every site moment agrees within \qty{0.007}{\bohrmagneton}, and their total energies, both computed with VASP 6.5.1, differ by \qty{0.5}{\meV} per \num{360}-atom supercell.
Because this test keeps the geometry fixed, it does not exclude a \CeIV-like state with its own relaxed structure.

\Cref{tbl:esi:dmu} summarizes the vacancy-induced changes in the local moments by element for all six compositions (response to oxygen-vacancy formation in the \num{90} defective supercells).
In every composition, the changes occur on Ce and Mn, with smaller induced moments on O, whereas individual Ti and Ca moments change by less than \qty{0.05}{\bohrmagneton}; Ce and Mn are therefore the redox-active cations throughout the composition range studied.
Within a supercell, individual Mn sites can change in opposite directions, some consistent with reduction and others with oxidation, so the summed site-level magnitudes of the Mn changes (\qtyrange[range-phrase = --, range-units = single]{1.1}{2.7}{\bohrmagneton}) exceed the net Mn change, as also found in the earlier study;\cite{wexler_multiple_2023} for Ce, the summed site-level magnitudes (\qtyrange[range-phrase = --, range-units = single]{0.86}{1.03}{\bohrmagneton}) are closer to the net change.
These moment changes are computational indicators of the redox response rather than a measure of electron transfer.

\begin{table}[ht]
    \caption{
Per-composition counts of \CeIII and \CeIV sites in the bulk \num{360}-atom SQS supercells, assigned from PAW-sphere magnetic moments using $|\mu_{\ce{Ce}}| > \qty{0.4}{\bohrmagneton}$ as the threshold for $4f^1$ occupation.
The single low-moment site occurs in CCTM2112.
    }
    \label{tbl:esi:ce-counts}
    \centering
    \small
    \begin{tabular*}{\columnwidth}{@{\extracolsep{\fill}}lccccc}
        \hline
        Composition & \XCe & \XMn & $N_{\ce{Ce}}$ & $N_{\CeIII}$ & $N_{\CeIV}$ \\
        \hline
        \ce{Ca_{0.75}Ce_{0.25}Ti_{0.50}Mn_{0.50}O3}  & 0.25 & 0.50 & 18 & 18 & 0 \\
        \ce{Ca_{0.71}Ce_{0.29}Ti_{0.42}Mn_{0.58}O3}  & 0.29 & 0.58 & 21 & 21 & 0 \\
        \ce{Ca_{0.67}Ce_{0.33}Ti_{0.33}Mn_{0.67}O3}  & 0.33 & 0.67 & 24 & 23 & 1 \\
        \ce{Ca_{0.63}Ce_{0.37}Ti_{0.25}Mn_{0.75}O3}  & 0.37 & 0.75 & 27 & 27 & 0 \\
        \ce{Ca_{0.67}Ce_{0.33}Ti_{0.42}Mn_{0.58}O3}  & 0.33 & 0.58 & 24 & 24 & 0 \\
        \ce{Ca_{0.67}Ce_{0.33}Ti_{0.25}Mn_{0.75}O3}  & 0.33 & 0.75 & 24 & 24 & 0 \\
        \hline
    \end{tabular*}
\end{table}

\begin{table}[ht]
    \caption{
Vacancy-induced changes in the PAW-sphere moment magnitudes by element, averaged over the \num{15} oxygen vacancies of each composition (\unit{\bohrmagneton} per vacancy).
Net: signed change in $|\mu|$ summed over all sites of the element.
Summed: sum of the site-level magnitudes $|\Delta|\mu||$ over all sites of the element.
Ti and Ca changes are below \qty{0.05}{\bohrmagneton} at every site and are omitted.
    }
    \label{tbl:esi:dmu}
    \centering
    \small
    \begin{tabular*}{\columnwidth}{@{\extracolsep{\fill}}lccccccc}
        \hline
        & & & \multicolumn{2}{c}{Ce} & \multicolumn{2}{c}{Mn} & O \\
        \cline{4-5} \cline{6-7} \cline{8-8}
        Composition & \XCe & \XMn & Net & Summed & Net & Summed & Summed \\
        \hline
        \ce{Ca_{0.75}Ce_{0.25}Ti_{0.50}Mn_{0.50}O3}  & 0.25 & 0.50 & 0.94 & 1.00 & 0.67 & 1.10 & 0.44 \\
        \ce{Ca_{0.71}Ce_{0.29}Ti_{0.42}Mn_{0.58}O3}  & 0.29 & 0.58 & 0.76 & 0.88 & 0.70 & 1.36 & 0.55 \\
        \ce{Ca_{0.67}Ce_{0.33}Ti_{0.33}Mn_{0.67}O3}  & 0.33 & 0.67 & 0.70 & 0.89 & 0.65 & 2.15 & 0.70 \\
        \ce{Ca_{0.63}Ce_{0.37}Ti_{0.25}Mn_{0.75}O3}  & 0.37 & 0.75 & 0.55 & 0.86 & 1.13 & 2.04 & 0.68 \\
        \ce{Ca_{0.67}Ce_{0.33}Ti_{0.42}Mn_{0.58}O3}  & 0.33 & 0.58 & 0.86 & 1.03 & 0.52 & 1.91 & 0.71 \\
        \ce{Ca_{0.67}Ce_{0.33}Ti_{0.25}Mn_{0.75}O3}  & 0.33 & 0.75 & 0.70 & 1.02 & 0.57 & 2.66 & 0.87 \\
        \hline
    \end{tabular*}
\end{table}

\section{Density of states and metallic character}\label{sec:esi:dos}

\begin{figure}[ht]
    \centering
    \includegraphics[width=\textwidth]{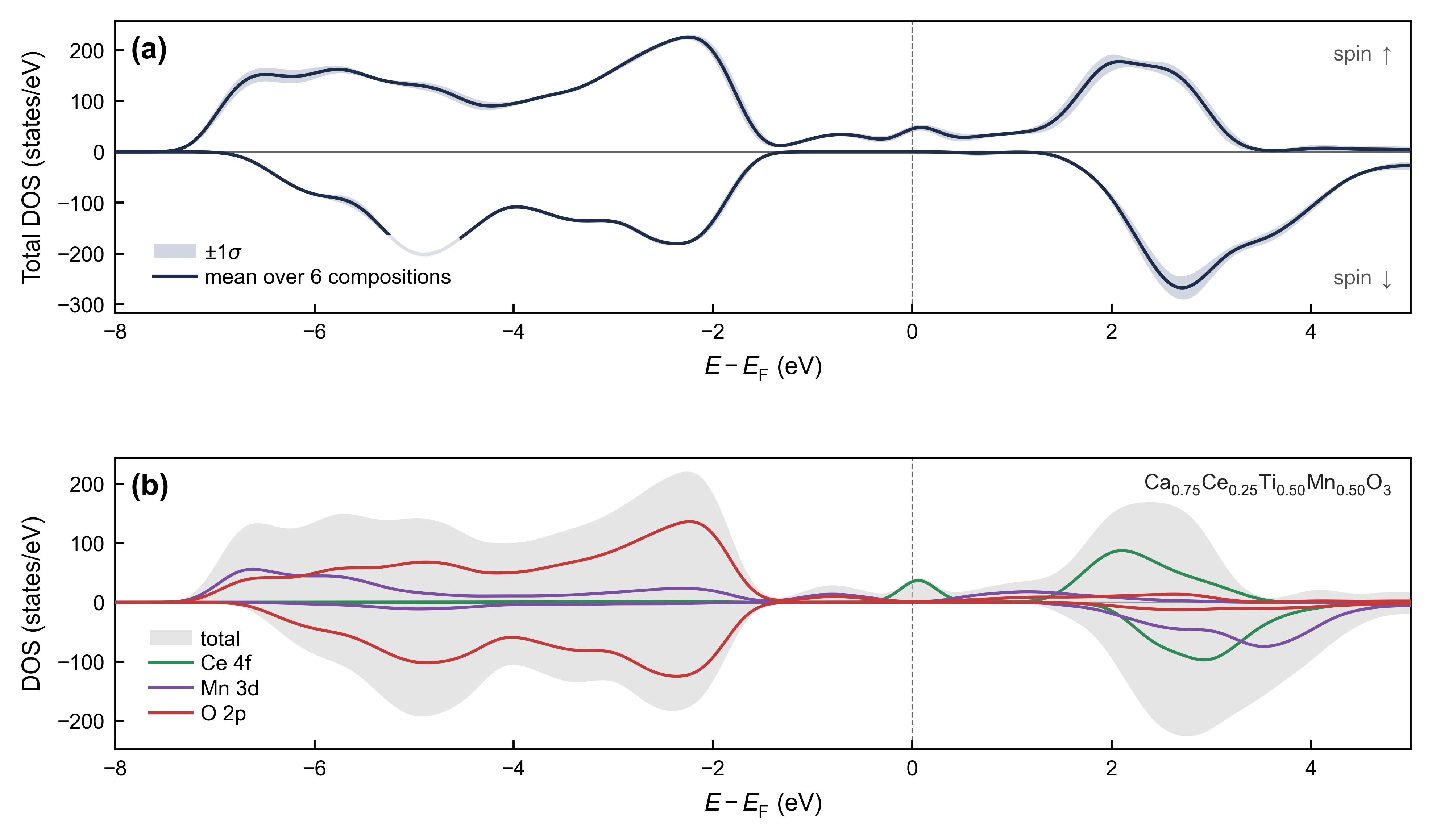}
    \caption{
Spin-resolved density of states of CCTM perovskites at the $\Gamma$ point.
(a) Total DOS averaged across the six bulk \num{360}-atom SQS supercells (dark line: mean; shaded envelope: $\pm 1\sigma$ across compositions).
(b) Element- and orbital-resolved decomposition for \ce{Ca_{0.75}Ce_{0.25}Ti_{0.50}Mn_{0.50}O3}: total DOS (gray fill), Ce $4f$ (green), Mn $3d$ (purple), and O $2p$ (red).
The dashed vertical line in each panel marks the Fermi level.
    }
    \label{fgr:esi:dos}
\end{figure}

\Cref{fgr:esi:dos}a shows the spin-resolved total density of states for the six bulk CCTM compositions, computed at the supercell zone center on the \num{360}-atom $\Gamma$-relaxed cells.
The pattern is consistent across the composition range: in each composition, the spin-up channel carries a finite DOS at the Fermi level (\qtyrange[range-phrase = --, range-units = single]{0.1}{0.2}{\states\per\eV\per\atom}), while the spin-down channel is gapped around $E_\mathrm{F}$.
The compositional spread of the total DOS is small (\cref{fgr:esi:dos}a, $\pm 1\sigma$ envelope), indicating that the metallic character is a family-wide feature rather than a property of any single composition.
At the SCAN+$U$ level adopted here, CCTM perovskites are therefore metallic and fully spin-polarized at $E_\mathrm{F}$ across the composition range studied.

\Cref{fgr:esi:dos}b decomposes the DOS of \ce{Ca_{0.75}Ce_{0.25}Ti_{0.50}Mn_{0.50}O3} by element and orbital character.
The states crossing $E_\mathrm{F}$ are dominated by partially filled Ce $4f$ orbitals, with a smaller Mn $3d$ contribution.
The valence band is principally O $2p$ with hybridized Mn $3d$ character (\qtyrange[range-units = single]{-7}{-2}{\eV}); the conduction band (\qtyrange[range-phrase = --, range-units = single]{1}{4}{\eV}) carries both Mn $3d$ and unoccupied Ce $4f$ weight.
That the Ce $4f$ band spans $E_\mathrm{F}$ rather than localizing fully below it reflects the bandwidth of the dense Ce sublattice relative to $U_{\ce{Ce}} = \qty{2.0}{\eV}$,\cite{saigautam_evaluating_2018} and is consistent with the delocalized-reduction picture: electrons released on vacancy formation are accommodated across multiple partially filled Ce $4f$ sites rather than via discrete \CeIV $\rightarrow$ \CeIII transitions.

\section{Crystal-feature model parity}\label{sec:esi:cfm-parity}

The crystal-feature model (CFM) introduced in the main text approximates the oxygen-vacancy formation energy \Ev at each symmetry-distinct site as a linear combination of two physically motivated descriptors of the local environment: the mean crystal bond dissociation energy \Eb and the mean crystal reduction potential \Vr of the six first-nearest-neighbor cations (four A-site, two B-site) of the vacant oxygen.
The per-cation crystal bond dissociation energies and crystal reduction potentials follow the definitions of ref.~\citenum{wexler_factors_2021} (eqn~(2) and (5) therein); see \cref{sec:esi:atomic-props} for the values used here and for the averaging convention.
We fit the regression
\begin{equation}
    \Ev = \bb \, \Eb + \br \, \Vr + c
    \label{eqn:cfm}
\end{equation}
on the full density functional theory (DFT) dataset of \num{90} symmetry-distinct oxygen sites across the six CCTM compositions of main-text \cref{M-tbl:cctm_compositions}.
Ref.~\citenum{wexler_factors_2021} additionally included the per-composition energy above the convex hull \Ehull as a descriptor in their parent model.
We omit it here because our composition set contains only six unique \Ehull values across the narrow range \qtyrange[range-phrase = --, range-units = single]{18}{39}{\meV\per\atom}, which is insufficient to constrain \bhull (see the descriptor-subset comparison at the end of this section).

We evaluate \cref{eqn:cfm} under two complementary cross-validation schemes.
Random \qty{50}{\percent}/\qty{50}{\percent} train--test splits, repeated \num{1000} times following the protocol of ref.~\citenum{wexler_factors_2021}, measure prediction accuracy at held-out sites within the sampled composition space.
Leave-one-group-out (LOGO) cross-validation with composition as the group, in which each of the six compositions is held out in turn and the model is fit on the remaining five, measures generalization to compositions outside the training set; this is the relevant criterion for subsequent screening of the per-cycle oxygen-exchange capacity \Dd across composition space.\cite{witmanMatFoldSystematicInsights2025}
For each split, we fit a Huber regressor on the training partition and compute the mean absolute error (MAE) on the held-out partition; for random splits, we report the mean and standard deviation of MAE across the \num{1000} splits, and for LOGO, the mean and standard deviation across the six folds.
Reported coefficients are obtained from a single Huber fit (the scikit-learn \texttt{HuberRegressor}\cite{pedregosa_scikitlearn_2011} with its default settings, $\epsilon = 1.35$ and $\alpha = \num{e-4}$) to the full \num{90}-site dataset; \qty{95}{\percent} confidence intervals are obtained by bootstrap resampling of sites with replacement (\num{2000} resamples).
An ordinary least squares fit on the same data yields coefficients within these bootstrap confidence intervals and MAE within \qty{0.001}{\eV}.

\begin{figure}[ht]
    \centering
    \includegraphics[width=\textwidth]{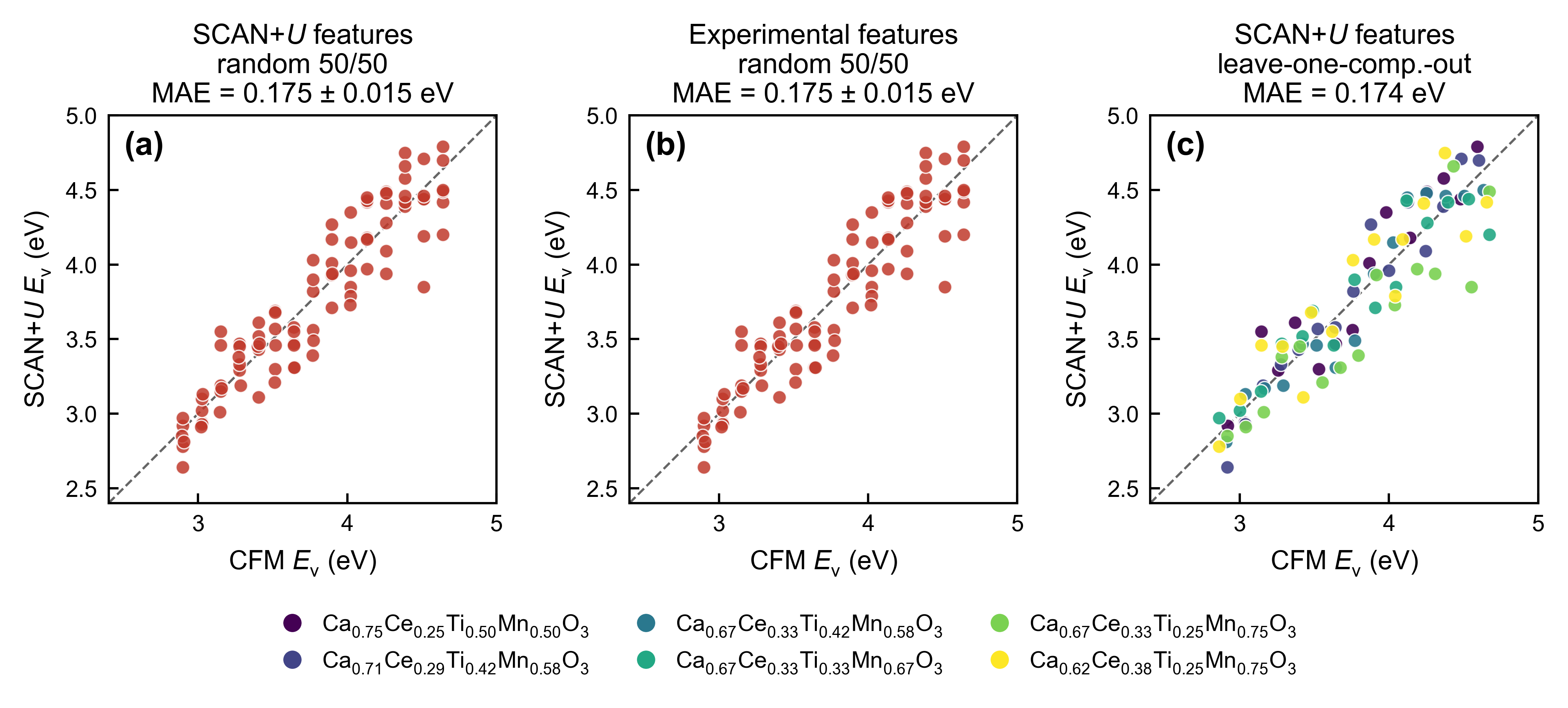}
    \caption{
Parity plots comparing SCAN+$U$ oxygen-vacancy formation energies of CCTM perovskites with crystal-feature-model predictions, using per-cation features from (a) the SCAN+$U$ binary-oxide calculations and (b) the compilation of experimental binary-oxide thermochemistry, both reported in ref.~\citenum{wexler_factors_2021}, each under random \qty{50}{\percent}/\qty{50}{\percent} train--test splits; and (c) the SCAN+$U$ features under leave-one-composition-out cross-validation, with points colored by composition (legend).
The dashed line indicates perfect agreement; all three panels share the same axis range to allow direct visual comparison.
Panels (a) and (b) yield MAE = \qty{0.175 \pm 0.015}{\eV} across \num{1000} random \qty{50}{\percent}/\qty{50}{\percent} train--test splits; panel (c) yields MAE = \qty{0.174 \pm 0.042}{\eV} under leave-one-composition-out cross-validation.
Fitted coefficients (full-data Huber fit; bootstrap \qty{95}{\percent} confidence intervals in brackets), SCAN+$U$ features: $\bb = +\num{2.00}~[+\num{1.78}, +\num{2.23}]$, $\br = -\num{1.34}~[-\num{1.64}, -\num{1.04}]$, $c = -\num{4.26}~[-\num{4.94}, -\num{3.58}]$.
Experimental features: $\bb = +\num{1.52}~[+\num{1.31}, +\num{1.74}]$, $\br = -\num{1.57}~[-\num{1.92}, -\num{1.22}]$, $c = -\num{3.85}~[-\num{4.52}, -\num{3.16}]$.
    }
    \label{fgr:cfm-parity}
\end{figure}

\Cref{fgr:cfm-parity} shows parity plots of SCAN+$U$ \Ev against the CFM-predicted \Ev for two parameterizations of the model: one in which the per-cation $E_{\mathrm{b},n}$ and $V_{\mathrm{r},n}$ are taken from the SCAN+$U$ binary-oxide calculations of ref.~\citenum{wexler_factors_2021} (\cref{fgr:cfm-parity}a), and one in which they are taken from the compilation of experimental binary-oxide thermochemistry in the same work (\cref{fgr:cfm-parity}b).
Both parameterizations yield MAE = \qty{0.175 \pm 0.015}{\eV} across the \num{1000} random splits, and MAE = \qty{0.174 \pm 0.042}{\eV} under LOGO.
The random-split and LOGO MAEs are equal within the across-fold standard deviation, indicating that the model generalizes to held-out compositions essentially as well as to held-out sites within sampled compositions, supporting the use of the CFM for composition-space screening of \Dd.
The MAE is small relative to the spread of \Ev across the dataset (\qty{2.15}{\eV}, main-text \cref{M-fgr:3}), corresponding to a relative error of approximately \qty{8}{\percent} of the \Ev range.

The identical MAEs of the two parameterizations, despite slightly different fitted \bb and \br values, arise because the SCAN+$U$ and experimental $E_{\mathrm{b},n}$ and $V_{\mathrm{r},n}$ tabulated by ref.~\citenum{wexler_factors_2021} agree closely across the cations present in CCTM (MAE of \qty{0.04}{\eV} between SCAN+$U$ and experimental $E_\mathrm{b}$ across their perovskite-oxide composition space); any feature-scale difference between the two sources is compensated by a corresponding adjustment in \bb and \br.
The agreement indicates that the CFM prediction is insensitive to whether the per-cation features are supplied from SCAN+$U$ calculations or from experimental thermochemistry, within their binary-oxide reference set.

The fitted coefficient signs match the extended Born--Haber picture of vacancy formation, applicable to either parameterization.
A positive \bb yields higher predicted \Ev at sites whose nearest-neighbor cations have stronger crystal bonds (larger $E_\mathrm{b}$), because vacancy formation breaks four A--O and two B--O bonds in proportion to their crystal bond dissociation energies.
A negative \br yields lower predicted \Ev at sites whose nearest-neighbor cations are more easily reduced (less negative $V_\mathrm{r}$), because the two electrons left behind by oxygen removal localize on the cation sublattice at lower energy cost when the surrounding cations are more reducible.

\Cref{tbl:cfm-subsets} reports the LOGO MAE for the CFM under selected subsets of the candidate descriptors \{\Eb, \Vr, \Ehull\} and supports the choice of \Eb and \Vr alone in \cref{eqn:cfm}.
Removing \Ehull from the parent three-descriptor model changes the LOGO MAE from \qty{0.175 \pm 0.047}{\eV} to \qty{0.174 \pm 0.042}{\eV}, a difference well inside the across-fold standard deviation.
Refitting the three-descriptor model on the full \num{90}-site dataset yields $\bhull = -\num{4.77}$ with a bootstrap \qty{95}{\percent} confidence interval of $[-\num{13.13}, +\num{3.74}]$ that spans zero, consistent with there being only six unique \Ehull values across the dataset to constrain it.
By contrast, removing \Eb individually raises the LOGO MAE to \qty{0.25}{\eV} and removing \Vr raises it to \qty{0.36}{\eV}; \Ehull on its own yields a LOGO MAE of \qty{0.50}{\eV}, comparable to the mean absolute deviation of \Ev across the dataset (\qty{0.49}{\eV}) and therefore no better than predicting the dataset mean for every site.

\begin{table}[ht]
    \caption{
Performance of the crystal-feature model under selected subsets of the candidate descriptors \{\Eb, \Vr, \Ehull\}, evaluated by leave-one-composition-out cross-validation (LOGO MAE reported as mean $\pm$ standard deviation across the six folds).
\Eb and \Vr each contribute substantively to the prediction; including \Ehull does not change the LOGO MAE within fold-to-fold variance, and \Ehull is therefore excluded from \cref{eqn:cfm}.
\textsuperscript{\emph{a}}~Descriptor set used in \cref{eqn:cfm}.
    }
    \label{tbl:cfm-subsets}
    \centering
    \begin{tabular}{lc}
        \toprule
        Descriptors & LOGO MAE (\unit{\eV}) \\
        \midrule
        \Eb, \Vr, \Ehull & \num{0.175 \pm 0.047} \\
        \Eb, \Vr\textsuperscript{\emph{a}} & \num{0.174 \pm 0.042} \\
        \Eb~~only & \num{0.253 \pm 0.028} \\
        \Vr~~only & \num{0.364 \pm 0.059} \\
        \Ehull~~only & \num{0.496 \pm 0.044} \\
        \bottomrule
    \end{tabular}
\end{table}

\section{Vacancy--vacancy electrostatic upper bound}\label{sec:esi:vv-electrostatic}

The statistical-mechanical model of main-text Methods (formalized in \cref{sec:esi:thermo}) treats oxygen vacancies as independent point defects: configurational entropy is taken in its ideal-solution form, and the per-vacancy reduction enthalpy is approximated by the dilute-limit DFT value \Ev.
This functional form is justified, beyond its agreement with experimental $\delta(T,\POtwo)$, when the vacancy--vacancy interaction $V_{\mathrm{vv}}$ at experimentally accessed vacancy densities is small compared with the thermal energy $\kB T$ under reduction conditions.
We bound $V_{\mathrm{vv}}$ from above using Thomas--Fermi screening in the metallic host (\cref{sec:esi:dos}).

In the metallic host, two notional $+2e$ point charges (an upper bound on any residual local charge a neutral vacancy might carry, corresponding to the formal Kr\"oger--Vink ${V_\mathrm{O}^{\bullet\bullet}}$ defect) interact via a Yukawa-screened potential
\begin{equation}\label{eqn:tf-screened}
    V_{\mathrm{vv}}(r) = \dfrac{(2e)^2}{4\pi\varepsilon_0\, r}\,\exp\!\left(-r / \lambda_\mathrm{TF}\right),
    \qquad
    \lambda_\mathrm{TF} = \sqrt{\dfrac{\varepsilon_0}{e^2\, g(E_\mathrm{F})}},
\end{equation}
where $g(E_\mathrm{F})$ is the Fermi-level density of states per unit volume.
Using the smoothed mean total DOS at $E_\mathrm{F}$ across the six bulk compositions (\cref{fgr:esi:dos}a, \qty{\sim 45}{\states\per\eV} per \num{360}-atom supercell at $V_{\mathrm{cell}} \approx \qty{4080}{\angstrom\cubed}$) gives $\lambda_\mathrm{TF} \approx \qty{0.71}{\angstrom}$.
Reducing $g(E_\mathrm{F})$ to \qty{\sim 20}{\states\per\eV} per supercell as a conservative-low sensitivity case extends $\lambda_\mathrm{TF}$ to \qty{1.06}{\angstrom}.

\begin{figure}[ht]
    \centering
    \includegraphics[width=0.6\columnwidth]{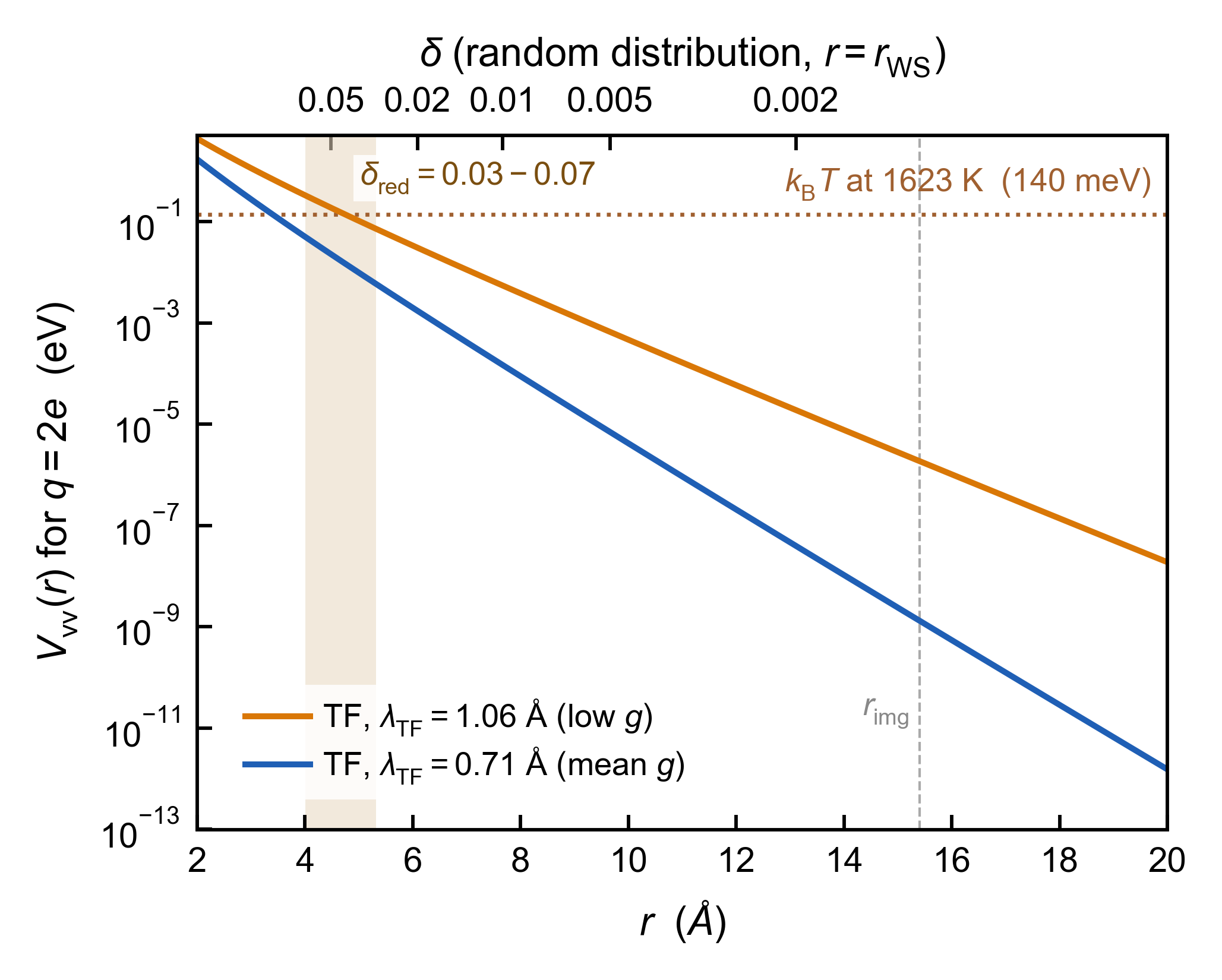}
    \caption{
Thomas--Fermi screened vacancy--vacancy electrostatic interaction $V_{\mathrm{vv}}(r)$ (\cref{eqn:tf-screened}) for a hypothetical $(2e)^2$ point-charge pair, at the family-mean $g(E_\mathrm{F})$ from \cref{fgr:esi:dos}a (blue, $\lambda_\mathrm{TF} = \qty{0.71}{\angstrom}$) and at a conservative-low $g(E_\mathrm{F})$ (orange, $\lambda_\mathrm{TF} = \qty{1.06}{\angstrom}$).
Shaded vertical band: range of Wigner--Seitz radii (\cref{eqn:rWS}) for the reduction-side oxygen off-stoichiometry $\delta_\mathrm{red} \approx \numrange[range-phrase = \text{--}]{0.03}{0.07}$ accessed by TGA (\cref{sec:esi:tga}).
Top axis: corresponding $\delta$ at $r = r_\mathrm{WS}$.
Dotted horizontal line: thermal energy $\kB T$ at the TGA reduction temperature (\qty{1623}{\kelvin}).
Light dashed vertical line: vacancy--vacancy periodic-image separation in the $(3\times3\times2)$ SQS supercells, $r_\mathrm{img} \approx \qty{15.4}{\angstrom}$, at which the screened interaction falls to \qty{\sim 1}{\nano\eV}, indicating that the screened electrostatic contribution of periodic images to single-vacancy \Ev is negligible.
    }
    \label{fgr:esi:vv-electrostatic}
\end{figure}

The vacancy--vacancy separation relevant to the statistical-mechanical model is set by the experimental oxygen off-stoichiometry $\delta$.
Across the reduction conditions accessed by thermogravimetric analysis (TGA; \cref{sec:esi:tga}, peak temperature \qty{\sim 1623}{\kelvin}), $\delta_\mathrm{red}$ spans approximately \numrange[range-phrase = --]{0.03}{0.07}.
For vacancies randomly distributed on the oxygen sublattice at density $n_\mathrm{v} = 3\delta / V_\mathrm{fu}$ ($V_\mathrm{fu} \approx \qty{56.6}{\angstrom\cubed}$), the Wigner--Seitz radius is
\begin{equation}\label{eqn:rWS}
    r_\mathrm{WS} = \left(\dfrac{3}{4\pi n_\mathrm{v}}\right)^{1/3}
                  = \left(\dfrac{V_\mathrm{fu}}{4\pi\delta}\right)^{1/3},
\end{equation}
giving $r_\mathrm{WS} \approx \qty{5.3}{\angstrom}$ at $\delta_\mathrm{red} = 0.03$ and $\qty{4.0}{\angstrom}$ at $\delta_\mathrm{red} = 0.07$.
At these separations, \cref{eqn:tf-screened} yields $V_{\mathrm{vv}} \approx \qtyrange[range-phrase = \text{--}, range-units = single]{6}{50}{\meV}$ at the family-mean $\lambda_\mathrm{TF}$ (\cref{fgr:esi:vv-electrostatic}, blue curve).
Comparison with $\kB T \approx \qty{140}{\meV}$ at the reduction temperature places the worst-case Yukawa interaction a factor of approximately \numrange[range-phrase = --]{3}{20} below the thermal scale across the experimentally accessed range, supporting the noninteracting-vacancy approximation of the statistical-mechanical model.
The conservative-low $\lambda_\mathrm{TF}$ sensitivity (orange curve) lies on both sides of $\kB T$ over the same $r$ range; even here, the interaction remains within a factor of a few of $\kB T$, and the corrections discussed next reduce it further.

Three considerations indicate that this estimate overstates the actual interaction.
First, the $(2e)^2$ prefactor assumes the full Kr\"oger--Vink charge state; the neutral $V_\mathrm{O}^0$ defects treated in the formation-energy framework carry zero formal charge in the metallic host, so the actual leading-order monopole--monopole interaction vanishes and the residual contribution is dominated by higher multipoles and elastic terms, both of which are typically meV-scale at the relevant separations.
Second, Thomas--Fermi captures the leading-order metallic screening but neglects the wavevector dependence of the dielectric function (Lindhard / random-phase-approximation corrections) and the local depletion of $g(E_\mathrm{F})$ in the immediate neighborhood of the vacancy, both of which further suppress $V_{\mathrm{vv}}$ at the short distances relevant here.
Third, a random distribution of vacancies is an upper bound on the mean-field interaction: any short-range configurational repulsion between like-charged defects would increase $r_\mathrm{WS}$ relative to the random value used in \cref{eqn:rWS}.
Each of these corrections reduces $V_{\mathrm{vv}}$ further below the Yukawa estimate plotted in \cref{fgr:esi:vv-electrostatic}.

These bounds, together with the empirical agreement between the statistical-mechanical model and TGA $\delta(T, \POtwo)$ isotherms, provide a physical justification for the ideal-solution functional form of the model under experimentally accessed reduction conditions.
At higher $\delta$ than experimentally probed (e.g.,~$\delta \gtrsim 0.10$, where $r_\mathrm{WS} < \qty{3.6}{\angstrom}$ and $V_{\mathrm{vv}}$ approaches $\kB T$ even at the family-mean $\lambda_\mathrm{TF}$), interactions become nonnegligible and a more elaborate treatment (a cluster expansion, a Bragg--Williams model, or Wagner-type defect chemistry) would be required.

\section{Powder X-ray diffraction}\label{sec:esi:xrd}

\begin{figure}[ht]
    \centering
    \includegraphics[width=0.6\textwidth]{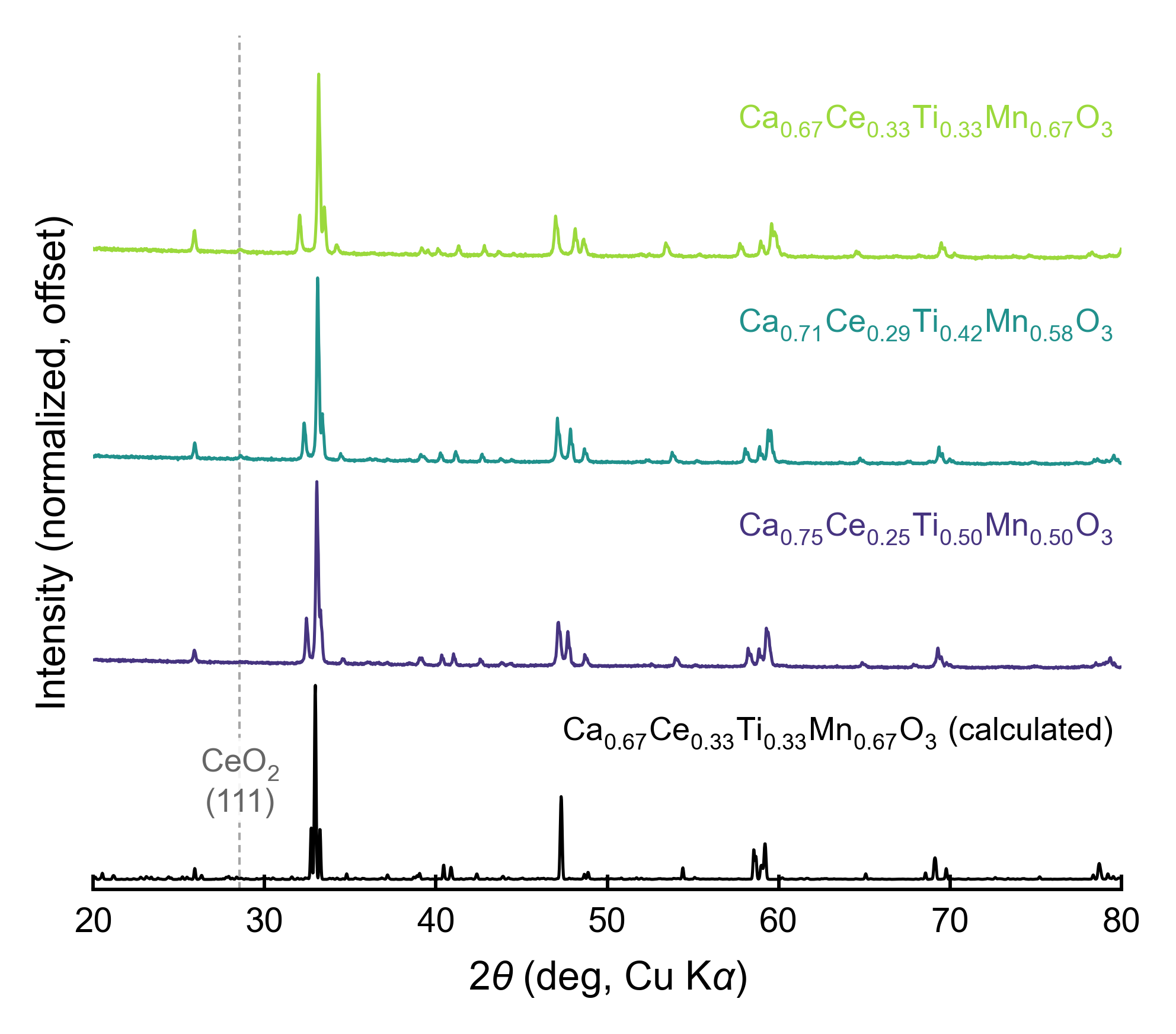}
    \caption{
Powder X-ray diffraction patterns of the three CCTM compositions synthesized for the TGA and LSFR experiments (top three traces, ordered top-to-bottom by descending Mn content) and a calculated reference pattern for CCTM2112 (bottom trace).
The dashed vertical line at $2\theta = \qty{28.55}{\degree}$ marks the position of the \ce{CeO2}(111) reflection, the strongest reflection of \ce{CeO2} and the expected indicator of cerium exsolution, observed as a minor \ce{CeO2} second phase in a prior refinement of a higher-Ce CCTM composition.\cite{wexler_multiple_2023}
The calculated reference was generated from the \num{360}-atom $3 \times 3 \times 2$ SQS supercell of CCTM2112 using the \texttt{XRDCalculator} module of \textsc{pymatgen} with Cu~K$\alpha_1$ radiation (\qty{1.54056}{\angstrom}), convolved with a Gaussian profile of FWHM \qty{0.10}{\degree}; experimental patterns were collected with Cu~K$\alpha$ radiation (see main-text Methods).
    }
    \label{fgr:xrd}
\end{figure}

We characterized the phase purity of the three CCTM compositions synthesized for the TGA and laser-heated stagnation flow reactor (LSFR) experiments by powder X-ray diffraction; full chemical formulas appear in main-text \cref{M-tbl:cctm_delta_delta}, and synthesis and acquisition conditions are described in main-text Methods.
\Cref{fgr:xrd} stacks the three experimental patterns with a calculated reference pattern for CCTM2112.
All three patterns reproduce the peak positions and relative intensities of the calculated orthorhombic perovskite reference, with no \ce{CeO2}(111) reflection at $2\theta = \qty{28.55}{\degree}$\cite{wexler_multiple_2023} (dashed vertical line) and no additional reflections attributable to secondary phases.
No crystalline secondary phase was detected in any of the three samples, which are therefore majority single-phase orthorhombic perovskites.
Laboratory XRD does not exclude nanocrystalline or amorphous \ce{CeO2}, whose reflections would be broadened into the background.

The calculated reference was generated from the same \num{360}-atom $3 \times 3 \times 2$ SQS supercell of CCTM2112 used in the DFT calculations (main-text Methods) with the \texttt{XRDCalculator} module of \textsc{pymatgen}\cite{ong_python_2013} using Cu~K$\alpha_1$ radiation (\qty{1.54056}{\angstrom}); the calculated line pattern was convolved with a Gaussian of FWHM \qty{0.10}{\degree} for visual comparison with the experimental traces.

\begin{figure}[ht]
    \centering
    \includegraphics[width=0.6\textwidth]{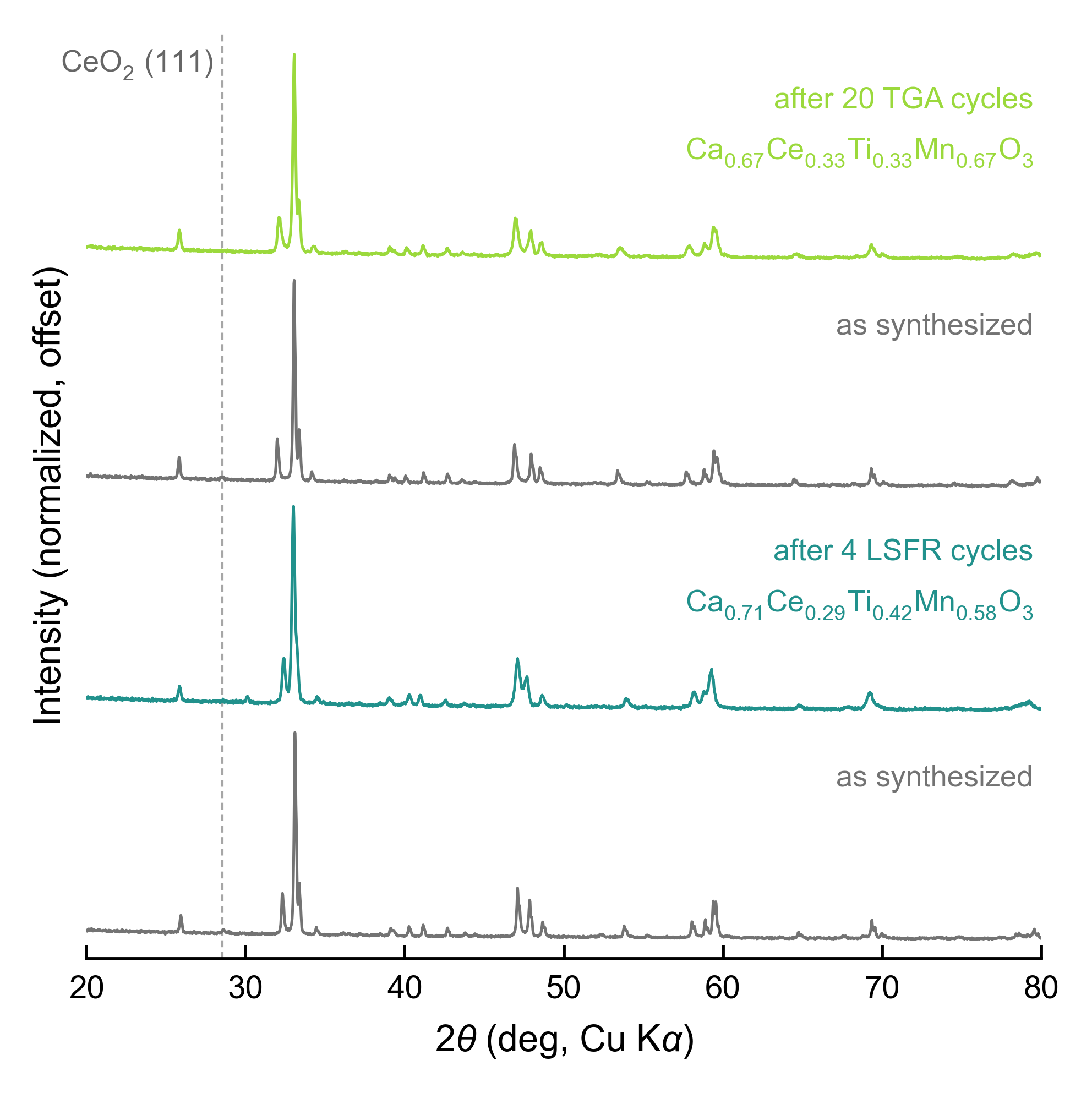}
    \caption{
Powder X-ray diffraction patterns of \ce{Ca_{0.71}Ce_{0.29}Ti_{0.42}Mn_{0.58}O3} ($\XCe = 0.29$) as synthesized and after the four LSFR cycles of main-text \cref{M-fgr:6} (bottom two traces), and of CCTM2112 (\ce{Ca_{0.67}Ce_{0.33}Ti_{0.33}Mn_{0.67}O3}) as synthesized and after \num{20} TGA cycles (Protocols~A, B, and~C, followed by further cycling; top two traces).
Each trace is normalized to its maximum and offset vertically; experimental patterns were collected with Cu~K$\alpha$ radiation.
The dashed vertical line at $2\theta = \qty{28.55}{\degree}$ marks the position of the \ce{CeO2}(111) reflection.
    }
    \label{fgr:esi:xrd-postcycling}
\end{figure}

\Cref{fgr:esi:xrd-postcycling} compares the patterns of two samples recovered after cycling with their as-synthesized patterns: \ce{Ca_{0.71}Ce_{0.29}Ti_{0.42}Mn_{0.58}O3} ($\XCe = 0.29$) after the four LSFR cycles of main-text \cref{M-fgr:6}, and CCTM2112 after \num{20} TGA cycles (Protocols~A, B, and~C, followed by further cycling).
Neither post-cycling pattern shows secondary phases, decomposition, or structural changes relative to the corresponding as-synthesized pattern.

\section{Thermogravimetric analysis}\label{sec:esi:tga}

\begin{figure}[htbp]
    \centering
    \includegraphics[width=0.6\textwidth]{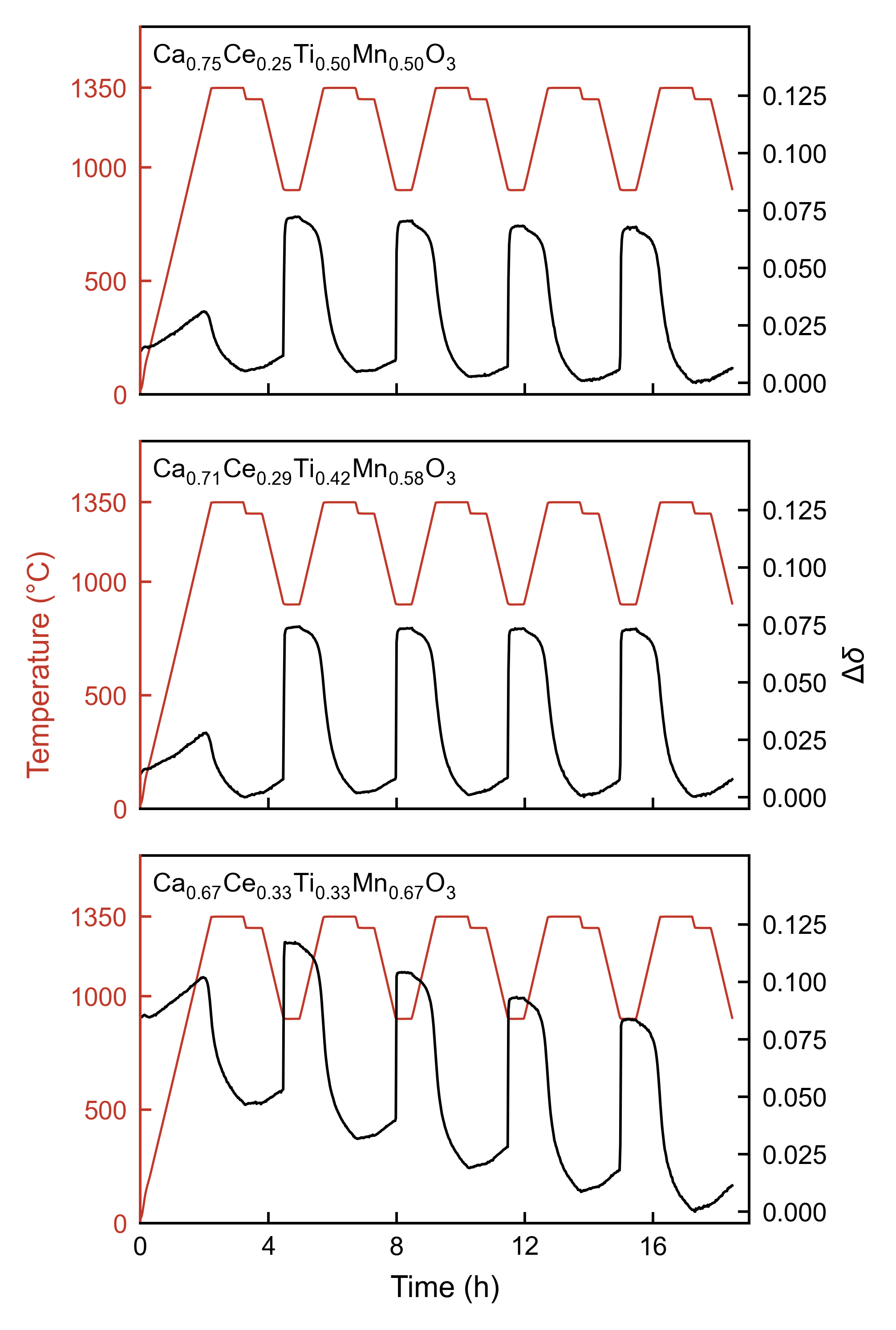}
    \caption{
Thermogravimetric Protocol~B cycling traces for the three CCTM compositions, ordered top-to-bottom by ascending Mn content.
Each panel shows temperature (red, left axis) and the change in oxygen content, plotted relative to the most-reduced state of the run (black, right axis; the per-cycle excursion of this trace equals \Dd), over five reduction--reoxidation cycles; reduction holds at \qty{1350}{\degreeCelsius} for \qty{1}{\hour} followed by \qty{1300}{\degreeCelsius} for \qty{0.5}{\hour} under flowing \ce{Ar}, and reoxidation holds at \qty{900}{\degreeCelsius} for \qty{0.5}{\hour} under \qty{20}{\percent}~\ce{O2}/\ce{Ar}.
\Dd was computed from the measured mass change as described in main-text Methods.
Cycle-to-cycle drift in the end-of-reduction and end-of-reoxidation $\delta$ values is most pronounced for CCTM2112 (bottom panel); \Dd values reported in main-text \cref{M-tbl:cctm_delta_delta} average over cycles \numrange[range-phrase = --]{2}{5}.
    }
    \label{fgr:tga}
\end{figure}

\Cref{fgr:tga} shows the temperature and oxygen-content cycling traces obtained on the three CCTM compositions during Protocol~B cycling.
Protocol~B uses a two-step reduction at \qty{1350}{\degreeCelsius} (\qty{1}{\hour}) followed by \qty{1300}{\degreeCelsius} (\qty{0.5}{\hour}) under flowing \ce{Ar}, and reoxidation at \qty{900}{\degreeCelsius} (\qty{0.5}{\hour}) under \qty{20}{\percent}~\ce{O2}/\ce{Ar}; full Protocol~A and Protocol~C conditions are described in main-text Methods.
Acquisition was performed on a NETZSCH STA 449~F3 Jupiter thermogravimetric analyzer with \qty{80}{\mg} of CCTM powder loaded in a calcia-stabilized zirconia crucible and heated at \qty{10}{\degreeCelsius\per\minute} to the reduction temperature.

As in main-text Methods, cycle~\num{1} is a preconditioning (break-in) cycle and is excluded from the reported averages, because the as-synthesized material is more oxidized than the steady-state cycled state.
The per-cycle oxygen-exchange capacity is approximately constant across cycles for all three compositions, with a cycle-to-cycle standard deviation no larger than \num{0.0013} (CCTM2112) over cycles \numrange[range-phrase = --]{2}{5}.
Throughout this section the end-of-step $\delta$ values are referenced to the as-loaded (initial) state, so they report the change in oxygen content relative to that state, not the absolute deficiency in \ce{ABO_{3-\delta}}; $\delta$ therefore decreases (becomes negative) as the sample loses oxygen on reduction. The traces plotted in \cref{fgr:tga} show the same referenced $\delta$ shifted so that the most-reduced point of each run is zero. This referenced $\delta$ is distinct from the per-cycle oxygen-exchange capacity \Dd reported in main-text \cref{M-tbl:cctm_delta_delta}, which is a positive quantity.
For CCTM2112 (bottom panel), the end-of-reduction $\delta$ becomes progressively more negative across cycles (from \num{-0.038} in cycle~\num{1} to \num{-0.084} in cycle~\num{5}) and the end-of-reoxidation $\delta$ drifts correspondingly, indicating that the \qty{0.5}{\hour} reoxidation dwell at \qty{900}{\degreeCelsius} under \qty{20}{\percent}~\ce{O2}/\ce{Ar} is insufficient to re-establish the same oxidized state cycle-to-cycle; the per-cycle excursion is essentially unaffected by this drift.
\ce{Ca_{0.75}Ce_{0.25}Ti_{0.50}Mn_{0.50}O3} and \ce{Ca_{0.71}Ce_{0.29}Ti_{0.42}Mn_{0.58}O3} (top and middle panels) show negligible cycle-to-cycle drift in both the reduction and reoxidation endpoints, consistent with steady-state cycling under Protocol~B.
\Dd values reported in main-text \cref{M-tbl:cctm_delta_delta} are averaged over cycles \numrange[range-phrase = --]{2}{5}.

\FloatBarrier

\section{Finite-temperature thermodynamic formalism\label{sec:esi:thermo}}

This section provides the formal link between the \qty{0}{\kelvin} DFT vacancy formation energy \Ev (tabulated throughout the main text) and the equilibrium oxygen deficiency $\delta(T, P)$ used to predict cycle-level performance.
Four components are needed: (i) the free-energy decomposition that allows a temperature-independent \Ev to appear in a finite-temperature occupation expression (\cref{sec:esi:thermo:derivation}); (ii) the equilibrium occupation derived from a two-state lattice-gas model (\cref{sec:esi:thermo:occupation}); (iii) the gas-phase chemical-potential references that supply \muO under reduction and water-splitting conditions (\cref{sec:esi:thermo:references}); and (iv) a numerical example carrying the evaluation from \Ev through to \Dd for CCTM2112 (\cref{sec:esi:thermo:example}).

\subsection{From the \texorpdfstring{\qty{0}{\kelvin}}{0 K} DFT energy to the finite-temperature free energy}\label{sec:esi:thermo:derivation}

Removing one oxygen from a stoichiometric supercell to a gas-phase \ce{O2} reservoir is written
\begin{equation}
    \ce{ABO3}\,(\text{pristine}) \to \ce{ABO}_{3 - 3/N_{\ce{O}}}\,(\text{defective}) + \tfrac{1}{2}\,\ce{O2(g)},
    \label{eqn:vac-reaction}
\end{equation}
for a supercell containing $N_{\ce{O}}$ oxygen atoms.
The associated free-energy change, neglecting vibrational contributions from the solid (harmonic-phonon calculations across oxides show the solid-state vibrational entropy of vacancy formation to be the smaller contribution to the total reduction entropy, with a high-temperature dependence that largely counterbalances that of the \ce{O2} gas entropy\cite{baldassarri_vibrational_2024}), is
\begin{equation}
    \Delta G_{\mathrm{v}}(T, \POtwo) \approx E_{\mathrm{def}}^{\mathrm{DFT}} - E_{\mathrm{prist}}^{\mathrm{DFT}} + \tfrac{1}{2}\,\mu_{\ce{O2}}(T, \POtwo),
    \label{eqn:DeltaGv-raw}
\end{equation}
where $E_{\mathrm{def}}^{\mathrm{DFT}}$ and $E_{\mathrm{prist}}^{\mathrm{DFT}}$ are the DFT total energies of the defective and pristine supercells.
Splitting the gas-phase chemical potential into a DFT reference and an ideal-gas thermal contribution,
\begin{equation}
    \mu_{\ce{O2}}(T, \POtwo) = E_{\ce{O2}}^{\mathrm{DFT}} + \Delta\mu_{\ce{O2}}(T, \POtwo),
    \label{eqn:muO2-split}
\end{equation}
where $E_{\ce{O2}}^{\mathrm{DFT}}$ is the SCAN total energy of an isolated \ce{O2} molecule computed with the same settings as the solids and
\begin{equation}
    \Delta\mu_{\ce{O2}}(T, \POtwo) = G_{\ce{O2}}^{\circ}(T) + \kB T \ln\!\bigl(\POtwo / P^{\circ}\bigr),
    \label{eqn:DeltamuO2}
\end{equation}
with $G_{\ce{O2}}^{\circ}(T)$ the standard Gibbs energy at temperature $T$ from the tables of \cref{sec:esi:thermo:references} and $P^{\circ} = \qty{1}{\bar}$.
Substituting \cref{eqn:muO2-split,eqn:DeltamuO2} into \cref{eqn:DeltaGv-raw} and grouping DFT terms gives
\begin{equation}
    \Delta G_{\mathrm{v}}(T, \POtwo) = \Ev + \muO(T, \POtwo),
    \label{eqn:DeltaGv-final}
\end{equation}
with
\begin{equation}
    \Ev = E_{\mathrm{def}}^{\mathrm{DFT}} - E_{\mathrm{prist}}^{\mathrm{DFT}} + \tfrac{1}{2}\,E_{\ce{O2}}^{\mathrm{DFT}},
    \label{eqn:Ev-DFT}
\end{equation}
the temperature-independent \qty{0}{\kelvin} DFT vacancy formation energy (the quantity tabulated throughout the main text) and
\begin{equation}
    \muO(T, \POtwo) \equiv \tfrac{1}{2}\,\Delta\mu_{\ce{O2}}(T, \POtwo),
    \label{eqn:muO-def}
\end{equation}
the per-atom shift of the oxygen chemical potential relative to the $E_{\ce{O2}}^{\mathrm{DFT}}/2$ reference.
\Cref{eqn:muO-def} is the quantity denoted \muO in main-text Methods \cref{M-eqn:thermal-reduction-muO-split}: the $E_{\ce{O2}}^{\mathrm{DFT}}/2$ reference is implicit there because it cancels with the corresponding term in \Ev (\cref{eqn:Ev-DFT}).
Positive \Ev corresponds to endothermic vacancy formation at \qty{0}{\kelvin}; more negative \muO (lower \POtwo or higher $T$) lowers $\Delta G_{\mathrm{v}}$ and favors vacancy formation, consistent with the expectation that reducing atmospheres promote oxide reduction.

\subsection{Equilibrium occupation from a two-state lattice-gas model}\label{sec:esi:thermo:occupation}

With each oxygen site treated as a distinguishable position on the lattice in equilibrium with the gas-phase reservoir at chemical potential \muO, every site has two states: occupied by \ce{O^2-} (formation energy $0$, particle number $n_{\ce{O}} = 1$) or vacant (formation energy \Ev, $n_{\ce{O}} = 0$), as in standard ideal-solution defect-equilibrium treatments.\cite{lany_chemical_2024}
The single-site grand partition function is then
\begin{equation}
    \Xi = \exp\!\left( \muO / \kB T \right) + \exp\!\left( -\Ev / \kB T \right),
    \label{eqn:partition}
\end{equation}
and the probability that the site is vacant is
\begin{equation}
    n(\Ev, T, \muO) = \frac{\exp(-\Ev / \kB T)}{\Xi}
    = \frac{1}{1 + \exp\!\left[ (\Ev + \muO) / \kB T \right]},
    \label{eqn:occupation}
\end{equation}
the quantity appearing in main-text Methods \cref{M-eqn:occupation}.

The CCTM perovskites are configurationally disordered, and distinct local nearest-neighbor environments $\mathbf{x} = (\xCe, \xMn)$ carry distinct \Ev values.
Each oxygen has four A-site and two B-site nearest-neighbor cations, so \xCe takes values \numlist[input-symbols = {/}]{0;1/4;1/2;3/4;1} and \xMn takes values \numlist[input-symbols = {/}]{0;1/2;1}, yielding \num{15} symmetry-distinct local environments.
Under an ideal-solution approximation for the bulk cation distribution, the probability of finding a given $\mathbf{x}$ at bulk composition $\mathbf{X} = (\XCe, \XMn)$ is binomial,
\begin{equation}\label{eqn:p-binomial}
    P(\mathbf{x} \mid \mathbf{X}) = \binom{4}{4 \xCe} \XCe^{4 \xCe} (1 - \XCe)^{4 (1 - \xCe)} \binom{2}{2 \xMn} \XMn^{2 \xMn} (1 - \XMn)^{2 (1 - \xMn)},
\end{equation}
where the factors of \num{4} and \num{2} are the A- and B-site coordination numbers of each oxygen in the perovskite structure.
The bulk oxygen deficiency is the binomial-weighted average of \cref{eqn:occupation} over the \num{15} environments,
\begin{equation}\label{eqn:delta-avg}
    \delta(T, \muO) = 3 \sum_{\mathbf{x}} P(\mathbf{x} \mid \mathbf{X}) \, n\!\left(\Ev(\mathbf{x}, \mathbf{X}), T, \muO\right),
\end{equation}
where the prefactor of \num{3} accounts for the three oxygen atoms per \ce{ABO3} formula unit.
This expression relies on the independent-site (dilute-vacancy) approximation, whose validity at experimentally accessed $\delta$ is discussed in \cref{sec:esi:vv-electrostatic}, and on the ideal-solution approximation for the bulk cation distribution.\cite{park_accurate_2023}
The SCAN+$U$ and CFM entries of main-text \cref{M-tbl:cctm_delta_delta} are obtained by evaluating \cref{eqn:delta-avg} directly using DFT- and CFM-predicted per-environment \Ev, respectively.
The dGNN entries (main-text \cref{M-fgr:5}b and \cref{M-tbl:cctm_delta_delta}) are obtained by evaluating the corresponding average empirically over the \num{216} oxygen sites of an SQS supercell (\cref{sec:esi:dgnn:screening}); the two estimators agree to within \qty{\pm 4}{\percent} on \Dd across the six DFT compositions of main-text \cref{M-tbl:cctm_compositions}, well inside the spread between the CFM and dGNN.

\subsection{Gas-phase chemical-potential references}\label{sec:esi:thermo:references}

Thermochemical functions for \ce{O2(g)}, \ce{H2(g)}, and \ce{H2O(g)} were obtained from the NIST Chemistry WebBook,\cite{linstrom_nist} which tabulates Shomate-equation coefficients derived from the NIST-JANAF Thermochemical Tables.\cite{chase_nistjanaf_1998}
The Shomate parameterization expresses the standard heat capacity, enthalpy, and entropy as polynomials in $t = T / \qty{1000}{\kelvin}$:
\begin{align}
    C_P^{\circ}(T) &= A + B t + C t^2 + D t^3 + E / t^2, \label{eqn:shomate-Cp} \\
    H^{\circ}(T) - H^{\circ}(\qty{298.15}{\kelvin}) &= A t + \tfrac{B}{2} t^2 + \tfrac{C}{3} t^3 + \tfrac{D}{4} t^4 - E / t + F - H, \label{eqn:shomate-H} \\
    S^{\circ}(T) &= A \ln t + B t + \tfrac{C}{2} t^2 + \tfrac{D}{3} t^3 - E / (2 t^2) + G, \label{eqn:shomate-S}
\end{align}
where $C_P^{\circ}$ and $S^{\circ}$ are in \unit{\J\per\mol\per\kelvin} and $H^{\circ}$ in \unit{\kJ\per\mol}.
The standard Gibbs energy is then
\begin{equation}
    G_i^{\circ}(T) = \Delta_{\mathrm{f}} H_i^{\circ}(\qty{298.15}{\kelvin}) + \bigl[ H_i^{\circ}(T) - H_i^{\circ}(\qty{298.15}{\kelvin}) \bigr] - T\, S_i^{\circ}(T),
    \label{eqn:G-std}
\end{equation}
with $\Delta_{\mathrm{f}} H_i^{\circ}(\qty{298.15}{\kelvin}) = 0$ for \ce{O2} and \ce{H2} in their reference states and $\Delta_{\mathrm{f}} H_{\ce{H2O}}^{\circ}(\qty{298.15}{\kelvin}) = \qty{-241.826}{\kJ\per\mol}$ for \ce{H2O(g)}.
The chemical potential of species $i$ at partial pressure $P_i$ follows from
\begin{equation}
    \mu_i(T, P_i) = G_i^{\circ}(T) + \kB T \ln\!\bigl( P_i / P^{\circ} \bigr).
    \label{eqn:mu-pressure}
\end{equation}
The expressions for \muO under the reduction and water-splitting limits (main-text Methods \cref{M-eqn:thermal-reduction-muO-split,M-eqn:water-splitting-muO-split}) follow from \cref{eqn:mu-pressure} applied to the appropriate species labels.
\Cref{tbl:shomate} lists the coefficient sets used; the temperature ranges shown bracket the reduction (\qtyrange[range-phrase = --, range-units = single]{1123}{1623}{\kelvin}) and water-splitting (\qty{1123}{\kelvin}) conditions of this work, and the corresponding coefficient sets are the only ones evaluated.

\begin{table}[ht]
    \caption{
Shomate-equation coefficients for the gas-phase species used in this work, taken from the NIST Chemistry WebBook (NIST Standard Reference Database 69).\textsuperscript{\emph{a}}
The coefficient sets shown are the only ones evaluated; low-temperature sets are not used at the cycling conditions of this work.
    }
    \label{tbl:shomate}
    \centering
    \small
    \begin{tabular}{l c r r r r r r r r}
        \toprule
        Species & $T$ range (\unit{\kelvin}) & $A$ & $B$ & $C$ & $D$ & $E$ & $F$ & $G$ & $H$ \\
        \midrule
        \ce{O2}  & 700--2000  & 30.032 & 8.773   & $-3.988$ & 0.788    & $-0.742$ & $-11.325$  & 236.166 & 0 \\
        \ce{H2}  & 1000--2500 & 18.563 & 12.257  & $-2.860$ & 0.268    & 1.978    & $-1.147$   & 156.288 & 0 \\
        \ce{H2O} & 500--1700  & 30.092 & 6.833   & 6.793    & $-2.534$ & 0.082    & $-250.881$ & 223.397 & $-241.826$ \\
        \bottomrule
    \end{tabular}
    \\[2pt]
    \footnotesize
    \textsuperscript{\emph{a}}~Units of the resulting quantities: $C_P^{\circ}$ in \unit{\J\per\mol\per\kelvin}, $H^{\circ}$ in \unit{\kJ\per\mol}, $S^{\circ}$ in \unit{\J\per\mol\per\kelvin}.
\end{table}

\subsection{Worked example: CCTM2112 under cycling conditions}\label{sec:esi:thermo:example}

To illustrate the complete procedure, we evaluate \cref{eqn:occupation,eqn:delta-avg} for CCTM2112 at the screening conditions of main-text Methods.

\emph{Reduction.}
At $\Tred = \qty{1623.15}{\kelvin}$ (\qty{1350}{\degreeCelsius}) and $\POtwo = \qty[exponent-mode = scientific]{1e-4}{\bar}$, evaluation of \cref{eqn:shomate-H,eqn:shomate-S,eqn:G-std} with the \ce{O2} row of \cref{tbl:shomate} gives $G_{\ce{O2}}^{\circ}(\Tred) = \qty{-378.5}{\kJ\per\mol}$ (equivalently, \qty{-3.92}{\eV} per molecule).
Inserting this into \cref{eqn:mu-pressure} and halving per \cref{eqn:muO-def},
\begin{equation*}
    \muO(\Tred, \POtwo) = \tfrac{1}{2}\bigl[ G_{\ce{O2}}^{\circ}(\Tred) + \kB \Tred \ln(\POtwo / P^{\circ}) \bigr] = \qty{-2.61}{\eV}.
\end{equation*}

\emph{Water splitting.}
At $\Tox = \qty{1123.15}{\kelvin}$ (\qty{850}{\degreeCelsius}) and $\PHtwoO / \PHtwo = \num{1000}$, the same procedure applied to \ce{H2O} and \ce{H2} gives $G_{\ce{H2O}}^{\circ}(\Tox) = \qty{-4.95}{\eV}$ and $G_{\ce{H2}}^{\circ}(\Tox) = \qty{-1.72}{\eV}$ per molecule, so that
\begin{equation*}
    \muO(\Tox, \PHtwoO, \PHtwo) = G_{\ce{H2O}}^{\circ}(\Tox) - G_{\ce{H2}}^{\circ}(\Tox) + \kB \Tox \ln(\PHtwoO / \PHtwo) = \qty{-2.56}{\eV}.
\end{equation*}

\emph{Occupation and oxygen-exchange capacity.}
CCTM2112 has $\XCe = 1/3$ and $\XMn = 2/3$.
Substituting into \cref{eqn:p-binomial} gives the binomial probability of each of the \num{15} local environments, dominated by $(\xCe, \xMn) \in \{(1/4, 1/2),\, (1/4, 1),\, (1/2, 1/2),\, (1/2, 1)\}$ (i.e., one to two Ce among the four A-site neighbors and one to two Mn among the two B-site neighbors) with $P \in [0.13, 0.20]$ each.
Across the \num{15} sites, the SCAN+$U$ \Ev values span \qtyrange[range-phrase = --, range-units = single]{2.97}{4.44}{\eV} with binomial-weighted mean $\langle \Ev \rangle = \qty{3.49}{\eV}$ (\cref{tbl:esi:cctm2112-env}).
Substituting $\langle \Ev \rangle$ into \cref{eqn:occupation} at the two conditions above gives $n_{\mathrm{red}} = \num{1.8e-3}$ and $n_{\mathrm{ox}} = \num{6.6e-5}$, hence $\delta_{\mathrm{red}} \approx \num{5.4e-3}$, $\delta_{\mathrm{ox}} \approx \num{2.0e-4}$, and $\Dd \approx \num{5.2e-3}$.
This single-mean estimate systematically underestimates \Dd because $n(\Ev)$ is strongly nonlinear in \Ev and the site distribution is broad: environments with $\Ev < \langle \Ev \rangle$ contribute disproportionately.
Evaluating \cref{eqn:delta-avg} over the binomial-weighted environment distribution instead gives $\delta_{\mathrm{red}} = \num{0.052}$, $\delta_{\mathrm{ox}} = \num{8.7e-3}$, and $\Dd = \num{0.044}$, an order of magnitude larger than the single-mean estimate.
This is the value reported in the SCAN+$U$ column of main-text \cref{M-tbl:cctm_delta_delta} for CCTM2112; the CFM and dGNN entries follow the related estimators described after \cref{eqn:delta-avg}.

\begin{table}[ht]
    \centering
    \small
    \caption{
The \num{15} symmetry-distinct oxygen first-nearest-neighbor environments of CCTM2112 ($\XCe = 1/3$, $\XMn = 2/3$): local A-site Ce fraction $\xCe$, local B-site Mn fraction $\xMn$, binomial environment probability $P(\mathbf{x} \mid \mathbf{X})$ from \cref{eqn:p-binomial}, and the representative-site SCAN+$U$ oxygen-vacancy formation energy \Ev.
The binomial-weighted mean is $\langle \Ev \rangle = \qty{3.49}{\eV}$; these are the inputs to the worked example.
    }
    \label{tbl:esi:cctm2112-env}
    \begin{tabular*}{\columnwidth}{@{\extracolsep{\fill}}cccc}
        \hline
        $\xCe$ & $\xMn$ & $P(\mathbf{x} \mid \mathbf{X})$ & \Ev (\unit{\eV}) \\
        \hline
        0   & 0   & 0.022 & 4.43 \\
        0   & 1/2 & 0.088 & 3.69 \\
        0   & 1   & 0.088 & 2.97 \\
        1/4 & 0   & 0.044 & 4.28 \\
        1/4 & 1/2 & 0.176 & 3.46 \\
        1/4 & 1   & 0.176 & 3.02 \\
        1/2 & 0   & 0.033 & 4.42 \\
        1/2 & 1/2 & 0.132 & 3.90 \\
        1/2 & 1   & 0.132 & 3.15 \\
        3/4 & 0   & 0.011 & 4.44 \\
        3/4 & 1/2 & 0.044 & 3.71 \\
        3/4 & 1   & 0.044 & 3.47 \\
        1   & 0   & 0.001 & 4.20 \\
        1   & 1/2 & 0.005 & 3.85 \\
        1   & 1   & 0.005 & 3.52 \\
        \hline
    \end{tabular*}
\end{table}

\section{Defect graph neural network model\label{sec:esi:dgnn}}

The defect graph neural network (dGNN) provides a complementary, data-driven counterpart to the linear crystal-feature model of \cref{sec:esi:cfm-parity}: it uses the same SCAN+$U$ vacancy formation energies but learns the structure--composition--\Ev map without a prescribed functional form, so nonlinear contributions that lie outside the linear form of \cref{eqn:cfm} can be captured empirically.
This section documents the architecture, training and inference protocol, and validation against SCAN+$U$ used to generate the screening predictions of main-text \cref{M-fgr:5}b and the dGNN entries of main-text \cref{M-tbl:cctm_delta_delta}.

\subsection{Architecture and training protocol}\label{sec:esi:dgnn:method}

The dGNN follows the architecture of Witman et al.,\cite{witman_defect_2023} a crystal graph convolutional neural network (CGCNN)\cite{xie_crystal_2018} adapted for site-resolved defect-property prediction.
The model takes a relaxed defect-free host structure and the index of a single crystallographic site as input, and outputs the relaxed vacancy formation energy at that site, without ever constructing or relaxing a defective supercell.
Atomic-site nodes are initialized with one-hot encodings of nine elemental properties (Mendeleev number, atomic weight, melting temperature, covalent radius, electronegativity, ground-state volume per atom, ground-state band gap, ground-state magnetic moment, and space group number) concatenated with a one-hot encoding of the site oxidation state inferred from the host DFT calculation.
Edges are initialized via a radial-basis-function expansion of interatomic distances within a cutoff radius.
After $T = \num{2}$ CGCNN convolution steps, the post-convolution feature vector at the site to be vacated is concatenated with global host descriptors (band gap, effective electron mass, and formation enthalpy of the pristine host) and passed through a single fully connected layer to produce a scalar \Ev.
The model contains \num{\sim 2000} learnable parameters in total; the convolution function, complete node and edge featurizations, and node-level pooling step are specified in Witman et al.\cite{witman_defect_2023}
For the disordered CCTM SQS supercells, the site oxidation states required as input features are inferred from the SCAN+$U$-computed local magnetic moments under the constraint of overall charge neutrality, following the convention of Witman et al.\cite{witman_defect_2023} This assignment shares the physical basis of the analysis in \cref{sec:esi:magmoments} but is carried out within the dGNN featurization of Witman et al., rather than with the explicit moment cutoffs tabulated there.

Training uses $K$-fold cross-validation with $K = \num{10}$, following Witman et al.\cite{witman_defect_2023}
The training set is partitioned into $K$ disjoint folds; for each fold, an independent model is trained, with \qty{10}{\percent} of the remaining training data held out as a within-fold validation set for early stopping based on the mean absolute error (MAE).
At inference, all $K$ models predict \Ev for a given input structure; the ensemble mean is reported as the prediction and the ensemble standard deviation as a heuristic uncertainty estimate, used both to flag high-uncertainty individual predictions and to define the upper- and lower-bound envelopes on the screening surface of main-text \cref{M-fgr:5}b.
The loss, optimizer, learning rate, batch size, and maximum number of epochs follow Witman et al.,\cite{witman_defect_2023} whose architecture and training are implemented in the open-source \texttt{cgcnndefect} \textsc{PyTorch} code (\url{https://github.com/mwitman1/cgcnndefect}); the learning rate was selected to minimize the within-fold validation MAE. The complete training configuration is provided in that code release and in the supplementary information of Douglas et al.\cite{douglas_largescale_2026}

\subsection{Pretraining and fine-tuning datasets}\label{sec:esi:dgnn:datasets}

Two datasets are used: a Base dataset for pretraining, and a fine-tuned (FT) dataset that augments the Base data with a fraction of the CCTM vacancy formation energies computed in this work, a transfer-learning strategy for extending defect-property models to chemistries absent from the pretraining set.\cite{witman_transferlearning_2026}

The Base dataset is the augmented training set assembled by Douglas et al.,\cite{douglas_largescale_2026} merging (i) the binary and ternary oxide vacancy database of Witman et al.\cite{witman_defect_2023} (\num{15} cations across seven crystal systems), (ii) additional quaternary oxides and alloy structures from the validation set of the same study, (iii) the \ce{ABO3} perovskite vacancy database of ref.~\citenum{wexler_factors_2021}, and (iv) additional binary and ternary oxides newly computed by Douglas et al.\cite{douglas_largescale_2026}
The combined Base dataset contains \num{1859} neutral vacancy formation energies (\num{1047} oxygen and \num{812} cation) across \num{253} unique host structures, as compiled in the training-data deposit of Douglas et al.;\cite{douglas_largescale_2026} their Fig.~2a gives the per-cation structure count.
The CCTM chemical system lies outside this collection: although all four CCTM cations (\ce{Ca}, \ce{Ce}, \ce{Ti}, \ce{Mn}) appear individually in the Base data, no four-component \ce{Ca-Ce-Ti-Mn-O} oxide is present, and the B-site \ce{Ti}/\ce{Mn} disorder of CCTM produces nearest-neighbor cation environments that are absent from the Base data, making Base predictions on CCTM a compound-wise extrapolation test.

The FT dataset extends the Base dataset with a randomly drawn \qty{20}{\percent} of the \num{90} CCTM oxygen-vacancy formation energies (\num{18} sites).
The remaining \num{72} CCTM sites are held out from training and serve as the validation set in \cref{sec:esi:dgnn:validation}.

\subsection{Inference and screening}\label{sec:esi:dgnn:screening}

To screen across the $(\XCe, \XMn)$ composition domain, SQS supercells were generated at \num{100} grid points spanning the same domain as the six DFT compositions of main-text \cref{M-tbl:cctm_compositions}, using the SQS construction protocol described in main-text Methods, with three independent realizations at each grid point and a single cell at the four endpoint compositions, which are configurationally ordered.
At each grid point the dGNN predicts \Ev for every oxygen site of the supercell, yielding an \Ev distribution at that composition.
The distribution is then averaged via \cref{eqn:delta-avg} of \cref{sec:esi:thermo} to produce $\delta(T, \muO)$, and the reduction- and oxidation-step occupations are differenced to give \Dd at the cycling conditions of main-text Methods.
The \num{216}-site supercell average evaluates \cref{eqn:delta-avg} empirically, with the empirical site distribution serving as a finite-supercell approximation to the binomial environment distribution; this approximation is the source of the few-percent spread between the dGNN and SCAN+$U$ estimators noted in \cref{sec:esi:thermo:occupation}.
The \Dd surface of main-text \cref{M-fgr:5}b is the result of this calculation evaluated on the \num{100}-point grid; the dGNN entries of main-text \cref{M-tbl:cctm_delta_delta} for off-grid compositions are obtained by linear interpolation of $\Dd(\XCe, \XMn)$ between the four nearest grid points.

\subsection{Validation and predicted \texorpdfstring{\Ev}{Ev} distributions}\label{sec:esi:dgnn:validation}

The accuracy of the dGNN architecture has been characterized on the Base dataset via nested $(K = \num{10}, L = \num{10})$-fold cross-validation by Douglas et al.,\cite{douglas_largescale_2026} who report expected MAEs for oxygen-vacancy formation energies of \qty{0.38}{\eV} under structure-wise hold-out (whole structures withheld from training, with their constituent cations still represented elsewhere in the training set) and \qty{0.75}{\eV} under element-wise hold-out (all structures containing a target cation withheld from training).
CCTM is out of sample by compound but not by cation, so the relevant generalization regime for Base predictions on CCTM falls between these two bounds, and most closely corresponds to the structure-wise regime.

\begin{figure}[htbp]
    \centering
    \includegraphics[width=\linewidth]{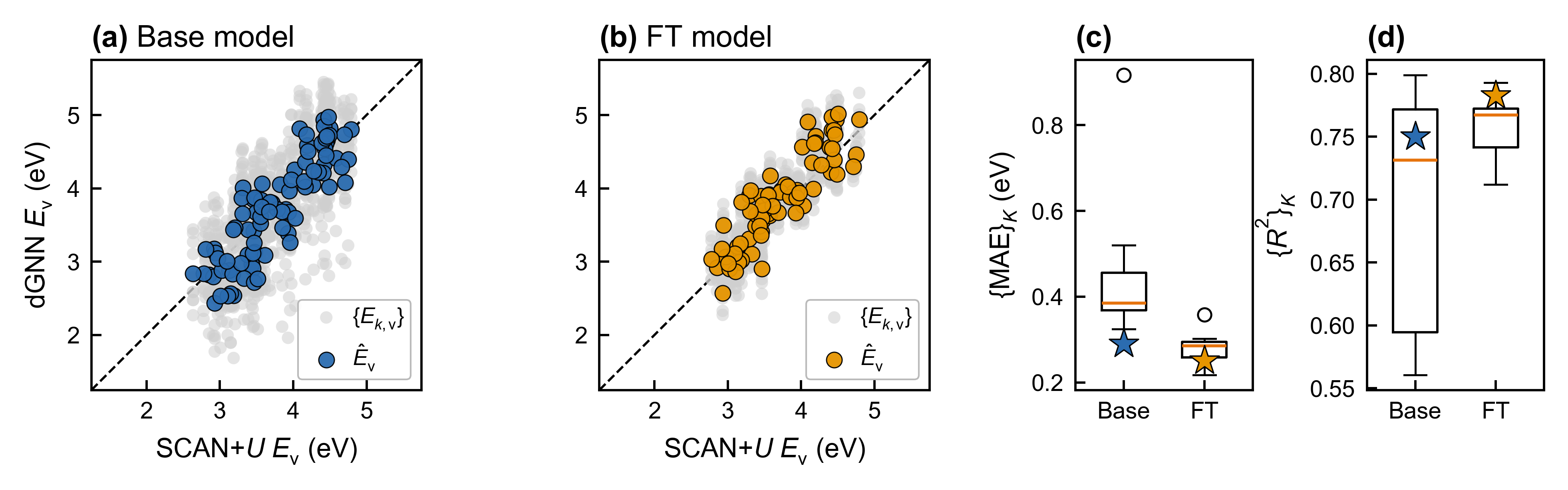}
    \caption{
Performance of dGNN \Ev predictions on the CCTM compositions.
(a)~Base model: SCAN+$U$~vs.~dGNN-predicted \Ev for the \num{90} CCTM oxygen-vacancy sites; gray points are individual $k$-fold predictions $\{E_{k,\mathrm{v}}\}$, blue points the ensemble mean $\hat{E}_\mathrm{v}$.
(b)~Fine-tuned (FT) model: same, evaluated on the \num{72}-site hold-out (the \num{18} FT-training sites are excluded), with ensemble means in orange.
(c)~Per-fold MAE distributions; stars mark the overall ensemble-mean MAE on each model's evaluation set.
(d)~Per-fold Pearson $R^2$ distributions; stars mark the ensemble-mean overall $R^2$.
Dashed lines in (a, b) are the identity.
    }
    \label{fgr:esi:dgnn:parity}
\end{figure}

\Cref{fgr:esi:dgnn:parity}(a, b) compares dGNN-predicted with SCAN+$U$ \Ev for the Base model (evaluated on all \num{90} CCTM sites) and the FT model (evaluated on the \num{72}-site hold-out, excluding the \num{18} FT-training sites).
Panels (c, d) summarize the per-fold MAE and Pearson $R^2$ distributions across the $K = \num{10}$ ensemble for each model.
The Base model achieves an MAE of \qty{0.29}{\eV} on the full \num{90}-site CCTM set with a mean ensemble standard deviation of \qty{0.44}{\eV}, below the structure-wise hold-out expectation of \qty{0.38}{\eV} and thus consistent with the structure-wise hold-out regime.
Fine-tuning reduces the MAE to \qty{0.25}{\eV} on the \num{72}-site hold-out and tightens the mean ensemble standard deviation to \qty{0.16}{\eV}; the FT model overpredicts across the full \Ev range (slightly more at high \Ev), producing a positive overall bias of \qty{+0.14}{\eV}.

\begin{figure}[htbp]
    \centering
    \includegraphics[width=\linewidth]{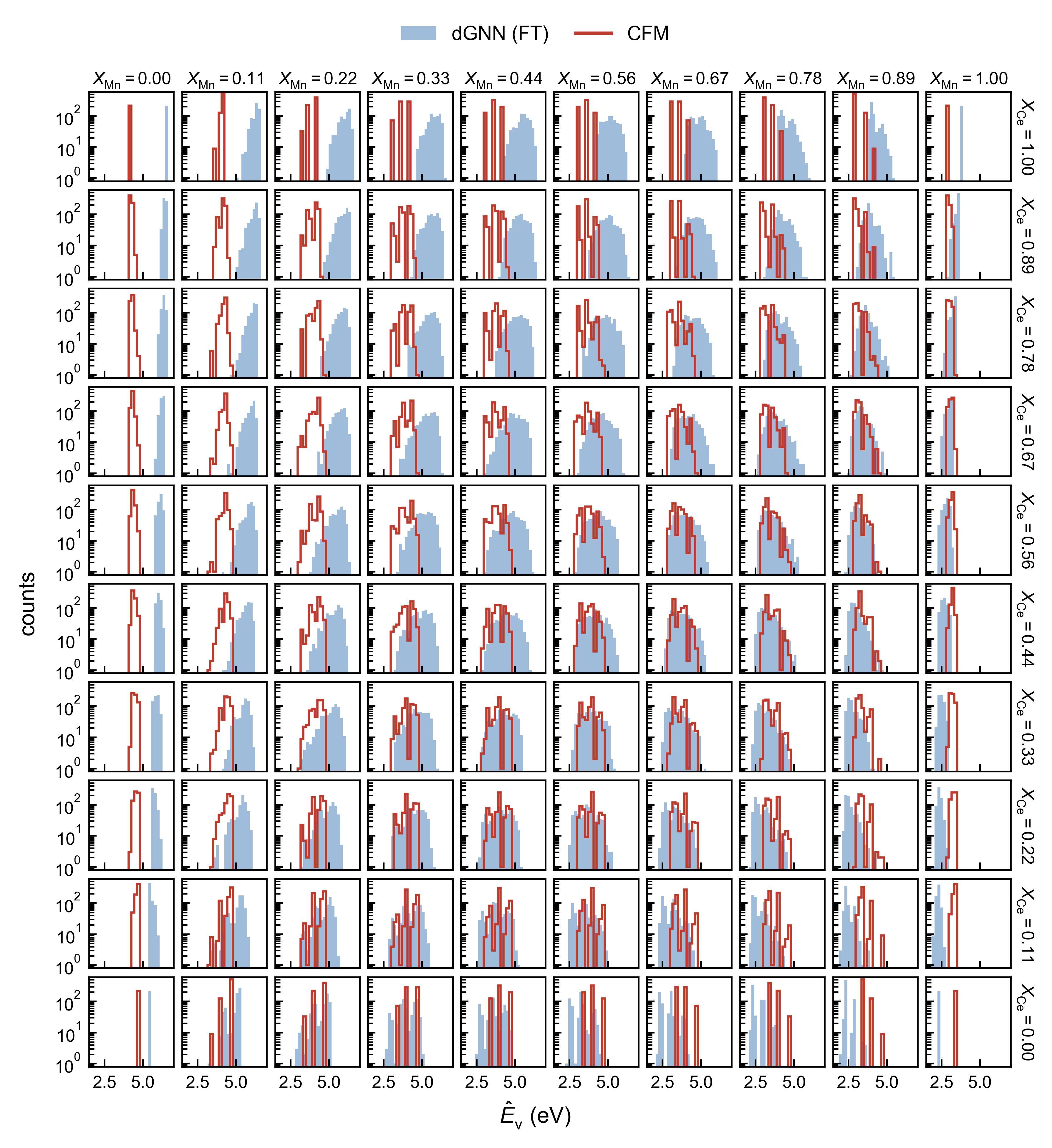}
    \caption{
Per-site \Ev distributions across the $(\XCe, \XMn)$ screening grid (\num{100} points): fine-tuned dGNN ensemble-mean predictions $\hat{E}_\mathrm{v}$ (blue, filled) and CFM predictions (red, outlined).
Each subpanel is a log-scale histogram aggregated over all SQS realizations (\num{\sim 648} oxygen sites at interior grid points, \num{216} at the four endpoint compositions).
CFM evaluation uses the SCAN+$U$-feature parameterization ($\bb = +\num{2.00}$, $\br = -\num{1.34}$, $c = -\num{4.26}$) with fractional \ce{Mn} assignment from bulk charge balance, $\nMn = 4 - \XCe / \XMn$.
The CFM is evaluated on the full grid; in the $\XCe > \XMn$ half-plane $\nMn$ extrapolates below \num{3} and CFM values there reflect extrapolation by the fitted model outside the charge-balanced regime.
    }
    \label{fgr:esi:dgnn:vs:cfm:distributions}
\end{figure}

\Cref{fgr:esi:dgnn:vs:cfm:distributions} compares the per-site \Ev distributions of the fine-tuned dGNN and the CFM across the \num{100}-point composition grid used for screening.
The dGNN distribution mean shifts smoothly with composition, but its spread and the position of its lower tail are nonmonotonic across the $(\XCe, \XMn)$ domain; the nonmonotonic ridge along the \ce{CaTiO3}--\ce{CaMnO3} and \ce{CaMnO3}--\ce{CeMnO3} edges of main-text \cref{M-fgr:5}b is traceable to broader dGNN distributions and lower-\Ev tails in those regions, both of which raise the count-weighted occupation average of \cref{eqn:delta-avg} at fixed mean \Ev.

The CFM and dGNN distributions track one another within the CFM's training composition window ($\XCe \in [\num{0.25}, \num{0.375}]$, $\XMn \in [\num{0.5}, \num{0.75}]$), but the linear CFM compresses predictions into a much narrower range across the broader grid (\qty{2.88}{\eV} to \qty{4.65}{\eV},~vs.~\qty{1.98}{\eV} to \qty{6.50}{\eV} for the dGNN).
The deviation is most pronounced in the $\XCe > \XMn$ half-plane, where the bulk-level \MnIII/\MnIV assignment underlying \cref{eqn:cfm} extrapolates to formally unphysical values (the bulk-charge-balance Mn oxidation state $\nMn = 4 - \XCe/\XMn$ falls below \num{3}; \cref{sec:esi:atomic-props}).
Because the CFM depends on the nearest-neighbor composition alone, its predictions at any one composition take at most \num{15} discrete \Ev values, one per nearest-neighbor cation environment, which appear as the sharp peaks within each subpanel; the dGNN, which incorporates structural context beyond the nearest-neighbor shell of each oxygen site, represents continuous site-to-site variation that the CFM by construction does not.

\begin{figure}[htbp]
    \centering
    \includegraphics[width=\linewidth]{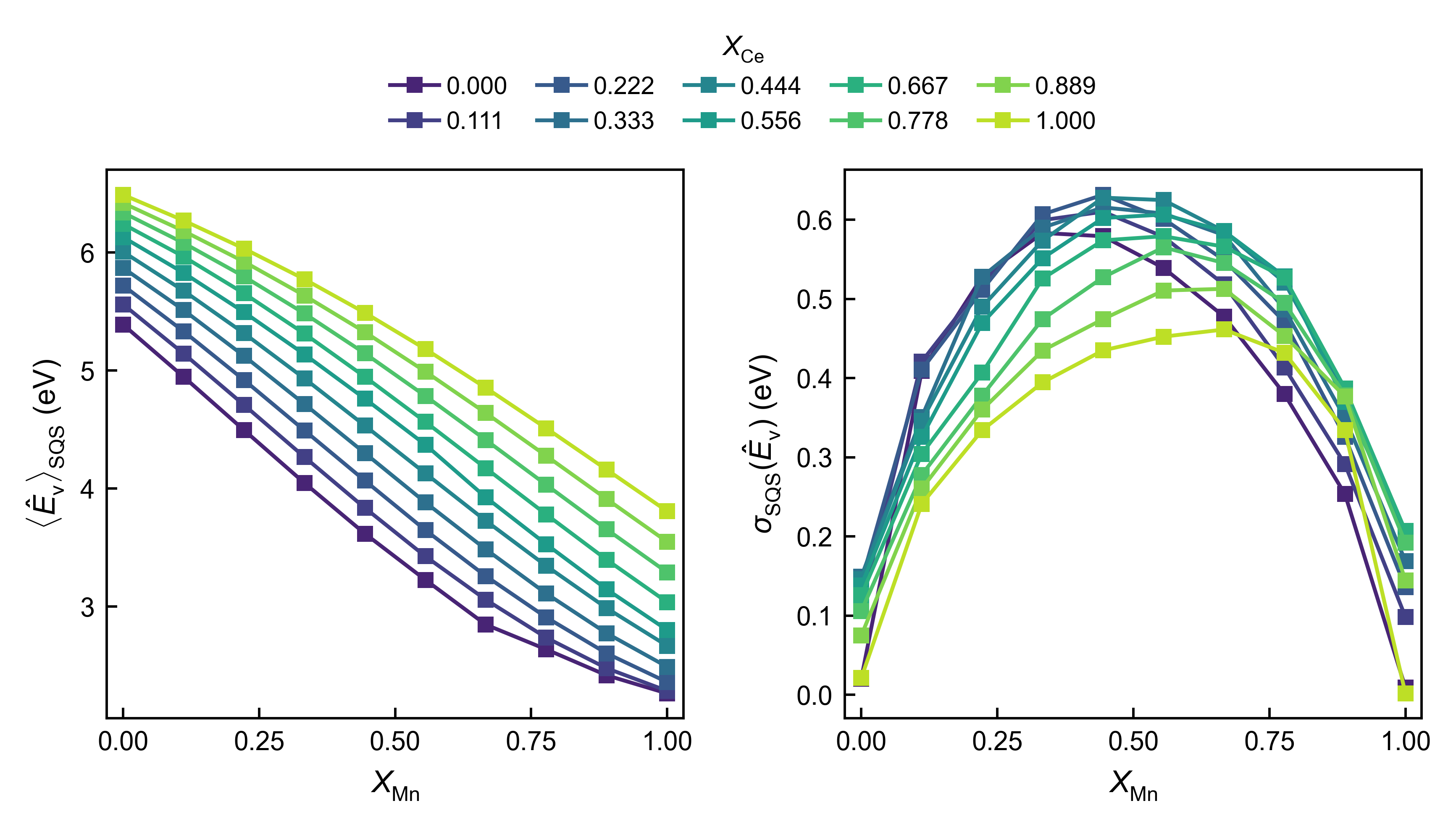}
    \caption{
Summary of per-site fine-tuned dGNN \Ev distributions across the screening grid.
(a)~Mean $\langle \hat{E}_\mathrm{v} \rangle_\mathrm{SQS}$ and (b)~standard deviation $\sigma_\mathrm{SQS}(\hat{E}_\mathrm{v})$~vs.~\XMn, one line per \XCe.
    }
    \label{fgr:esi:dgnn:summary}
\end{figure}

\Cref{fgr:esi:dgnn:summary} condenses the dGNN screening data into line traces.
The SQS-averaged mean \Ev (panel a) decreases monotonically with \XMn for all \XCe, consistent with \ce{Mn} being the reducible cation.
The SQS standard deviation (panel b) peaks at $\XMn \approx \num{0.5}$, where mixed A-/B-site disorder is maximal; the peak is largest at $\XCe \approx \numrange[range-phrase = \text{--}]{0.3}{0.5}$ and shallowest at the $\XCe = \num{1}$ endpoint, tracing out the same broader-distribution regions that drive the ridge in main-text \cref{M-fgr:5}b.

\FloatBarrier

\section{Configurational sensitivity of bulk energetics under SQS sampling}\label{sec:esi:sqs-sensitivity}

The default special quasirandom structure (SQS) objective $Q$ in \textsc{icet}, the package used for SQS generation (main-text Methods), does not constrain which subset of the fifteen symmetry-distinct oxygen first-nearest-neighbor (1NN) environments is realized in a finite supercell.
Different stochastic runs at the same composition can therefore produce different environment subsets and yield different total energies, which introduces an uncontrolled variance into bulk quantities such as \Ehull.
The coverage-constrained protocol used in this work (main-text Methods) removes this environment-subset freedom by penalizing supercells that do not realize all fifteen 1NN environments.

For CCTM2112, the coverage-constrained protocol yields $\Ehull = \qty{33}{\meV\per\atom}$, within the \qtyrange[range-phrase = --, range-units = single]{33}{35}{\meV\per\atom} reported for a standard \textsc{icet} SQS realization of the same composition in ref.~\citenum{wexler_multiple_2023}.
The agreement indicates that the configurational sensitivity of \Ehull at this composition is at most a few \unit{\meV\per\atom}, below the resolution of a single-realization comparison.
The objective $Q$ attains the same minimum value in the coverage-constrained cells as in standard cells: the objective is degenerate, with multiple equally optimal solutions, and the coverage constraint restricts the optimization to the subset of solutions realizing all fifteen 1NN environments.
Coverage is therefore obtained at no cost in the SQS objective $Q$.
Full coverage is also a prerequisite for sampling the complete distribution of local properties, including the lower tail of the site-resolved \Ev distribution used throughout this work.

All six compositions in this work were generated under the coverage-constrained protocol, so cross-composition \Ehull comparisons are internally consistent.

\section{Brillouin-zone sampling convergence}\label{sec:esi:bz-convergence}

\begin{figure}[htbp]
    \centering
    \includegraphics[width=\columnwidth]{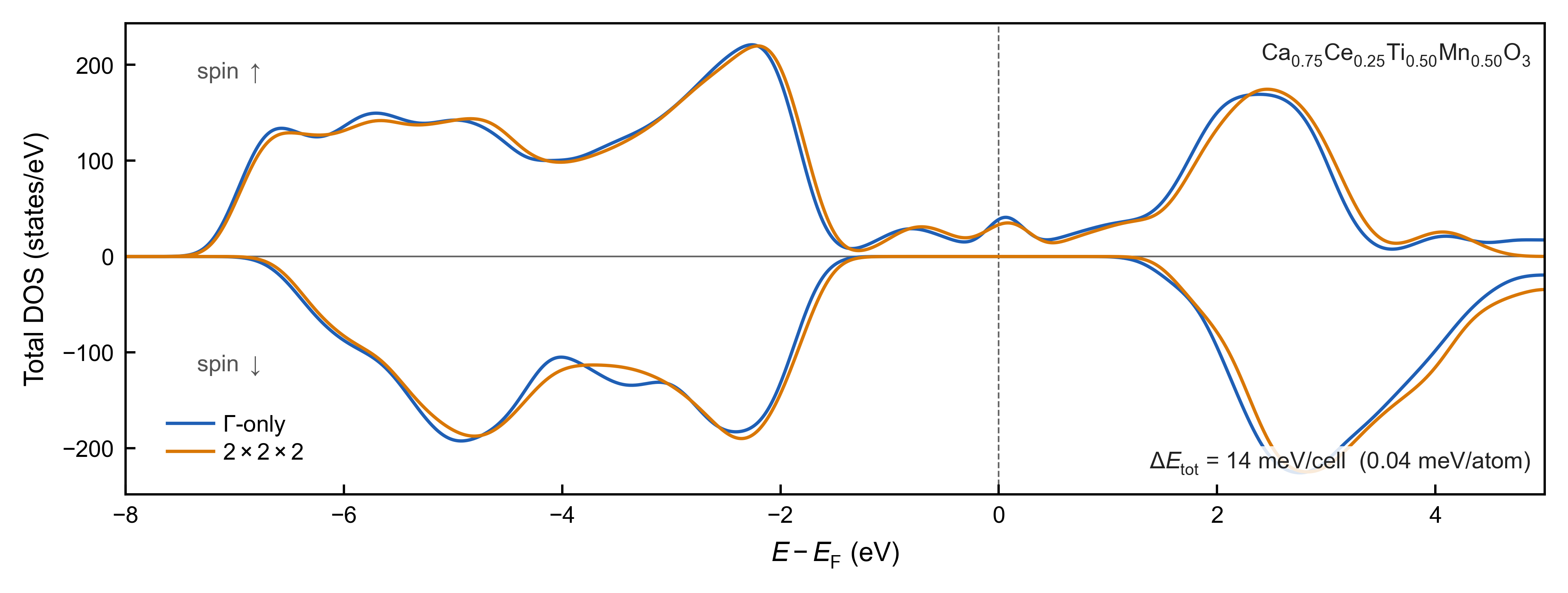}
    \caption{
Spin-resolved total density of states for \ce{Ca_{0.75}Ce_{0.25}Ti_{0.50}Mn_{0.50}O3} at $\Gamma$-only (blue) and at a $\Gamma$-centered $2 \times 2 \times 2$ Monkhorst--Pack mesh (orange), both evaluated as single-point self-consistent calculations on the same $\Gamma$-relaxed geometry.
The dashed vertical line marks the Fermi level.
    }
    \label{fgr:esi:bz-convergence}
\end{figure}

The production calculations sample the Brillouin zone at the supercell zone center ($\Gamma$-only), a choice justified by the size of the \num{360}-atom SQS supercells.
We tested this choice on the $\XCe = \num{0.25}$ composition (\ce{Ca_{0.75}Ce_{0.25}Ti_{0.50}Mn_{0.50}O3}) by performing single-point self-consistent calculations at a $\Gamma$-centered $2 \times 2 \times 2$ Monkhorst--Pack mesh on the $\Gamma$-relaxed geometries.
For the pristine supercell, the denser mesh lowers the total energy by \qty{14}{\meV} per supercell (\qty{0.04}{\meV\per\atom}), one to two orders of magnitude below conventional convergence thresholds for transition-metal oxides.\cite{wexler_factors_2021}
\Cref{fgr:esi:bz-convergence} overlays the corresponding spin-resolved DOS at the two grid densities; the spectral features track closely, with no qualitative change introduced under denser sampling.

Vacancy formation energies \Ev are differences of total energies between a defective supercell, the pristine reference, and the oxygen chemical-potential reference (\cref{sec:esi:thermo}).
The chemical-potential reference is independent of the supercell $k$-grid, so the $k$-sampling sensitivity of \Ev is set by the defective- and pristine-cell total energies alone.
We quantified it directly for the highest-frequency vacancy environment in this composition ($\xCe = \num{0.25}$, $\xMn = \num{0.5}$), recomputing \Ev at the $2 \times 2 \times 2$ mesh on the $\Gamma$-relaxed geometries: \Ev decreases from \qty{3.47}{\eV} ($\Gamma$-only) to \qty{3.41}{\eV} ($2 \times 2 \times 2$), a shift of \qty{62}{\meV}.
Here \xCe and \xMn denote the Ce and Mn fractions among the vacancy's four A-site and two B-site nearest-neighbor cations, in contrast to the bulk mole fractions \XCe and \XMn (main-text \cref{M-fgr:4}).
Because both values use the same $\Gamma$-relaxed geometries, this difference isolates the $k$-sampling contribution at fixed geometry.
Any additional geometry-response contribution, from re-relaxing each cell at the denser mesh, is expected to be smaller still: geometry optimization is variational, so each cell's response is second order in its small mesh-induced displacements, and the two responses partially cancel in \Ev as the fixed-geometry errors do.

This shift exceeds the \qty{14}{\meV} pristine-cell error, so the $k$-sampling errors of the defective and pristine cells do not fully cancel: the defective cell carries the larger error (\qty{77}{\meV} per supercell).
The \qty{62}{\meV} residual is nonetheless small relative to the within-composition site-resolved spread (standard deviation \qty{0.53}{\eV}) and the inter-composition spread in \Ev (\qtyrange[range-phrase = --, range-units = single]{0.1}{1.0}{\eV}), and it does not move this environment out of the \qtyrange[range-phrase = --, range-units = single]{3.4}{3.9}{\eV} target window.
Because the residual is well below the variations examined in the analysis, $\Gamma$-only production sampling does not change the reported trends.

\FloatBarrier

\section{Atomic properties used in CFM descriptors}\label{sec:esi:atomic-props}

The descriptors \Eb and \Vr in \cref{eqn:cfm} are computed at each oxygen vacancy site as unweighted averages over the six first-nearest-neighbor cations of the vacant oxygen, comprising four A-site cations and two B-site cations:
\begin{equation}
    \Eb = \frac{1}{6} \sum_{n=1}^{6} E_{\mathrm{b},n}, \quad \Vr = \frac{1}{6} \sum_{n=1}^{6} V_{\mathrm{r},n}.
    \label{eqn:cfm-averaging}
\end{equation}
The per-cation crystal bond dissociation energy $E_{\mathrm{b},n}$ is the cohesive energy of the ground-state polymorph of the binary oxide of cation $n$ in its assigned oxidation state, divided by the number of \ce{O^{2-}}--cation bonds per formula unit\cite{wexler_factors_2021}.
The per-cation crystal reduction potential $V_{\mathrm{r},n}$ (in \unit{\V} relative to \qty[input-symbols = {/}]{1/4}{\mol} of \ce{O2(g)}) is derived from the reduction energy of the binary oxide containing cation $n$ to the binary oxide containing cation $n$ in its next-lower oxidation state (ref.~\citenum{wexler_factors_2021}).
This formulation modifies the original perovskite CFM of ref.~\citenum{wexler_factors_2021}, which used a sum over all six bonds for \Eb and the maximum of the per-cation values among the nearest neighbors for \Vr, by replacing both with arithmetic means over the six neighbors; the change makes both descriptors site-resolved within the disordered SQS supercell and sensitive to the local cation environment at each symmetry-distinct oxygen site.

\Cref{tbl:esi:bond-energies,tbl:esi:reduction-potentials} list the per-cation $E_{\mathrm{b},n}$ and $V_{\mathrm{r},n}$ values used in the two parameterizations of the parity comparison in \cref{fgr:cfm-parity}.
Both columns of each table are taken from ref.~\citenum{wexler_factors_2021}: the SCAN+$U$ columns from their SCAN+$U$ binary-oxide calculations, the experimental columns from their compilation of binary-oxide thermochemistry (their Fig.~4 and 5, and ref.~107--111 therein).
The A-site cations and Ti are assigned fixed oxidation states (\CaII, \CeIII, \TiIV), whereas Mn is assigned a composition-dependent fractional valence $\nMn = 4 - \XCe/\XMn$ (\numrange[range-phrase = --]{3.43}{3.56} across the six compositions) when selecting $E_{\mathrm{b},n}$ and $V_{\mathrm{r},n}$; its descriptor values are obtained by linearly interpolating the \MnIII and \MnIV binary-oxide references (weights $4 - \nMn$ and $\nMn - 3$, respectively). These assignments serve only to select the binary-oxide reference values and do not otherwise reflect the local electronic state inferred from the SCAN+$U$ calculations.

\begin{table}[ht]
    \centering
    \small
    \caption{
Per-cation crystal bond dissociation energies $E_{\mathrm{b},n}$ used in the \Eb descriptor, following ref.~\citenum{wexler_factors_2021}, eqn~(2).
Both columns are taken from the same work: the SCAN+$U$ column from their SCAN+$U$ binary-oxide calculations, the experimental column from their compilation of binary-oxide thermochemistry.
The \MnIV (\ce{MnO2}) and \MnIII (\ce{Mn2O3}) rows are the endpoints of the fractional-\ce{Mn} interpolation described above.
    }
    \label{tbl:esi:bond-energies}
    \begin{tabular*}{\columnwidth}{@{\extracolsep{\fill}}lcccc}
        \hline
        & & & \multicolumn{2}{c}{$E_{\mathrm{b},n}$ (\unit{\eV})} \\
        \cline{4-5}
        Cation & Oxidation state & Reference oxide & SCAN+$U$ & Experimental \\
        \hline
        Ca  & +2  & \ce{CaO}     & 1.88 & 1.83 \\
        Ce  & +3  & \ce{Ce2O3}   & 2.48 & 2.52 \\
        Ti  & +4  & \ce{TiO2}    & 3.16 & 3.30 \\
        Mn  & +4  & \ce{MnO2}    & 2.25 & 2.25 \\
        Mn  & +3  & \ce{Mn2O3}   & 1.97 & 1.96 \\
        \hline
    \end{tabular*}
\end{table}

\begin{table}[ht]
    \centering
    \small
    \caption{
Per-cation crystal reduction potentials $V_{\mathrm{r},n}$ used in the \Vr descriptor, given for the reduction of the assigned cation to its next-lower stable oxidation state.
Following ref.~\citenum{wexler_factors_2021} (eqn~(5)), values are normalized per electron transferred and expressed relative to \qty[input-symbols = {/}]{1/4}{\mol} of \ce{O2(g)}.
Both columns are taken from the same work: the SCAN+$U$ column from their SCAN+$U$ binary-oxide calculations, the experimental column from their compilation of binary-oxide thermochemistry.
The \ce{Mn^{4+} -> Mn^{3+}} and \ce{Mn^{3+} -> Mn^{2+}} rows are the endpoints of the fractional-\ce{Mn} interpolation described above.
    }
    \label{tbl:esi:reduction-potentials}
    \begin{tabular*}{\columnwidth}{@{\extracolsep{\fill}}lcc}
        \hline
        & \multicolumn{2}{c}{$V_{\mathrm{r},n}$ (\unit{\V}~vs.~\ce{O2})} \\
        \cline{2-3}
        Reduction & SCAN+$U$ & Experimental \\
        \hline
        \ce{Ca^{2+} -> Ca^{0}}    & $-3.28$ & $-3.29$ \\
        \ce{Ce^{3+} -> Ce^{0}}    & $-2.95$ & $-3.11$ \\
        \ce{Ti^{4+} -> Ti^{3+}}   & $-1.91$ & $-1.90$ \\
        \ce{Mn^{4+} -> Mn^{3+}}   & $-0.40$ & $-0.43$ \\
        \ce{Mn^{3+} -> Mn^{2+}}   & $-1.02$ & $-0.98$ \\
        \hline
    \end{tabular*}
\end{table}

\bibliographystyle{rsc}
\bibliography{references}